%% file: jls5.tex
\documentstyle[12pt,a4wide,titlepage,epsfig,amssymb]{article}

\newcommand{\be}{\begin{equation}}
\newcommand{\ee}{\end{equation}}
\newcommand{\E}{{\rm E}}
\newcommand{\constant}{{\rm constant}}
\newcommand{\sign}{{\rm sign}}

\newcommand{\Real}{{\sf Re}}
\newcommand{\rp}{\right)}
\newcommand{\lp}{\left(}
\newcommand{\bea}{\begin{eqnarray}}
\newcommand{\eea}{\end{eqnarray}}
\newcommand{\dj}{Dow Jones Average}

\begin{document}

\title{Crashes as Critical Points}
\author{Anders Johansen$^1$, Olivier Ledoit$^2$ and Didier Sornette$^{1,3,4}$\\
$^1$ Institute of Geophysics and
Planetary Physics\\ University of California, Los Angeles, California 90095\\
$^2$ Anderson Graduate School of Management\\
University of California, Box
90095-1481, 110 Westwood Plaza\\Los Angeles CA 90095-1481\\
$^3$ Department of Earth and Space Science\\
University of California, Los Angeles, California 90095\\
$^4$ Laboratoire de Physique de la Mati\`{e}re Condens\'{e}e\\ CNRS UMR6622 and
Universit\'{e} de Nice-Sophia Antipolis\\ B.P. 71, Parc
Valrose, 06108 Nice Cedex 2, France}

\thispagestyle{empty}
\maketitle

\newpage

\begin{abstract}
We study a rational expectation model of bubbles and crashes. The model
has two components\,: (1) our key assumption is
that a crash may be caused by {\em local} self-reinforcing imitation
between noise traders. If
the tendency for noise traders to imitate their nearest neighbors increases
up to a certain point called the ``critical'' point, all noise traders may
place
the same order (sell) at the same time, thus causing a crash. The interplay
between the progressive strengthening of imitation and the ubiquity of
noise is characterized
by the hazard rate, i.e. the probability per unit time that the crash will
happen
in the next instant if it has not happened yet.
(2) Since the crash is not a certain deterministic outcome of the bubble, it
remains rational for traders to remain invested provided they are
compensated by
a higher rate of growth of the bubble for taking the risk of a crash.
Our model distinguishes between the end of the bubble and the time of the
crash\,:
the rational expectation constraint has the specific
implication that the date of the crash must be random. The theoretical death
of the bubble
is not the time of the crash because the crash
could happen at any time before, even though this is not very likely. The death
of the bubble is the most probable time for the crash. There also
exists a finite probability of attaining the end of the bubble without crash.
Our model has specific
predictions about the presence of certain critical log-periodic patterns in
pre-crash prices, associated with the deterministic components of the
bubble mechanism.
We provide empirical evidence showing that these patterns were indeed present
before the crashes of 1929, 1962 and 1987 on Wall Street and the 1997 crash
on the Hong Kong Stock Exchange. These results are compared with statistical
tests on synthetic data.

\end{abstract}

\newpage
\pagenumbering{arabic}

\section{Introduction}

Stock market crashes are momentous financial events that are fascinating to
academics and practitioners alike. Within the efficient markets literature,
only the revelation of a dramatic piece of information can cause a crash, yet
in reality even the most thorough {\em post-mortem} analyses are typically
inconclusive as to what this piece of information might have been. For traders,
the fear of a crash is a perpetual source of stress, and the onset of the
event itself always ruins the lives of some of them.

Our hypothesis is that stock market crashes are caused by the slow buildup
of long-range correlations leading to a collapse of the stock market in one
critical instant. The use of the word ``critical'' is not purely literary
here: in
mathematical terms, complex dynamical systems can go through so-called
``critical''
points, defined as the explosion to infinity of a normally well-behaved
quantity.
As a matter of fact, as far as nonlinear dynamical systems go, the
existence of
critical points is the rule rather than the exception. Given the puzzling and
violent nature of stock market  crashes, it is worth investigating whether
there could possibly be a link between stock market crashes and critical
points.
The theory of critical phenomena has been recently applied to other
economic models
\cite{A96}.

On what one might refer to as the ``microscopic'' level, the stock market has
characteristics with strong analogies to the field of statistical physics and
complex systems. The individual trader has only $3$ possible actions (or
``states''): selling, buying or waiting. The transformation from one of these
states to another is furthermore a discontinuous process due to a threshold
being
exceeded, usually the price of the stock. The transition involves another
trader
and the process is irreversible, since a trader cannot sell the stock
he bought back to the same trader at the same price. In general, the
individual traders only have information on the action of a limited number
of other traders and generally only see the cooperative response of the
market as a whole in terms of an increase or decrease in the value of the
market. It is thus natural to think of the stock market in terms of complex
systems with analogies to dynamically driven out-of-equilibrium systems
such as earthquakes, avalanches, crack propagation {\it etc.}
A non-negligible difference however involves the reflectivity mechanism\,:
the ``microscopic'' building blocks, the traders, are conscious of their
action. This has been masterly captured by Keynes under the parable of the
beauty contest.

In this paper we develop a model based on the interplay between
economic theory and its rational expectation postulate on one hand
 and statistical physics on the other hand. We find the following major
points.
First, it is entirely possible to build a dynamic model of the stock market
exhibiting well-defined critical points that lies within the strict confines
of rational expectations and is also intuitively appealing. Furthermore, the
mathematical properties of such a critical point are largely independent of
the specific model posited, much more so in fact than the ``regular''
(non-critical) behavior, therefore our key predictions should be relatively
robust to model misspecification. In this spirit, we explore several versions
of the model.
Second, a statistical analysis of the stock
market index in this century shows that the largest crashes may very well be
outliers. This is supported by an extensive analysis of an artificial index.
Third, the predictions made by our critical model are strongly borne out in
several stock market crashes in this century, most notably the U.S.~stock
market crashes of 1929 and 1987 and the 1997 Hong-Kong stock market crash.
Indeed, it is possible to identify clear signatures of near-critical behavior
many years before these crash and use them to ``predict'' (out of sample) the
date where the system will go critical, which happens to coincide very closely
with the realized crash date. Again, we compare our results with an extensive
analysis of an artificial index.

Lest this sound like voodoo science, let us reassure the reader that the
ability to predict the critical date is perfectly consistent with the behavior
of the rational agents in our model: they all know this date, the crash may
happen
anyway, and they are unable to make any abnormal risk-ajusted profits by
using this
information.

The first proposal for a connection between crashes and critical point was
made by Sornette et al. (1996) which identified
precursory patterns as well as characteristic
oscillations of relaxation and aftershock signatures on the Oct. 1987 crash.
These results were later confirmed independently (Feigenbaum and Freund, 1996)
for the 1929 and 1987 crashes and it was in addition
 pointed out that the log-frequency of the observed
log-periodic oscillations seems to decrease as the time of crash is approached.
This was substantiated by the nonlinear renormalization group analysis of
Sornette and Johansen (1997) on the 1929 and 1987 crashes.
Other groups have also taken up the idea
(Vandewalle et al., 1998a) and applied
it to the analysis of the US stock market in 1997
(Feigenbaum and Freund, 1998; Gluzman and Yukalov, 1998; Laloux et al., 1998;
Vandewalle et al., 1998b). These papers
present some of the empirical results that
we provide below. These papers were mostly empirical and did not present any
detailled theoretical model and were based on qualitative arguments
concerning the
analogies between the stock market and critical phenomena in statistical
physics discussed above. By contrast, we provide a rational economic model
explaining how such patterns might arise. This has some fundamental
implications, for example we see that the date of the crash must be random,
therefore it needs not coincide with the date where the critical point is
attained.
This was not apparent in the purely ``physical'' approach taken by the earlier
papers. Furthermore, the empirical study conducted in our paper is
signicantly more detailed
than the earlier ones and includes the most up-to date data as well as
extensive
back-testing to assess the extent to which our empirical evidence for stock
market
crashes as critical points is credible. Sornette and Johansen (1998)
explore a different modelling route using a dynamical rupture analogy on a
hierarchical network of trader connections, with the aim of identifying
conditions
for critical behavior to appear.

The next section presents and solves a rational model of bubbles
and crashes. The third section argues that the mathematical properties so
obtained are general signatures common to all critical phenomena. The fourth
section discusses the empirical evidence for the model. The last section
concludes.

\section{Model}

\subsection{Price Dynamics}
\label{sub:price}

We consider a purely speculative asset that pays no dividends. For
simplicity, we ignore the interest rate, risk aversion, information asymmetry,
and the market-clearing condition. In this dramatically stylised framework,
rational expectations are simply equivalent to the familiar martingale
hypothesis:
\be
\label{eq:martingale}
\forall t'>t\qquad\E_t[p(t')] = p(t)
\ee
where $p(t)$ denotes the price of the asset at time $t$ and $\E_t[\cdot]$
denotes the expectation conditional on information revealed up to time $t$.
Since we will uphold Equation (\ref{eq:martingale}) throughout the paper,
we place ourselves firmly on the side of the efficient markets hypothesis
(Fama; 1970, 1991). If we do not allow the asset price to fluctuate under
the impact of noise,
the solution to Equation (\ref{eq:martingale}) is a constant: $p(t) = p(t_0)$,
where $t_0$ denotes some initial time. Since the asset pays no dividend, its
fundamental value is $p(t)=0$. Thus a positive value of $p(t)$ constitutes
a speculative bubble. More generally, $p(t)$ can be interpreted as the price in
excess of the fundamental value of the asset.

In conventional economics, markets are assumed to be efficient if all
available information is reflected in current market prices. This
constitutes a first-order approximation, from which to gauge and quantify any
observed departures from market efficiency. This approach is based on the
theory
of rational expectation, which was an answer to the previous dominent
theory, the so-called Adaptive
Expectation theory. According to the latter, expectations of
future inflation, for example, were forecast as being an average of past
inflation rates. This effectively meant that changes in expectations would
come slowly with the changing data. In contrast, rational expectation
theory argues
that the aggregate effect of people's forecasts of any variable future behavior
is more than just a clever extrapolation but is close
to an optimal forecast from using all available data. This is true
notwithstanding the fact that individual people forecasts may be relatively
limited\,: the near optimal result of the aggregation process reflects
a subtle cooperative adaptative organization of the market
(Potters et al., 1998). Recently, the
assumption of rational expectations has been called into question by many
economists. The idea of heterogeneous expectations has become of increasing
interest to specialists. Shiller (1989), for example, argues that most
participants in the stock market are not ``smart investors'' (following the
rational expectation model) but rather follow trends and fashions.
Here, we take the view that the forecast of a given trader may be sub-optimal
but the aggregate effect of all forecasts on the market price leads
to its rational expected determination. The no-arbitrage condition
resulting from the
rational expectation theory is more than a useful
idealization, it describes a self-adaptive dynamical state of the market
resulting from the incessant actions of the traders (arbitragers).
It is not the out-of-fashion
equilibrium approximation sometimes described but rather
embodies a very subtle cooperative organization of the market.
Our distinction between individual sub-optimal traders and near-optimal
market is at the basis of our model constructed in two steps.

Now let us introduce an exogenous probability of crash. Formally, let $j$
denote a jump process whose value is zero before the crash and one
afterwards. The cumulative distribution function (cdf) of the time of the
crash is called $Q(t)$, the probability density function (pdf) is
$q(t) = dQ/dt$ and the hazard rate is $h(t) = q(t)/[1-Q(t)]$. The hazard
rate is the
probability
per unit of time that the crash will happen in the next instant if it has not
happened yet. Assume for simplicity that, in case of a crash, the price drops
by a fixed percentage $\kappa\in(0,1)$. Then the dynamics of the asset price
before the crash are given by:
\be
\label{eq:crash}
dp = \mu(t)\,p(t)\,dt-\kappa p(t)dj
\ee
where the time-dependent drift $\mu(t)$ is chosen so that the price process
satisfies the martingale condition, i.e.~$\E_t[dp] = \mu(t)p(t)dt-\kappa
p(t)h(t)dt = 0$. This yields: $\mu(t) = \kappa h(t)$. Plugging it into Equation
(\ref{eq:crash}), we obtain a ordinary differential equation whose solution
is:
\be
\label{eq:price}
\log\left[\frac{p(t)}{p(t_0)}\right] = \kappa\int_{t_0}^th(t')dt'
\qquad\mbox{before the crash}.
\ee
The higher the probability of a crash, the faster the price must increase
(conditional on having no crash) in order to satisfy the martingale condition.
Intuitively, investors must be compensated by the chance of a higher return
in order to be induced to hold an asset that might crash. This is the only
effect that we wish to capture in this part of the model. This effect is
fairly
standard, and it was pointed out earlier in a closely related model of bubbles
and crashes under rational expectations by Blanchard (1979, top of p.389). It
may go against the naive preconception that price is adversely affected by
the probability of the crash, but our result is the only one consistent with
rational expectations. Notice that price is driven by the hazard rate of crash
$h(t)$, which is so far totally unrestricted.

A few additional points deserve careful attention. First, the crash is
modelled
as an exogenous event: nobody knows exactly when it could happen, which
is why the rational traders cannot earn abnormal risk-ajusted profits by
anticipating it.
Second, the probability of a crash itself is an exogenous variable that must
come from outside this first part of the model.  There may or may not be
feedback loop whereby
prices would in turn affect either the arrival or the probability of a crash.
This may not sound totally satisfactory, but it is hard to see how else to
obtain crashes in a rational expectations model: if rational agents could
somehow trigger the arrival of a crash they would choose never to do so, and
if they could control the probability of a crash they would always choose
it to be zero. In our model, the crash is a random event whose probability is
driven by external forces, and {\em once this probability is given} it is
rationally reflected into prices.

\subsection{The Crash}
\label{sub:crash}

The goal of this section is to explain the macro-level probability of a
crash in
terms of micro-level agent behavior, so that we can derive specific
implications pinning down the hazard rate of crash $h(t)$.

We start by a discussion in naive terms. A crash happens when a large group
of agents place sell orders simultaneously. This group of agents must create
enough of an imbalance in the order book for market makers to be unable to
absorb the other side without lowering prices substantially. One curious fact
is that the agents in this group typically do not know each other. They did
not convene a meeting and decide to provoke a crash. Nor do they take orders
from a leader. In fact, most of the time, these agents disagree with one
another, and submit roughly as many buy orders as sell orders (these are all
the times when a crash {\em does not} happen). The key question is: by what
mechanism did they suddenly manage to organise a coordinated sell-off?

We propose the following answer: all the traders in the world are organised
into a network (of family, friends, colleagues, etc) and they influence each
other {\em locally} through this network. Specifically, if I am directly
connected with $k$ nearest neighbors, then there are only two forces that
influence my opinion: (a) the opinions of these $k$ people; and (b) an
idiosyncratic signal that I alone receive. Our working assumption here is that
agents tend to {\em imitate} the opinions of their nearest neighbors, not
contradict them. It is easy to see that force (a) will tend to create
order, while
force (b) will tend to create disorder. The main story that we are telling in
this paper is the fight between order and disorder. As far as asset prices are
concerned, a crash happens when order wins (everybody has the same
opinion: selling), and normal times are when disorder wins (buyers and
sellers disagree with each other and roughly balance each other out). We
must stress that this is exactly the opposite of the popular
characterisation of
crashes as times of chaos.

Our answer has the advantage that it does not require an overarching
coordination mechanism: we will show that macro-level coordination can
arise from micro-level imitation. Furthermore, it relies on a somewhat
realistic model of how agents form opinions. It also makes it easier to accept
that crashes can happen for no rational reason. If selling was a decision that
everybody reached independently from one another just by reading the
newspaper, either we would be able to identify unequivocally the triggering
news after the fact (and for the crashes of 1929 and 1987 this was not the
case), or we would have to assume that everybody becomes irrational in
exactly the same way at exactly the same time (which is distasteful). By
contrast, our reductionist model puts the blame for the crash simply on the
tendency for agents to imitate their nearest neighbors. We do not ask why
agents are influenced by their neighbors within a network: since it is a
well-documented fact (see e.g.~Boissevain and Mitchell, 1973), we take it as
a primitive assumption rather than as the conclusion of some other model of
behavior.\footnote{Presumably some justification for these imitative
tendencies can be found in evolutionary psychology. See Cosmides and
Tooby (1994) on the relationship between evolutionary psychology and
Economics.} Note, however, that there is no information in our model,
therefore what determines the state of an agent is pure noise (see Black,
1986, for a more general discussion of noise traders).

We first present a simple ``mean field'' approach to the imitation problem
and then turn to a more microscopic description.

\subsubsection{Macroscopic modelling}

In the spirit of ``mean field'' theory of collective systems (see for
instance (Goldenfeld, 1992)), the simplest way to describe an imitation
process is
to assume that the hazard rate $h(t)$ evolves according to the following
equation\,:
\be
{dh \over dt} = C~h^{\delta}~,~~~~~~~{\rm with}~\delta > 1~,
\label{azzer}
\ee
where $C$ is a positive constant.
Mean field theory amounts to embody the diversity of trader actions by a
single effective
representative behavior determined from an average interaction between the
traders.
In this sense, $h(t)$ is the collective result of the interactions between
trader.
The term $h^{\delta}$ in the r.h.s. of (\ref{azzer}) models that fact that the
hazard rate will increase or decrease due to the presence of {\em
interactions}
between the traders. The exponent $\delta > 1$ quantifies the effective number
equal to $\delta - 1$ of interactions felt by a typical trader. The condition
$\delta > 1$ is crucial to model interactions and is, as we now show, essential
to obtain a singularity (critical point) in finite time.
Indeed, integrating (\ref{azzer}), we get
\be
h(t) = ({h_0 \over t_c - t})^{\alpha}~,~~~~~~~~{\rm with}~\alpha \equiv {1
\over \delta - 1}~.
\ee
The critical time $t_c$ is determined by the initial conditions at some
origin of time.
The exponent $\alpha$ must lie between zero and one for an economic
reason\,: otherwise,
the price would go to infinity when approaching $t_c$ (if the bubble has
not crashed
in the mean time),
as seen from (\ref{eq:price}). This condition translates into $2 < \delta <
+\infty$\,:
for this theory to make sense, this means that a typical trader must be
connected to more than one
other trader.

It is possible to incorporate a feedback loop whereby prices affect the
probability
of a crash.  The higher the price, the
higher the hasard rate or the increase rate of the crash probability.
This process reflects the phenomenon of a self-fulfilling crisis,
a concept that has recently gained wide attention, in particular with
respect to
the crises that occurred during the past four years in seven countries --
Mexico, Argentina, Thailand, South Korea, Indonesia, Malaysia, and Hong Kong
(Krugman, 1998). They all have experienced severe
economic recessions, worse than anything the United States had seen since
the thirties. It is believed that this is due to the feedback process
associated with the gain and loss of confidence from market investors.
Playing the ``confidence game'' forced these countries into
macroeconomic policies that exacerbated slumps instead of relieving them
(Krugman, 1998).
For instance, when the Asian crisis struck, countries were told to raise
interest
rates, not cut them, in order to persuade some foreign investors to keep
their money in place and thereby limit the exchange-rate plunge.
In effect, countries were told to forget about
macroeconomic policy; instead of trying to prevent or even alleviate the
looming slumps in their economies, they were told to follow policies that
would actually deepen those slumps, all this for fear of speculators.
Thus, it is possible that a loss of confidence in a country can
produce an economic crisis that justifies that loss of confidence--
countries may be vulnerable to what economists call ``self-fulfilling
speculative attacks.'' If investors believe that a crisis may occur in
absence of certain actions, they are surely right,
because they themselves will generate that crisis.
Van Norden and Schaller (1994) have proposed a Markov regime switching
model of speculative behavior whose key feature is similar to ours, namely that
the overvaluation over the fundamental price increases the probability
and expected size of a stock market crash.

Mathematically, in the spirit of the mean field approach,
the simplest way to model this effect is to assume
\be
{dh \over dt} = D~p^{\mu}~,~~~~~~~{\rm with}~\mu > 0~.
\label{azqqzear}
\ee
$D$ is a positive constant. This equation, together with
(\ref{eq:price}), captures the self-fulfilling
feature of speculators\,: their lack of confidence, quantified by $h(t)$,
deteriorates as the market price departs increasingly from the fundamental
value.
As a consequence, the price has to increase further to remunerate
the investors for their increasing risk.
The fact that it is the rate of change of the hazard rate which is
a function of the price may account for panic psychology, namely that
the more one lives in a risky situation, the more one becomes accutely
sensitive
to this danger until one becomes prone to exageration.

Putting (\ref{azqqzear}) in (\ref{eq:price}) gives
\be
{d^2x \over dt^2} = \kappa D~e^{\mu x}~,
\ee
where $x \equiv \ln p$. Its solution is, for large $x$,
\be
x = {2 \over \mu} ~\ln \biggl( {\sqrt{\mu \kappa/2} \over t_c - t}\biggl)~.
\label{llskqllql}
\ee
The log-price exhibits a weak logarithmic singularity of the type proposed
by Vandewalle et al. (1998). This solution (\ref{llskqllql}) is the formal
analytic continuation of the general solution (\ref{eq:solution}) below for
$\beta \to 0$.

\subsubsection{Microscopic modelling}

Consider a network of agents: each one is indexed by an integer
$i=1,\dots,I$, and $N(i)$ denotes the set of the agents who are directly
connected to agent $i$ according to some graph. For simplicity, we assume
that agent $i$ can be in only one of two possible states: $s_i\in\{-1,+1\}$.
We could interpret these states as ``buy'' and ``sell'', ``bullish'' and
``bearish'',
``calm'' and ``nervous'', etc, but we prefer to keep the discussion about
imitation at a general level for now. We postulate that the state of trader
$i$ is determined
by:
\be
\label{eq:state}
s_i  =  \sign\left(K\sum_{j\in N(i)}s_j+\sigma \varepsilon_i\right)
\ee
where the $\sign(\cdot)$ function is equal to $+1$ (to $-1$) for positive
(negative) numbers, $K$ is a positive constant, and $\varepsilon_i$ is
independently
distributed according to the standard normal distribution. This equation
belongs to
the class of stochastic dynamical models of interacting particles
(Liggett, 1985, 1997), which have been much studied mathematically in the
context of
physics and biology.

In this model (\ref{eq:state}), the tendency
towards imitation is governed by $K$, which is called the coupling strength;
the tendency towards idiosyncratic behavior is governed by $\sigma$. Thus
the value of $K$ relative to $\sigma$ determines the outcome of the battle
between order and disorder, and eventually the probability of a crash. More
generally, the coupling strength $K$ could be heterogeneous across pairs of
neighbors, and it would not substantially affect  the properties of the model.
Some of the $K_{ij}$'s could even be negative, as long as the average of all
$K_{ij}$'s was strictly positive.

Note that Equation (\ref{eq:state}) only describes the state of an agent at
a given point in time. In the next instant, new $\varepsilon_i$'s are
drawn, new
influences propagate themselves to neighbors, and agents can change states.
Thus, the best we can do is give a statistical description of the states. Many
quantities can be of interest. In our view, the one that best describes the
chance that a large group of agent finds itself suddenly in agreement is
called
the {\em susceptibility} of the system. To define it formally, assume that a
global influence term $G$ is added to Equation (\ref{eq:state}):
\be
\label{eq:state2}
s_i  = \sign\left(K\sum_{j\in N(i)}s_j+\sigma\varepsilon_i+G\right).
\ee
This global influence term will tend to favour state $+1$ (state $-1$) if $G>
0$ (if $G<0$). Equation (\ref{eq:state}) simply corresponds to the special case
$G=0$: no global influence. Define the average state as
$M = (1/I)\sum_{i=1}^Is_i$. In the absence of global influence, it is easy to
show by symmetry that $\E[M]=0$: agents are evenly split between the two
states. In the presence of a positive (negative) global influence, agents
in the
positive (negative) state will outnumber the others: $\E[M]\times G\geq0$.
With this notation, the susceptibility of the system is defined as:
\be
\label{eq:susceptibility}
\chi = \left.\frac{d(E[M])}{dG}\right|_{G=0}
\ee
In words, the susceptibility measures the sensitivity of the average state
to a
small global influence. The susceptibility has a second interpretation as (a
constant times) the variance of the average state $M$ around its expectation
of zero caused by the random idiosyncratic shocks $\varepsilon_i$. Another
related
interpretation is that, if you consider two agents and you force the first
one to
be in a certain state, the impact that your intervention will have on the
second agent will be proportional to $\chi$. For these reasons, we believe
that the susceptibility correctly measures the ability of the system of agents
to agree on an opinion. If we interpret the two states in a manner relevant
to
asset pricing, it is precisely the emergence of this global synchronisation
from local imitation that can cause a crash.
Thus, we will characterise the behavior of the susceptibility, and we will
posit that the hazard rate of crash follows a similar process. We do not want
to assume a one-to-one mapping between hazard rate and susceptibility
because there are many other quantities that provide a measure of the degree
of coordination of the overall system, such as the correlation length
(i.e.~the
distance at which imitation propagates) and the other moments of the
fluctuations of the average opinion. As we will show in the next section, all
these quantities have the same generic behavior.

\subsection{Interaction Networks}

It turns out that, in the imitation model defined by Equation (\ref{eq:state}),
the structure of the network affects the susceptibility. We propose two
alternative network structures for which the behavior of susceptibility is
well understood. In the next section, we will show how the results we get for
these particular networks are in fact common to a much larger class of
models.

\subsubsection{Two-Dimensional Grid}
\label{subsub:grid}

As the simplest possible network, let us assume that agents are placed on a
two-dimensional grid in the Euclidean plane. Each agent has four nearest
neighbors: one to the North, one
to the South, the East and the West. The relevant parameter is
$K /\sigma$. It measures the tendency towards imitation relative to
the tendency towards idiosyncratic behavior. In the context of the alignment
of atomic spins to create magnetisation, this model is related to the
so-called
two-dimensional Ising model which has been solved explicitly by Onsager
(1944). There exists a critical point $K_c$ that determines the
properties of the system. When $K<K_c$, disorder reigns: the
sensitivity to a small global influence is small, the clusters of agents
who are
in agreement remain of small size, and imitation only propagates between
close neighbors. Formally, in this case, the susceptibility $\chi$ of the
system
is finite. When $K$ increases and gets close to $K_c$, order
starts to appear: the system becomes extremely sensitive to a small global
perturbation, agents who agree with each other form large clusters, and
imitation propagates over long distances. In the Natural Sciences, these are
the characteristics of so-called {\em critical} phenomena. Formally, in this
case the susceptibility $\chi$ of the system goes to infinity.  The
hallmark of
criticality is the {\em power law}, and indeed the susceptibility goes to
infinity according to a power law:
\be
\label{eq:power}
\chi\approx A(K_c-K)^{-\gamma}.
\ee
where $A$ is a positive constant and $\gamma>0$ is called the {\em critical
exponent} of the susceptibility (equal to $7/4$ for the 2-d Ising model).

We do not know the dynamics that drive the key parameter of the system
$K$. At this stage of the enquiry, we would like to just assume that it
evolves smoothly, so that we can use a first-order Taylor expansion around
the critical point. $K$ need not even be deterministic: it could very
well be a stochastic process, as long as it moves slowly enough. Let us call
$t_c$ the first time such that $K(t_c)=K_c$. Then prior to the
critical date $t_c$ we have the approximation: $K_c-
K(t)\approx\constant\times(t_c-t)$.
Using this approximation, we posit that the hazard rate of crash behaves in
the same way as the susceptibility (and all the other measures of
coordination between noise traders) in the neighborhood of the critical point.
This yields the following expression:
\be
\label{eq:hazard2}
h(t)\approx B\times(t_c-t)^{-\alpha}
\ee
where $B$ is a positive constant. The exponent $\alpha$ must lie between
zero and one for an economic reason: otherwise, the price would go to
infinity when approaching $t_c$ (if the bubble has not crashed yet). The
probability per unit of time of having a crash in the next instant conditional
on not having had a crash yet becomes unbounded near the critical date
$t_c$.

We stress that $t_c$ is not {\em the} time of the crash because the crash
could happen at any time before $t_c$, even though this is not very likely.
$t_c$ is the mode of the distribution of the time of the crash, i.e. the
most probable value. In addition, it is
easy to check that there exists a residual probability $1-Q(t_c)>0$ of
attaining the critical date without crash. This residual probability is
crucial
for the rational expectations hypothesis, because otherwise the whole model
would unravel because rational agents would anticipate the crash. Other than
saying that there is some chance of getting there, our model does not
describe what happens at and after the critical time. Finally, plugging
Equation (\ref{eq:hazard2}) into Equation (\ref{eq:price}) gives the
following law for price:
\be
\label{eq:solution}
\log[p(t)]\approx\log[p_c]-\frac{\kappa B}{\beta}\times(t_c-t)^{\beta}
\qquad\mbox{before the crash}.
\ee
where $\beta = 1-\alpha\in(0,1)$ and $p_c$ is the price at the critical
 time (conditioned on no crash having been triggered). We
see that the logarithm of the price before the crash also follows a power law.
It has a finite upper bound $\log[p_c]$. The slope of the logarithm of price,
which is the expected return per unit of time, becomes unbounded as we
approach the critical date. This is to compensate for an unbounded
probability of crash in the next instant.

\subsubsection{Hierarchical Diamond Lattice}

The stock market constitute an ensemble of inter-actors which differs in
size by many orders of magnitudes ranging from individuals to gigantic
professional investors, such as pension funds. Furthermore, structures at
even higher levels, such as currency influence spheres (US\$, DM, YEN ...),
exist and with the current globalization and de-regulation of the market
one may argue that structures on the largest possible scale, {\it i.e.},
the world economy, are beginning to form. This means that the structure
of the financial markets have features, which resembles that of hierarchical
systems and with ``traders'' on all levels of the market. Of course, this
does not imply that any strict hierarchical structure of the stock market
exists, but there are numerous examples of qualitatively hierarchical
structures in society. In fact, one may say that horizontal organisations
of individuals are rather rare. This means that the plane network used in
the previous section may very well represent a gross over-simplification.

\begin{figure}
\begin{center}
\epsfig{file=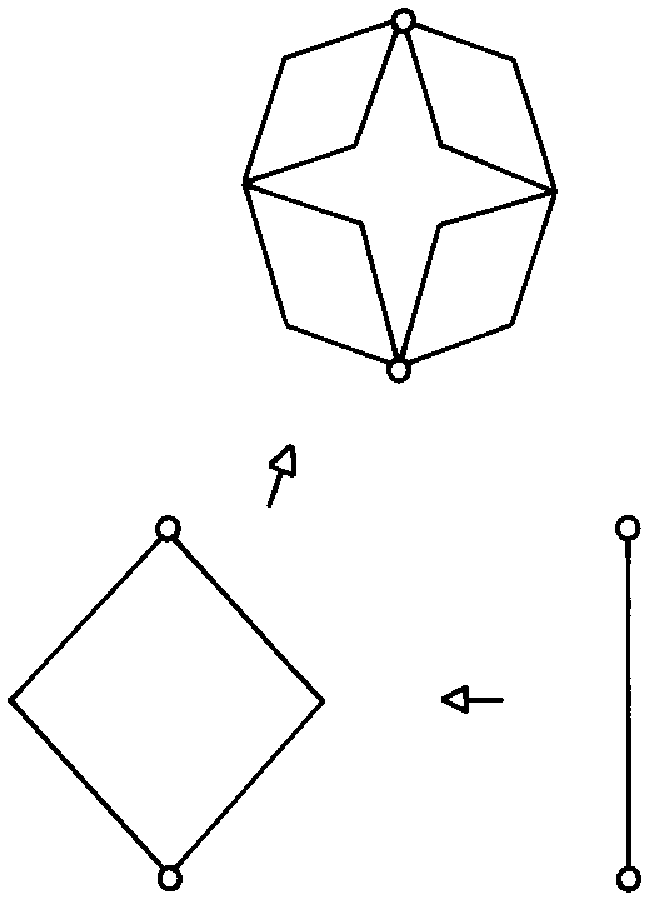}
\end{center}
\caption[]{\protect\label{fig:diamond} The first three steps of the recursive
construction of the hierarchical Diamond Lattice.}
\end{figure}

Another network structure for which our local imitation model has been
solved is the following. Start with a pair of traders who are linked to each
other. Replace this link by a diamond where the two original traders occupy
two diametrically opposed vertices, and where the two other vertices are
occupied by two new traders. This diamond contains four links. For each one
of these four links, replace it by a diamond in exactly the same way, and
iterate the operation. The result is a diamond lattice (see Figure
\ref{fig:diamond}). After $p$ iterations, we have and $\frac{2}{3}(2+4^p)$
traders and $4^p$ links between them. Most traders have only two
neighbors, a few traders (the original ones) have $2^p$ neighbors, and the
others are in between. Note that the least-connected agents have $2^{p-1}$
times fewer neighbors than the most-connected ones, who themselves have
approximately $2^p$ fewer neighbors than there are agents in total. This
may be a more realistic model of the complicated network of
communications between financial agents than the grid in the Euclidean
plane of Section \ref{subsub:grid}.

A version of this model was solved by Derrida et al.~(1983). The basic
properties are similar to the ones described above: there exists a critical
point
$K_c$; for $K<K_c$ the susceptibility is finite; and it
goes to infinity as $K$ increases towards $K_c$. The only
difference -- but it is an important one -- is that the critical exponent
can be a
complex number. The general solution for the susceptibility is a sum of
terms like the one in Equation (\ref{eq:power}) with complex exponents.
The first order expansion of the general solution is:
\begin{eqnarray}
\chi&\approx&\Real[A_0(K_c-K)^{-\gamma}
+A_1(K_c-K)^{-\gamma+i\omega}+\dots]\\
&\approx&A_0'(K_c-K)^\gamma+A_1'(K_c-
K)^\gamma\cos[\omega\log(K_c-K)+\psi]+\dots
\label{eq:fourier}
\end{eqnarray}
where $A_0'$, $A_1'$, $\omega$ and $\psi$ are real numbers, and
$\Real[\cdot]$ denotes the real part of a complex number. We see that the
power law is now corrected by oscillations whose frequency explodes as we
reach the critical time. These accelerating oscillations are called ``log-
periodic'', and $\frac{\omega}{2\pi}$ is called their ``log-frequency''.
There are many physical phenomena where they decorate the power law (see
Sornette (1998) for a review). Following the same steps as in Section
\ref{subsub:grid}, we can back up the hazard rate of a crash:
\be
\label{eq:hazard3}
h(t)\approx B_0(t_c-t)^{-\alpha}
+B_1(t_c-t)^{-\alpha}\cos[\omega\log(t_c-t)+\psi'].
\ee
Once again, the hazard rate of crash explodes near the critical date, except
that now it displays log-periodic oscillations. Finally, the evolution of the
price  before the crash and before the critical date is given by:
\be
\label{eq:complex}
\log[p(t)]\approx\log[p_c]-\frac{\kappa}{\beta}\left\{
B_0(t_c-t)^{\beta}
+B_1(t_c-t)^{\beta}\cos[\omega\log(t_c-t)+\phi]\right\}
\ee
where $\phi$ is another phase constant. The key feature is that oscillations
appear in the price of the asset just before the critical date. The local
maxima
of the function are separated by time intervals that tend to zero at the
critical
date, and do so in geometric progression, i.e.~the ratio of consecutive time
intervals is a constant
\be
\lambda \equiv e^{2 \pi \over \omega}~.
\ee
This is very useful from an empirical point
of view because such oscillations are much more strikingly visible in actual
data than a simple power law: a fit can ``lock in'' on the oscillations which
contain information about the critical date $t_c$. If they are present,
they can
be used to predict the critical time $t_c$ simply by extrapolating frequency
acceleration. Since the probability of the crash is highest near the critical
time, this can be an interesting forecasting exercise. Note that, for rational
traders in our model, such forecasting is useless because they already know
the hazard rate of crash $h(t)$ at every point in time (including at
$t_c$), and
they have already reflected this information in prices through Equation
(\ref{eq:price}).

\section{Generalisation}
\label{sec:general}

Even though the predictions of the previous section are quite detailed, we
will try to argue in this section that they are very robust to model
misspecification. We claim that models of crash that combine the following
features:
\begin{enumerate}
\item A system of noise traders who are influenced by their neighbors;
\item Local imitation propagating spontaneously into global cooperation;
\item Global cooperation among noise traders causing crash;
\item Prices related to the properties of this system;
\item System parameters evolving slowly through time;
\end{enumerate}
would display the same characteristics as ours, namely prices following a
power law in the neighborhood of some critical date, with either a real or
complex critical exponent. What all models in this class would have in
common is that the crash is most likely when the locally imitative system
goes through a {\em critical} point.

In Physics, critical points are widely considered to be one of the most
interesting
properties of complex systems. A system goes critical when local influences
propagate over long distances and the average state of the system becomes
exquisitely sensitive to a small perturbation, {\it i.e.} different parts of
the system becomes highly correlated. Another characteristic is that
critical systems are self-similar across scales: in our example, at the
critical
point, an ocean of traders who are mostly bearish may have within it several
islands of traders who are mostly bullish, each of which in turns surrounds
lakes of bearish traders with islets of bullish traders; the progression
continues all the way down to the smallest possible scale: a single trader
(Wilson, 1979). {\em Intuitively speaking, critical self-similarity is why
local imitation cascades through the scales into global coordination.}

Because of scale invariance (Dubrulle et al., 1997),
the behavior of a system near its critical point
must be represented by a power law (with real or complex critical exponent):
it is the only family of functions that are homogeneous, i.e.~they remain
unchanged (up to scalar multiplication) when their argument gets rescaled by
a constant. Mathematically, this means that scale invariance and critical
behavior is intimately associated with the following equation:
\be
\label{eq:scale}
F(\gamma x)  =  \delta F(x)
\ee
where $F$ is a function of interest (in our example, the susceptibility),
$x$ is
an appropriate parameter, and $\delta$ is a positive constant that describes
how the properties of the system change when we re-scale the whole system
by the factor $\gamma$. It is easy to verify that the general solution to
Equation (\ref{eq:scale}) is:
\be
\label{eq:rg}
F(x) = x^{\log(\delta)/\log(\gamma)}\,\pi\left[\frac{\log(x)}{\log(\gamma)
}
\right]
\ee
where $\pi$ is a periodic function of period one. Equation (\ref{eq:fourier}
)
is nothing but the terms of order 0 and 1 in the Fourier expansion of the
periodic function $\pi$.

In general, physicists study critical points by forming equations such as
(\ref{eq:scale}) to describe the behavior of the system across different
scales, and by analysing the mathematical properties of these equations. This
is known as {\em renormalisation group theory} (Wilson, 1979), and its
introduction was the major breakthrough in the understanding of critical
points (``renormalisation'' refers to the process of rescaling, and ``group''
refers to the fact that iterating Equation (\ref{eq:rg}) generates a similar
equation with the rescaling factor $\gamma^2$).\footnote{Rigorously
speaking, this only constitutes a {\em semi}-group because the inverse
operation is not defined.} Before renormalisation group theory was invented,
the fact that a system's critical behavior had to be correctly described at
all scales simultaneously prevented standard approximation methods from giving
satisfactory results. But renormalisation group theory turned this liability
into an asset by building its solution precisely on the self-similarity of the
system across scales. Let us add that, in spite of its conceptual elegance,
this method is nonetheless quite challenging mathematically.

For our purposes, however, it is sufficient to keep in mind that the key idea
of the paper is the following: the massive and unpredictable sell-off occuring
during stock market crashes comes from local imitation cascading through
the scales into global cooperation when a complex system approaches its
critical point. Regardless of the particular way in which this idea is
implemented, it will generate the same universal implications, which are the
ones in Equations (\ref{eq:solution}) and (\ref{eq:complex}).

Strictly speaking, these equations are approximations valid only in the
neighborhood of the critical point. Sornette and Johansen (1997) propose a
more general formula with additional degrees of freedom to better capture
behavior away from the critical point. The specific way in which these
degrees of freedom are  introduced is based on a finer analysis of the
renormalisation group theory that is equivalent to including the next term in
a systematic expansion around the critical point. It is given by:
\be
\label{eq:nonlinear}
\log\left[\frac{p_c}{p(t)}\right]\approx
\frac{(t_c-t)^\beta}{
\sqrt{1+\left(\frac{t_c-t}{\Delta_t}\right)^{2\beta}}}
\left\{B_0+B_1\cos\left[\omega\log(t_c-t)+\frac{\Delta_\omega}{2\beta}
\log\left(1+\left(\frac{t_c-
t}{\Delta_t}\right)^{2\beta}\right)+\phi\right]\right\}
\ee
where we have introduced two new parameters: $\Delta_t$ and
$\Delta_\omega$. Two new effects are included in this equation:
(i) far from the critical point the power law tapers off; and (ii) the
log-frequency shifts from
$\frac{\omega+\Delta_\omega}{2\pi}$ to $\frac{\omega}{2\pi}$ to as we
approach the critical point. Both transitions take place over the same time
interval of length $\Delta_t$. Thus, in addition to the critical regime
defined
by Equation (\ref{eq:complex}), we allow for a pre-critical regime where
prices oscillate around a constant level with a different log-periodicity
(see Sornette and Johansen, 1997, for the formal derivation). This
generalisation describe prices over a longer pre-crash period than the
original formula, but
it still captures exactly the same underlying phenomenon: critical
self-organisation. Note that for small $\frac{t_c-t}{\Delta_t}$ Equation
(\ref{eq:nonlinear}) boils down to Equation (\ref{eq:complex}).

Before we continue with the empirical results obtained from analysis of
stock market data, we would like to stress that replacing equation
(\ref{eq:complex}) with equation (\ref{eq:nonlinear}) {\em does not}
correspond to simply increasing the number of free variables in the function
describing the time evolution of the stock market index. Instead, we are
including the {\em next order term} in the analytical expansion of the
solution (\ref{eq:rg}) to equation (\ref{eq:scale}), which is something
quite different. As we shall see in section \ref{stockfit}, this has some
rather restrictive implications.

\section{Empirical Results}

\subsection{Large Crashes are Outliers}
\label{outlier}

\begin{figure}

\input{newestdd1}
\caption[]{\label{newdd1} Number of times a given level of draw down has been
observed in
this century in the Dow Jones Average. The bin-size is 1\%. A threshold of
1\% has been applied. The fit is equation (\protect\ref{zae}) with
$N_0 = 2789$ and DD$_c \approx 0.018$.}
\end{figure}
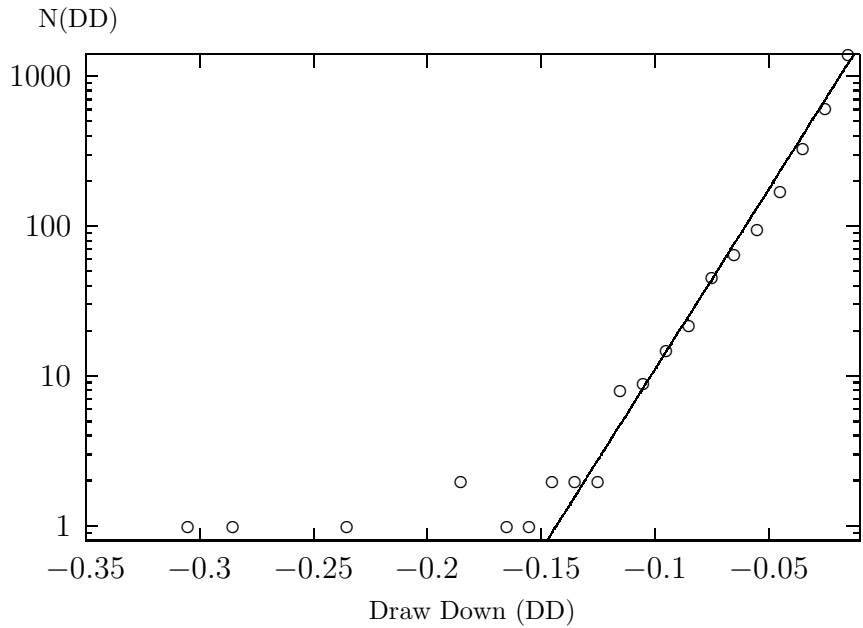

In figure \ref{newdd1}, we see the distribution of draw downs (continuous
decreases) in the closing value of the \dj \ larger than $1$\% in the period
1900-94. The distribution resembles very much that of an exponential
distribution while three events are standing out. The derivation of the
exponential distribution of draw downs is given in appendix A.
If we fit the distribution of draw downs DD larger
than $1$\% by an exponential law, we find
\be \label{zae}
N(\mbox{DD}) = N_0 ~e^{-\left|\mbox{DD}\right|/\mbox{DD}_c},
\ee
where the total number of draw downs is $N_0 = 2789$ and $DD_c
\approx 1.8$ \%, see figure \ref{newdd1}.

The important
point here is the presence of these three events that should not have occurred
at this high rate. This provides an empirical clue that large draw downs
and thus
crashes might
result from a different mechanism, that we attribute to the collective
destabilizing imitation process described above. Ranked,
the three largest crashes are the crash of 1987, the
crash following the outbreak of World War I\footnote{Why the out-break of
World War I should affect the Wall Street much more than the Japanese bombing
of Pearl Harbour or the out-break of the Korean War seems very odd. World War I
was largely an internal European affair and in the beginning something the
populations of the warring countries were very enthusiastic about.}
and the crash of 1929.

To quantify how much the three events deviates from equation (\ref{zae}),
we can calculate what accordingly would be the typical return time of
a draw down of an amplitude equal to or larger than the third largest crash
of $23.6 \%$. Equation (\ref{zae}) gives the number of draw downs equal to or
larger than DD and predicts the number of drawn downs equal to or larger
than $23.6\%$ per century to be $\approx 0.0056$. The typical return time of
a draw down equal to or larger than $23.6\%$ would then be the number of
centuries $n$ such that $0.0056 \cdot n \sim 1$, which yields $n \sim 180$
centuries. In contrast, Wall Street has sustained 3 such events in less
than a century. If we take the exponential distribution of draw downs
larger than $1$\% suggested by figure \ref{newdd1} at
face value, it suggest that the ``normal day-to-day behavior'' of the
stock market index is to a large extent governed by random processes but the
largest crashes are signatures of cooperative behavior with long-term
build-up of correlations between traders.

As an additional test of this hypothesis, we have applied a more sophisticated
null-hypothesis than that of an exponential distribution of draw downs and
used a GARCH(1,1) model estimated from the true Dow Jones Average, see
appendix \ref{appgarch}. The background for this model is the following
(Bollerslev et al., 1992).
First, the stock market index itself cannot follow a simple random (Brownian)
walk, since prices cannot become negative. In fact, one of the basic
notions of investment is that the gain should be proportional to the sum
invested. Using the logarithm of the stock market index instead of the index
itself takes care of both objections and allows us to keep the random walk
assumption for day-to-day variations in the stock return. However, the
logarithm of stock returns is not normally distributed, but follows a
fat-tailed distribution, such as the Student-t. Furthermore, the variance
around some average level should change as a function of time in a correlated
way, since large price movements tend to be followed by large price movements
and small price movements tend to follow small price movements.

>From this GARCH(1,1) model, we have generated $10.000$ independent data sets
corresponding in total to approximately {\em one million} years of
``garch-trading''
with a reset every century. Among these $10.000$ surrogate data sets only two
had $3$ draw downs above $22$\% and none had $4$\footnote{The first attempt of
``only'' $1000$ data sets did not contain any ``GARCH(1,1) centuries'' with 3
crashes larger than $22$\%.}. However, each of these data sets contained
crashes with a rather abnormal behavior in the sense that they were preceded
by a draw up of comparable size as the draw down. This means that in a million
years of garch-trading {\it not once} did $3$ asymmetric crashes occur in a
single century.

We note that the often symmetric behavior of artificial indices, {\it i.e.},
that large/small price movements in either direction are followed by
large/small price movements in either directions is not quite compatible
with what is seen in the Dow Jones Average \cite{LeBaron}. In general, we see
that the build-up of index is relatively slow and crashes quite rapid. In
other words, ``bubbles'' are slow and crashes are fast. This means that not
only must a long-term model of the stock market prior to large crashes be
highly non-linear but it must also have a ``time-direction'' in the sense
that the crash is ``attractive'' prior to the crash and ``repulsive'' after
the crash. This breaking of time-symmetry, in other word the fact that the time
series does not look statistically similar when viewed under reversal of the
time arrow, has been demonstrated by Arn\'eodo et al. (1998).
Using wavelets to decompose the volatility (standard deviation) of
intraday (S\&P500) return data across scales, Arn\'eodo et al. (1998) showed
that the two-point correlation functions of the
volatility logarithms across different time scales reveal
the existence of an asymmetric causal information cascade
from large scales to
fine scales and {\it not} the reverse.
Cont (1998) has recently confirmed this finding by using a three-point
correlation function suggested by Pomeau (1982), constructed in such a way that
it becomes non-zero under time asymmetry.

The analysis presented in this section suggests that it is {\it very} likely
that the largest crashes of this century have a different origin than the
smaller draw downs. We have suggested that these large crashes may be viewed
as critical points and the purpose of the next section is to quantify the
characteristic signatures associated with such critical points, specifically
that of a critical divergence and log-periodic oscillations.

\subsection{Fitting Stock Market Indices} \label{stockfit}

The simple power law in Equation (\ref{eq:solution}) is very difficult to
distinguish from a non-critical exponential growth over a few years when
data are noisy. Furthermore, systematic deviations from a monotonous rise
in the index is clearly visible to the naked eye. This is why all our
empirical efforts were focused on the log-periodic formulas.

Fitting the stock market with a complex formula like equation
(\protect\ref{eq:complex}) and especially equation (\protect\ref{eq:nonlinear})
involves a number of considerations, the most obvious being to secure that
the best possible  fit is obtained. Fitting a function to some data is nothing
but a minimisation algorithm of some cost-function of the data and the
fitting-function. However, with noisy data and a fitting-function with a large
number of degrees of freedom, many local minima of the cost-function exist
where the minimisation algorithm can get trapped. Specifically, the method
used here was a downhill simplex minimisation (Press {\it et al.~}, 1992)
 of the variance.
In order to reduce the number of free parameters of the fit, the $3$ linear
variables have been "slaved". This was done by requiring that the
cost-function has zero derivative with respect to $A,B,C$ in a minimum. If we
rewrite equations (\protect\ref{eq:complex}) and (\protect\ref{eq:nonlinear})
as $\log \lp p\lp t\rp \rp \approx A + B f\lp t \rp + C  g\lp t \rp$ then we
get $3$ linear equations in $A,B,C$
\bea \label{slaveeq}
\sum_i^N \lp
\begin{array}{c}\log \lp p\lp t_i\rp \rp \\ \log \lp p\lp t_i\rp \rp f\lp
t_i\rp \\
\log \lp p\lp t_i\rp \rp g\lp t_i\rp
\end{array} \rp =  \sum_i^N
\lp  \begin{array}{lcr} N &  f\lp t_i\rp  & g\lp t_i\rp \\
f\lp t_i\rp  &  f\lp t_i\rp^2 &  f\lp t_i\rp g\lp t_i\rp \\
g\lp t_i\rp & f\lp t_i\rp g\lp t_i\rp & g\lp t_i\rp^2
\end{array} \rp
\cdot \lp \begin{array}{c} A  \\  B  \\ C \end{array} \rp
\eea
to solve for each choice of values for the non-linear parameters. Equations
(\ref{slaveeq}) was solved using the LU decomposition algorithm in (Press
{\it et al.~}, 1992) thus expressing $A$, $B$ and $C$ as functions of the
remaining non-linear variables. Of these, the phase $\phi$ is just a time
unit and has no physical meaning\footnote{In fact, $A,B,C$ can be regarded
as simple units as well.}. By changing the time unit of the data (from f.ex.
days to months or years) it can be shown that $\beta$ and $\omega$ as well as
the timing of the crash represented by $t_c$ is independent of $\phi$. This
leaves $3$ (equation (\ref{eq:complex})) respectively $5$ (equation
(\ref{eq:nonlinear})) physical parameters controlling the fit, which
at least in the last case, is quite a lot. Hence, the fitting was
preceded by a so-called Taboo-search (Cvijovi\'c D.and J. Klinowski J. 1995)
where only $\beta$ and
$\phi$ was fitted for fixed values of the other parameters. All such scans,
which converged with $0<\beta <1$ was then re-fitted with all non-linear
parameters free. The rationale behind this restriction is, as explained in
section \ref{subsub:grid}, economical since $\beta < 0$ would imply that
the stock market index could go to infinity.

In the case of equation (\protect\ref{eq:complex}), the described procedure
always produced one or a few distinct minima. Unfortunately, in the case of
equation (\protect\ref{eq:nonlinear}) things were never that simple and
quite different sets of parameter values could produce more or less the same
value of the variance. This degeneracy of the variance as a function of
$\alpha ,t_c,\omega ,\Delta_t$ and $\Delta_\omega$ means that the error of
the fit alone is not enough to decide whether a fit is ``good or bad''.
Naturally, a large variance means that the fit is bad, but the converse is
not necessarily true. If we return to the derivation of equations
(\protect\ref{eq:complex}) and (\protect\ref{eq:nonlinear}), we note  three
things. First, that $\alpha$ is determining the strength of the singularity.
Second, that the frequency $\omega \approx 2\pi /\ln\lambda$ is determined by
the underlying hierarchical structure quantified by the allowed re-scaling
factors $\gamma^n$, as previously defined. Last, we recall that $\Delta_t$
in equation \ref{eq:nonlinear} describes a transition between two regimes
of the dynamics, {\i.e.}, far from respectively close to the singularity.
These
considerations have the following consequences. If we believe that
large crashes can be described as critical points and hence have the same
physical background, then $\beta$, $\omega$ (actually $\gamma$) and
$\Delta_t$ should have values which are comparable. Furthermore,
$\Delta_t$ should not be much smaller or much larger than the time-interval
fitted, since it describes a transition time. Realizing this, we can
use the values of $\beta $, $\omega$ and $\tau$ together with the error of
the fit to discriminate between good and bad fits. This provides us with
much stronger discriminating statistics than just the error of the fit.

\subsection{Estimation of Equations (\protect\ref{eq:complex})}
\label{sect1freq}

\begin{figure}
\input{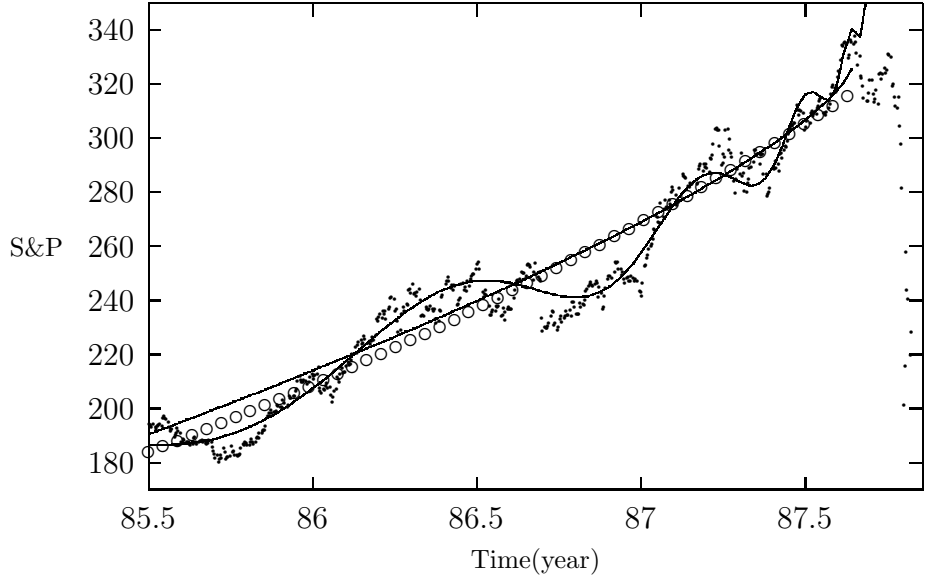}
\caption{\label{1freq87} The New York stock exchange index $S\&P500$ from
July $1985$ to the end of $1987$ corresponding to $557$ trading days.
The $\circ$ represent a constant return increase in terms an exponential with
a characteristic increase of $\approx 4$ years$^{-1}$ and
$\mbox{var}\lp F_{exp}
\rp \approx 113$. The best fit to a pure power-law gives
$A \approx 327$, $B  \approx -79$, $t_c  \approx 87.65$, $\alpha  \approx
0.7$ and $\mbox{var}_{pow} \approx 107$. The best fit to (\ref{eq:complex})
gives $A \approx 412$, $B \approx -165$, $t_c \approx 87.74$, $C \approx
12$, $\omega \approx 7.4$, $\phi =2.0$, $\alpha \approx 0.33$ and
$\mbox{var}_{lp} \approx 36$.}
\end{figure}

For the period before the crash of 1987, the optimal fit of Equation
(\ref{eq:complex}) is shown in Figure \ref{1freq87}, along with two trend
lines representing an exponential and a pure power law. The
log-periodic oscillations are so strong that they are visible even to the
naked eye. The estimation procedure yields a critical exponent of
$\beta = 0.57$. The position of the critical time $t_c$ is within a few days
of the actual crash date. This result was reported earlier by
Sornette et al.~(1996) and Feigenbaum and Freund (1996). The latter authors
also show a similar fit for the crash of 1929. See also Leland (1988) and
Rubinstein (1988) for discussions of the crash of 1987 along different lines.

The turmoil on the financial US market in Oct. 1997 fits into the framework
presented above. Detection of log-periodic structures and
prediction of the stock market turmoil at the end of october 1997 has been
made,
 based on an unpublished extension of the theory. This prediction has
been formally issued ex-ante on september 17, 1997,
to the French office for the  protection of
proprietary softwares and inventions under number registration 94781.
Vandewalle et al. have also issued an alarm, published in the Belgian
newspaper Tendances, 18. September 1997, page 26,
by H. Dupuis, entitled ``Un krach avant novembre'' (a krach before november).
It turned out that the crash did not really occur\,: only the largest dayly
loss
since Oct. 1987 was observed with ensuing large volatilities. This
has been argued by Laloux et al. (1998) as implying the ``death'' of the
theory.
In fact, this is fully consistent with our rational expectation model of a
crash\,:
the bubble expands, the market believes that the crash is more and more
probable,
the prices develop characteristic structures but the critical time passes
out without
much happening.
This corresponds to the finite probability $1- Q(t_c)$, mentionned above,
that no crash occurs over the whole time including the critical time of the
end of the bubble.
Feigenbaum and Freund have also analyzed, after the fact,
the log-periodic oscillations in the S\&P 500
and the NYSE in relation to the October 27'th ``correction''
seen on Wall Street.

\begin{figure}
\input{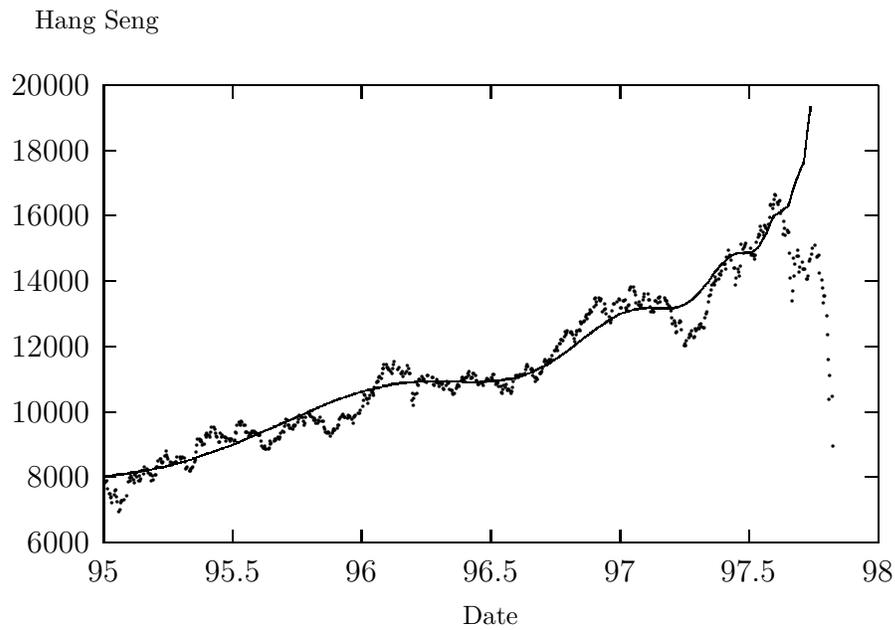}
\caption{\label{hongkong} The Hang Seng index of the Hong Kong stock exchange
approximately two and half years prior to the crash of October 1997. The best
fit to (\ref{eq:complex}) gives $A \approx 2.\cdot 10^4$, $B\approx -8240$,
$t_c \approx 97.74$, $C \approx -397$, $\omega \approx 7.5$, $\phi \approx
1.0$, $\alpha \approx 0.34$ and $r.m.s. \approx 436$. }
\end{figure}

In figure \ref{hongkong} we see the Hang Seng index fitted with
Equation (\ref{eq:complex}) approximately $2.5$ years prior to the recent
crash in October 1997. Rather remarkably the values obtained for the
exponent $\alpha$ and the frequency $\omega$ differs less $5$ \% from
the values reported for the 1987 crash on Wall Street by Sornette {\it et al.}
(1996). Note also the similar behavior of the two indices over the last few
months prior to the crashes.

\subsection{Estimation of Equation (\protect\ref{eq:nonlinear})}
\label{sect2freq}

The hope here is that we will be able to capture the evolution of the stock
price over a longer pre-crash period, say 7-8 years. For estimation purposes,
Equation (\ref{eq:complex}) was rewritten as:
\be
\label{eq:nonlinear2}
\log\left[p(t)\right] = A+B
\frac{(t_c-t)^\beta}{
\sqrt{1+\left(\frac{t_c-t}{\Delta_t}\right)^{2\beta}}}
\left\{1+C\cos\left[\omega\log(t_c-t)+\frac{\Delta_\omega}{2\beta}
\log\left(1+\left(\frac{t_c-t}{\Delta_t}\right)^{2\beta}\right)\right]\right
\}
\ee
We use the same least-squares method as above. We again concentrate away
the linear variables $A$, $B$ and $C$ in order to form an objective function
depending only on $t_c$, $\beta$, $\omega$ and $\phi$ as before, and the
two additional parameters $\Delta_t$ and $\Delta_\omega$. We can not put any
bounds on
the second frequency $\Delta_\omega$, but since $\Delta_t$ is, as mentioned,
a transition
time between two regimes we require it to full-fill $3<\tau<16$ years. As
before, the nonlinearity of the objective function creates multiple local
minima, and we use the preliminary grid search to find starting points for the
optimiser.

\begin{figure}
\begin{center}
\epsfig{file=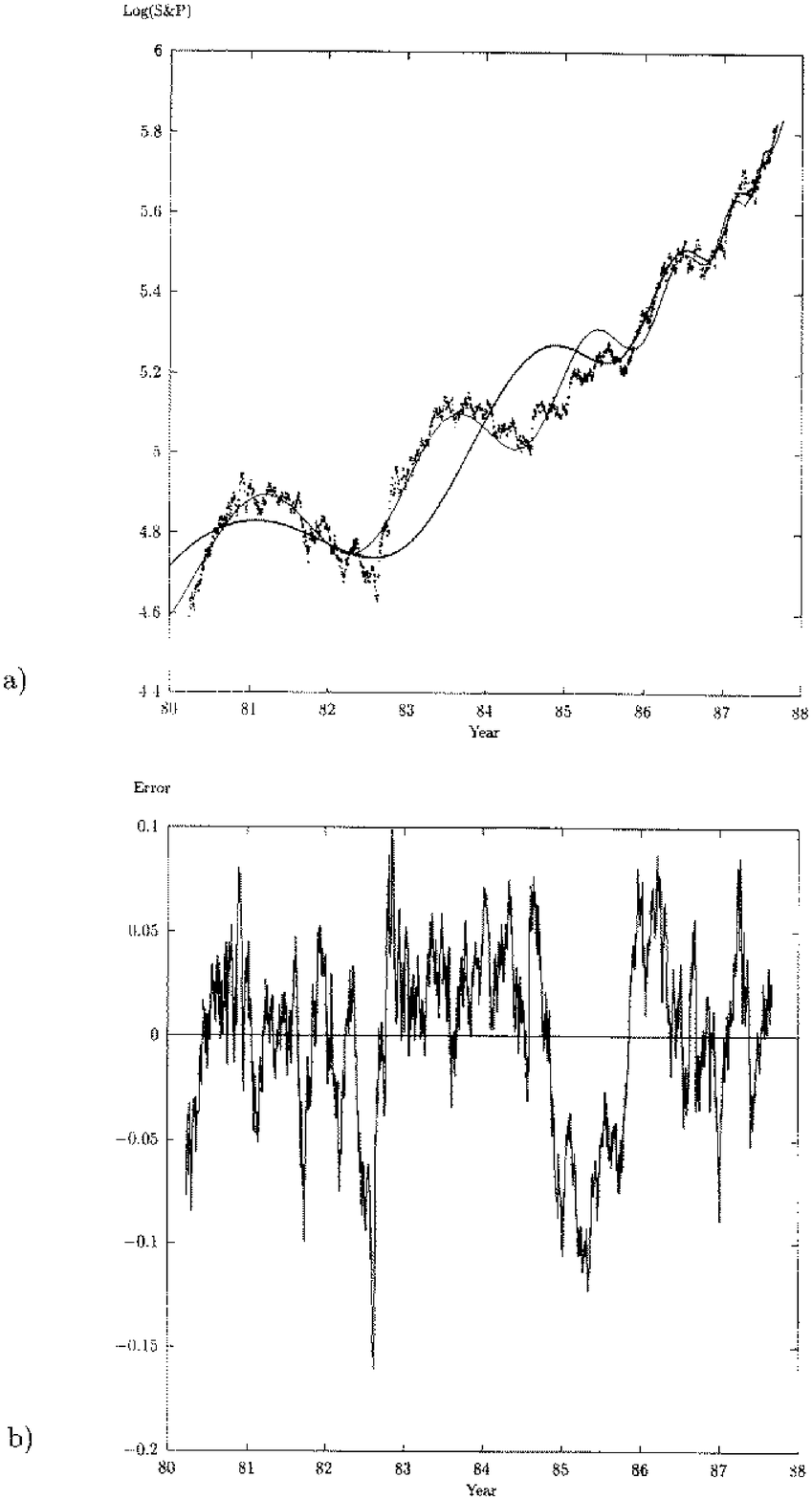,width=10cm}
\end{center}
\caption[]{\label{crashfit87}
a) The $S\&P500$ from January 1980 to September
1987 and best fit by (\protect\ref{eq:nonlinear}) (thin line) with $r.m.s.
\approx0.043$, $t_c \approx 1987.81$, $\alpha \approx 0.68$, $\omega
\approx 8.9$, $\Delta_\omega \approx 18$,
$\Delta_t \approx 11$ years, $A \approx 5.9$, $B \approx -0.38$,
$C \approx 0.043 $. The thick line
is the fit shown in figure \protect\ref{1freq87} extended to the full time
interval. The comparison with the thin line allows one to visualise the
frequency shift described by equation (\protect\ref{eq:nonlinear}).

b) The relative error of the fit.}
\end{figure}

\begin{figure}
\begin{center}
\epsfig{file=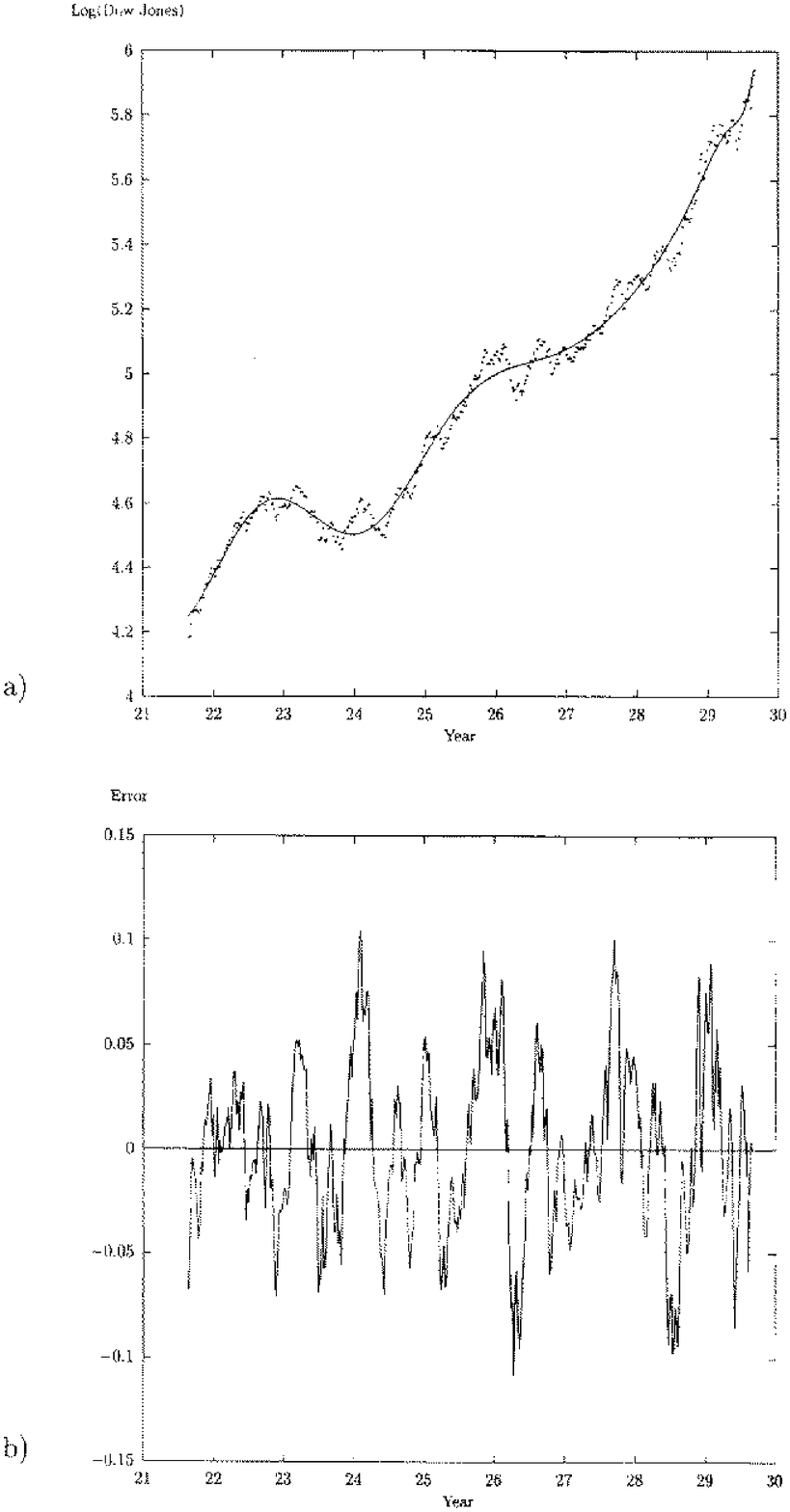,width=10cm}
\end{center}
\caption[]{\label{crashfit29}
a) The Dow Jones Average from June 1921 to September 1929 and best fit by
equation (\protect\ref{eq:nonlinear}). The parameters of the fit are
r.m.s.$\approx0.041$,
$t_c \approx 1929.84$ year, $\alpha \approx 0.63$, $\omega \approx 5.0$,
$\Delta_\omega \approx -70$,
$\Delta_t \approx 14$ years, $A \approx 61$, $B \approx -0.56$,
$C \approx 0.08$.

b) The relative error of the fit.}

\end{figure}

We fit the logarithm of the Dow Jones Industrial Average index prior to the
1929 crash and the $S\&P500$ index prior to the 1987 crash, both starting
approximately $8$ years prior to the crash thus extending the fitted region
from two years to almost eight years, see figures \ref{crashfit87}a and
\ref{crashfit29}a.

We see that the behavior of the logarithm of the index is well-captured by
equation (\protect\ref{eq:nonlinear}) over almost $8$ years with a general
error below $10$\%
for both crashes, see figures \ref{crashfit87}b and \ref{crashfit29}b.
Furthermore, the value for
$t_c$ is within two weeks of the true date for both crashes. The reason for
fitting the logarithm of the index and not the index itself can be seen as
an attempt to ``de-trend'' the index from the underlying exponential growth
on long time scales.

As we shall further elaborate in the next section, the small values
obtained for the $r.m.s.$ is of course encouraging but not decisive, since
we are fitting a function with considerable ``elasticity''. Hence, it's
quite reassuring that the parameter values for the ``physical variables''
$\alpha$, $\omega$ and $\Delta_t$ obtained for the two crashes are rather
similar
\begin{itemize}
\item $\alpha_{29} \approx 0.63$ compared to $\alpha_{87} \approx 0.68$
\item $\omega_{29} \approx 5.0$ compared to  $\omega_{87} \approx 8.9$ or
$\lambda_{29} \approx 3.5$ compared to  $\lambda_{87} \approx 2.0$.
\item $\Delta_{t_{29}} \approx 14$ years compared to
$\Delta_{t_{87}} \approx 11$ years
\end{itemize}

\subsection{Fitting the Model to Other Periods}

In order to investigate the significance of the results obtained for the
1929 and 1987 crashes, we picked at random $50$\footnote{Naturally, we
would have liked to use a much larger number, but these $50$ intervals
already correspond to $4$ centuries.} $400$-week intervals in the period
1910 to 1996 of the Dow Jones average and launched the fitting procedure
described above on these surrogate data sets. The approximate end-date of
the 50 data sets was
\newline

1951, 1964, 1950, 1975, 1979, 1963, 1934, 1960, 1936, 1958, 1985, 1884, 1967,
1943, 1991, 1982, 1972, 1928, 1932, 1946, 1934, 1963, 1979, 1993, 1960, 1935,
1974, 1950, 1970, 1980, 1940, 1986, 1923, 1963, 1964, 1968, 1975, 1929, 1984,
1944, 1994, 1967, 1924, 1974, 1954, 1956, 1959, 1926, 1947 and 1965.
\newline

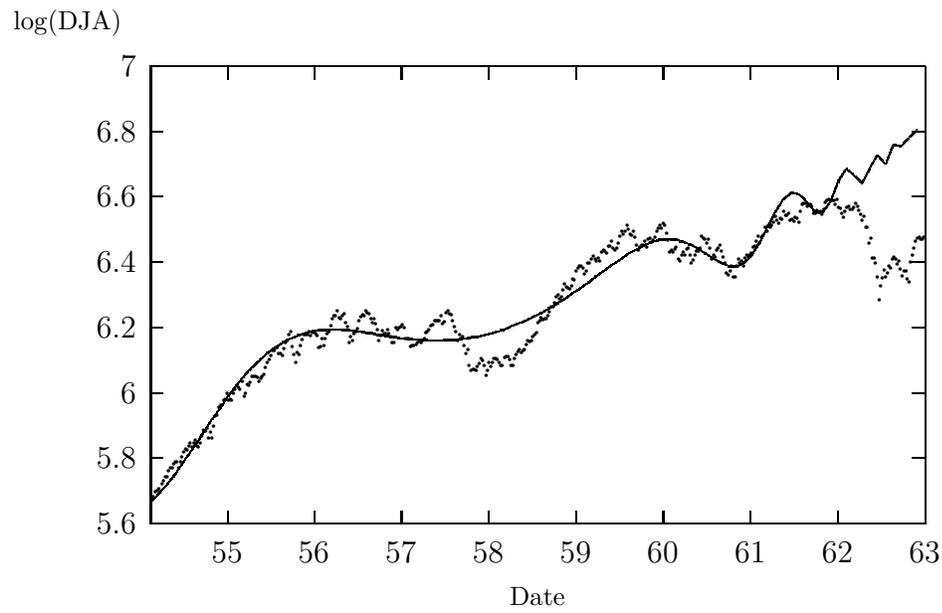
\begin{figure}
\input{crash62}
\caption{\label{crash62} The ``slow crash'' of 1962 and the best fit by
equation (\protect\ref{eq:nonlinear}).}
\end{figure}

The results were very encouraging. Of
the $11$ fits with a $r.m.s.$ comparable with the $r.m.s.$ of the two crashes,
only $6$ data sets produced values for $\alpha$, $\omega$ and $\Delta_t$
which were in the same range as the values obtained for the $2$ crashes,
specifically $0.45 < \alpha < 0.85$, $4 < \omega < 14$ ($1.6 < \lambda <4.8$)
and $3 < \Delta_t < 16$. All $6$ fits belonged to the periods prior to the
crashes of 1929, 1962 and 1987. The existence of a crash in 1962 was
before these results unknown to us and the identification of this crash
naturally strengthens our case. In figure \ref{crash62} we see the best fit
obtained for this crash and except for the ``postdicted'' time of the crash
$t_c \approx 1963.0$, which is rather off as seen in figure \ref{crash62}, the
values obtained for the parameters $\alpha_{62} \approx 0.83, \omega_{62}
\approx 13,
\Delta_t{_{62}} \approx 14$ years is again close to those of the 1929 and 1987
crashes.
A comment on the deviation between the obtained value for $t_c$ and the true
date of the crash is appropriate. As seen in figure \ref{crash62}, the crash of
1962 was anomalous in the sense that it was ``slow''. The stock market declined
approximately $30$\% in 3 months and not in less than one week as for the
other two crashes. One may speculate on the reason(s) for this and in
terms of the model presented here some external shock may have provoked
this slow crash before the stock market was ``ripe''. In fact within
our rational expectation model, a bubble that starts to ``inflate'' with
some theoretical critical time $t_c$ can be perturbed and not go to its
culmination due to the influence of external shocks. This does not prevent
the log-periodic
structures to develop up to the time when the course of the bubble evolution is
modified by these external shocks.

The results from fitting the surrogate data sets generated from the real
stock market index show that fits, which in terms of the fitting parameters
corresponds to the 3 crashes mentioned above, are not likely to occur
``accidentally''. Encouraged by this, we decided on a more elaborate
statistical test in order to firmly establish the significance of the
results presented above.

\subsection{Fitting a GARCH(1,1) Model}
\label{garchfit}

We generated 1000 surrogate date sets now of length 400 weeks using the
same GARCH(1,1) model used in section \ref{outlier}.
On each of these 1000 data sets we launched the same fitting routine as
for the real crashes.
Using the same parameter range for the $\alpha$, $\omega$ and $\Delta_t$
as for the 50 random intervals above showed that 66 of the best minima
had values in that range. All of these fits, except two, did not resemble
the true stock market index prior to the 1929 and 1987 crashes on Wall Street
very much, the reason primarily being that they only had one or two
oscillations. However, two fits looked rather much like that of the 1929
and 1987 crashes, see the two fit in figure 7. Note that one of the two fits
shown in figure 7 have parameter values right on the border of the allowed
intervals. This result correspond approximately to the usual 95 \% confidence
interval for {\em one} event. Opposed to this, we have here provided several
examples of log-periodic signatures before large stock market crashes.

\subsection{Fitting Truncated Stock Market Data}

An obvious question concerns how long time prior to the crash can one
identify the log-periodic signatures described in sections
\ref{sect1freq} and \ref{sect2freq}. There are several
reasons for this. Not only because
one would like to predict future crashes, but also to further test how
robust our results are. Obviously, if the log-period structure of the
data is purely accidental, then the parameter values obtained should depend
heavily on the size of the time interval used in the fitting. We have hence
applied the following procedure. For each of the two rapid crashes fitted
above, we have truncated the time interval used in the fitting by removing
points and re-launching the fitting procedure described above. Specifically,
the logarithm of the S\& P500 shown in figure \ref{crashfit87} was
truncated down
to an end-date of $\approx 1985$ and fitted using the procedure described
above. Then $\approx 0.16$ years was added consecutively and the fitting
was relaunched until the full time interval was recovered. In table
\ref{tabel87a}, we see the number of minima obtained for the different time
intervals. This number is to some extent rather arbitrary since it naturally
depends on the number of points used in the preliminary scan as well as the
size of the time interval used for $t_c$. Specifically, $40.000$ points were
used and $t_c$ was chosen $0.1$ years from the last data point used and $3$
years forward. What is more interesting is the number of ``physical minima''
as defined above and especially the values of $t_c ,\alpha ,\omega , \tau$
of these fits. The general picture to be extracted from this table, is that
a year or more before the crash, the data is not sufficient to give any
conclusive results at all.  This point correspond to the end of the 4'th
oscillation. Approximately a year before the crash, the fit begins to lock-in
on the date of the crash with increasing precision. In fact, in four of the
last
five time intervals, we can find a fit with a $t_c$, which differs from the
true
date of the crash by only a few weeks. In order to better investigate
this, we show in table \ref{tabel87b} the corresponding parameter values
for the other 3 physical variables $\beta ,\omega , \tau$. The scenario
resembles that for $t_c$ and we can conclude that our procedure is rather
robust up to approximately $1$ year prior to the crash. However, if one
wants to actually predict the time of the crash, a major obstacle is the
fact that our fitting procedure produces several possible dates for the
date of the crash even for the last data set. As a naive solution to this
problem we show in table \protect\ref{tabel87c} the average of the different
minima for $t_c ,\alpha ,\omega , \tau$. We see that the values for
$\beta ,\omega , \tau$ are within $20$\% of those for the best prediction,
{\em but} the prediction for $t_c$ has not improved significantly. The reason
for this is that the fit in general ``over-shoot'' the true day of the crash.

The same procedure was used on the logarithm of the Dow Jones index prior
to the crash of 1929 and the results are shown in tables \ref{tabel29a},
\ref{tabel29b} and \ref{tabel29c}. We see that we have to wait until
approximately $4$ month before the crash before the fit locks in on the
date of the crash, but from that point the picture is the same as for the
crash in 1987. The reason for the fact that the fit ``locks-in'' at a later
time for the 1929 is obviously the difference in the transition time
$\Delta_t$ for the two crashes which means that the index prior to the
crash of 1929 exhibits fewer distinct oscillations.

A general trend for the two crashes is that the $t_c$'s of the physical minima
tends to over-shoot the time of the crash. We have also tried to truncate the
data sets with respect to the starting point with 2 years and the effect is
quite similar. Even though $\beta ,\omega , \tau$ stay close to the values
obtained using the full interval, the fit starts to overshoot and $t_c$ moves
up around $88.2$ for the 1987 crash and $30.0$ for the 1929 crash. The reason
for this is rather obvious. The difference between a pure power law behavior
(equation \ref{eq:power}) and the discrete versions (equations \ref
{eq:complex}
and \ref {eq:nonlinear}) is that the pure power law is less constraining with
respect to determining $t_c$. This since $t_c$ is now determined not only
by the
over-all acceleration quantified by $\beta$ but also by the frequencies
$\omega$ and $\Delta_{\omega}$ in the cosine. This means that by truncating the
data we are in fact removing parts of the oscillations and hence decreasing
the accuracy in which we can determine $\omega$ and $\Delta_{\omega}$ and as a
consequence $t_c$.

To conclude this part of the analysis, we have seen that the three physical
variables
$\beta ,\omega$  and $\tau$ are reasonably robust with respect to
truncating the
data up to approximately a year. In contrast, this is not the case for the
timing of the crash $t_c$, which systematically over-shoot the actual date
of the
crash. We stress that these results are fully consistent with our rational
expectation model of
a crash. Indeed, recall that $t_c$ in the formulas
(\ref{eq:complex},\ref{eq:nonlinear})
is {\it not} the time of the crash but
the most probable value of the (skewed to the left) distribution of the
possible times of
the crash.
The occurrence of the crash is a random phenomenon which occurs with a
probability
that increases as time approaches $t_c$. Thus, we expect that fits will give
values of $t_c$ which are in general close to but {\it systematically}
later than the
real time of the crash. The phenomenon of ``overshot'' that has been clearly
documented above is thus fully consistent with the theory. This is one of the
major improvement over previous works brought by our combining the rational
expectation theory with the imitation model.

\section{Conclusion}

This paper has drawn a link between crashes in the stock market and critical
behavior of complex systems. We have shown how patterns that are typical
of critical behavior can arise in prices even when markets are rational.
Furthermore, we have provided some empirical evidence suggesting that these
characteristic patterns are indeed present, at certain times, in U.S. and
Hong Kong~stock
market indices. This supports our key hypothesis, which is that local
imitation between noise traders might cascade through the scales into large-
scale coordination and cause crashes. To sum up, the evidence we have
presented signifying that large financial
crashes are outliers to the distribution of draw downs and as a consequence
have their origin in cooperative phenomena are the following.

\begin{itemize}
\item The recurrence time for a sudden stock market drop equal or larger than
$23.6$\% was estimated to be 180 centuries. The stock market has sustained
three
such events in this century.
\item In a million years of GARCH(1,1)-trading with student-t noise with
four degrees
of freedom with a reset of every century,
      only twice did three draw downs larger than $22$\% occur and never four.
      However, three of these ``crashes'' were symmetric and none preceded by
      the log-periodic behavior seen in real crashes.
\item From 50 randomly sampled 400 weeks intervals of the Dow Jones Average
      in the period 1910-94, all fits that were found with parameters in
the range of those
      obtained for the 1929 and 1987 crashes could be related to those two
      crashes {\it or} to the slow crash of 1962. Furthermore, out of a 1000
      surrogate data sets generated from a GARCH(1,1) model, only 66 fits had
      parameter values comparable with that obtained for the crashes of 1929
      and 1987. This correspond approximately to the usual 95 \% confidence
      interval for {\em one} event. In contrast, we have here provided
      several examples of log-periodic signatures before large stock market
      crashes.
\item We have tested the prediction ability equation (\ref{eq:nonlinear}) by
      truncating the stock market indices with respect to the end-point of
      the data set. In general, our procedure tends to over-shoot the actual
      date of the crash and hence its prediction ability is not very
impressive.
      This is fully consistent with our rational expectation theory according
      to which the critical time $t_c$ is only the most probable value of
the distribution of
      times of the crash, which occurs (if it occurs) close to but before
$t_c$.
      However, these tests show that our fitting procedure is robust with
      respect to {\em all} variables in equation (\ref{eq:nonlinear}).
\item Three of the largest crashes of this century, the 1929 and 1987 crashes
      on Wall Street and the 1997 crash in Hong-Kong, have been preceded by
      log-periodic behavior well-described by equations
      (\protect\ref{eq:complex}) and (\protect\ref{eq:nonlinear}).
      In addition, prior to the slow crash of 1962 and the
      turmoil on Wall Street in late October 1997 log-periodic
      signatures have been observed with parameter values consistent with
      the three other crashes. The turbulence of Oct. 1997 in the US market
      was also preceded by a log-periodic structure that led to an estimation
      of $t_c$ close to the end of the year, with the other parameters
      consistent with the values found for the other crashes. No crash
occurred.
      This may be interpreted as a realization,
      compatible with our rational expectation theory of a crash,
      illustrating that there is a non-vanishing probability for no crash
to occur even
      under inflating bubble conditions.
\end{itemize}

\vskip 1cm
{\bf Acknowledgements}:  We thank M. Brennan, B.
Chowdhry, W.I. Newman, H. Saleur and P. Santa-Clara for useful
discussions. We also thank participants to a seminar at UCLA for their
feedback. Errors are ours.

\pagebreak

\appendix
\section{Derivation of the exponential distribution (\ref{zae}) of draw downs}

We use the dayly time scale as the unit of time.
Call $p_+$ (resp. $p_- = 1-p_+$) the probability for the stock market to go up
(resp. to go down) over a day. If $P_1(x)$ is the probability density
function (pdf)
of {\it negative} dayly price increments, the distribution $P(DD)$
of draw downs (continuous decreases)
is given by
\be
P(DD) = \sum_{n=1}^{\infty} p_+^2 ~p_-^n~P_1^{\otimes n}(DD)~,
\label{gqkjqk}
\ee
where $P_1^{\otimes n}$ denotes the distribution obtained by $n$
convolutions of
$P_1$ with itself, corresponding to the fact that $DD$ is the sum of $n$
random variables. The factor $p_+^2 ~p_-^n$ weights the probability that a
draw down results from consecutive $n$ losses starting after and finishing
before a gain.
Since $P_1(x)$ is defined for $-\infty < x \leq 0$, the relevant tool is
the Laplace
transform with argument $k$\,: the Laplace transform of
$P_1^{\otimes n}(DD)$ is the $n$th power of the Laplace transform ${\hat
P}_1(k)$ of
$P_1(x)$. When applied to (\ref{gqkjqk}), it gives after summation
\be
{\hat P}(k) = p_+^2 ~{p_- {\hat P}_1{k} \over 1 - p_- {\hat P}_1{k}}~.
\label{jfkklgld}
\ee
The dayly price distribution is closely approximated by
an exponential distribution $P_1(x) = {c \over 2} e^{-c |xl}$
(Bouchaud and Potters, 1997; Laherr\`ere and Sornette, 1998). Then,
${\hat P}_1(k) = {c/2 \over c + k}$.
Put in (\ref{jfkklgld}), this gives ${\hat P}(k) = {c \over 2}~{p_+^2 ~p_-
\over c(1-p_-) +k}$.
By inverse Laplace transform, we get
\be
P(DD) \propto e^{-c p_+ |x|}~.
\ee
We see that the typical amplitude $DD_c$ of ``normal'' draw downs is $1/p_+$
times the typical amplitude $1/c$ of the negative price variations. The
price distribution
of the Dow Jones is very close to being symmetric. Thus, $p_+$ is close to
$1/2$ and thus
$DD_C$ is about twice the typical dayly price variation.

\pagebreak

\section{GARCH(1,1) model of the stock market} \label{appgarch}

Estimating the five parameters of a GARCH(1,1) (Generalised Autoregressive
with Conditional Heteroskedasticity\footnote{When the errors of different
points have different variances but are uncorrelated with each other, then
the errors are said to exhibit heteroskedasticity. If the variances are
correlated, the heteroskedasticity is said to be conditional.}) model of
some process is a simple optimisation problem.

\subsection{Parameters}

The 5 parameters are related to the data $x_t$ as follows.
\bea \label{eqxt}
x_t &=& \mu+h_t\epsilon_t \\ \label{eqht}
h_t &=&\alpha+\beta\epsilon_{t-1}^2+\gamma h_{t-1}
\eea
where $\epsilon\lp t\rp$ belongs to a  Student-t distribution with $\kappa$
degrees of freedom and mean $0$ and variance $h\lp t\rp$. The interpretation
of the parameters are as follows:

\begin{description}
\item[$\mu$] Mean of the process
\item[$\kappa$] Number of degrees of freedom ({\it i.e}, ``fat-tailedness'')
     of the Student $t$-distribution
\item[$\alpha$] Controls the average level of volatility
\item[$\beta$] Controls the impact of short-term shocks to volatility
\item[$\gamma$] Controls the long-term memory of volatility
\end{description}

Furthermore, the range of the parameters are

\begin{description}
\item[$\mu$] Any real number.
\item[$\kappa$] Any number strictly above 2, so that the variance is finite;
the lower the $\kappa$ the fatter the tails; this number needs not be an
integer.
\item[$\alpha$] Between 0 and 1.
\item[$\beta$] Between 0 and 1.
\item[$\gamma$] Between 0 and 1.
\end{description} In addition there is the constraint $\alpha+\beta\leq1$.

\subsection{Initial values of the parameters}

These are the values recommended in order to start the optimisation.
\begin{description}
\item[$\mu$] Take the sample mean $m=\sum_{t=1}^Tx_t/t$.
\item[$\kappa$] Take 4.
\item[$\alpha$] Take 0.05 times the sample variance: $0.05\times v$, where
                the sample variance is defined by
                $v=\sum_{t=1}^T(x_t-m)^2/(t-1)$.
\item[$\beta$] Take 0.05.
\item[$\gamma$] Take 0.90.
\end{description}

The factor 0.05 for $\alpha$ comes from the fact that the average level
of the variance is equal to $\alpha/(1-\beta-\gamma)$, as can be seen by
replacing $h_t$ with its average level in equation (\ref{eqht}). For
$\beta=0.05$ and $\gamma=0.90$, this gives the factor 0.05.

\subsection{Objective function}

The five parameters are obtained by maximising the likelihood function $L$
over the parameter range defined above. This is a function of the 5
parameters and the data ($x_t$ for $t=1,\dots,T$), which
is computed in three steps.
\begin{enumerate}
\item Do a loop over
$t=1,\dots,T$ to compute the residuals $\epsilon_t=x_t-\mu$.
\item Do a
loop to compute recursively the conditional variance $h_t$ at time $t$.
Initialise the recursion by: $h_1=v$ (the sample variance computed above).
Iterate for $t=2,\dots,T$ with $h_t=\alpha+\beta\epsilon_{t-1}^2+\gamma
h_{t-1}$.
\item Compute the likelihood function as: \begin{eqnarray} L&=&
\log\left[\Gamma\left(\frac{\kappa+1}{2}\right)\right]
-\log\left[\Gamma\left(\frac{\kappa}{2}\right)\right]
-\frac{1}{2}\log(\kappa)\\ &&{}-\frac{1}{T}\sum_{t=1}^T\left[
\frac{\log(h_t)}{2}
+\frac{\kappa+1}{2}\log\left(1+\frac{\epsilon_t^2}{\kappa
h_t}\right)\right] \end{eqnarray} where $\Gamma$ denotes the gamma
function.
\end{enumerate}
This procedure yields the value of the logarithm of the likelihood, given
the data and the five parameters. Feed this likelihood function into an
optimiser and it will give the five GARCH(1,1) parameters.

Specifically for the GARCH(1,1) that generated the surrogate data used
in subsection \ref{garchfit}, we used $\mu = 4.38~10^{-4}$, $\gamma = 0.922$,
$\alpha = 2.19\cdot 10^{-5}$, $\beta = 0.044$ and the Student-t distribution
had $\kappa = 4$ degrees of freedom and was generated by a NAG-library
routine. Furthermore, the first 5000 data points was always discarded in
order to remove any sensitivity on the initial conditions. The logarithm of
the  index $\ln I$ was then calculated as
\bea
\ln I_{t+1} = \ln I_t + x_{t+1},
\eea
using equations \ (\ref{eqxt}) and (\ref{eqht}) and the arbitrary initial
condition $\ln I_0 = 2$.

\newpage \mbox{ }

\newpage \mbox{ }

\newpage \mbox{ }

\newpage \mbox{ }

\newpage \mbox{ }

\newpage \mbox{ }

\begin{figure}
\input{goodsurrfit1}
\caption{\label{goodsurrfit1} Surrogate data and best fit by equation
(\protect\ref{eq:nonlinear}) is $A\approx 6.6$, $B\approx -.34$, $C\approx
-.047$,
$\alpha \approx  0.74$, $t_c\approx  1957.7$,  $\phi\approx -.13$
$\omega \approx  9.1$, $\Delta_t \approx  4.1$,
$\Delta\omega \approx -2.1$ and $r.m.s. \approx  0.037$ }

\vspace{0.5cm}

\input{goodsurrfit2}
\caption{\label{goodsurrfit2} Surrogate data and best fit by equation
(\protect\ref{eq:nonlinear}) is $A\approx 5.1$, $B\approx -.24$, $C\approx
-.037$,
$\alpha \approx  0.85$, $t_c\approx  1957.7$,  $\phi\approx -3.9$
$\omega \approx  5.5$, $\Delta_t \approx  7.7$,
$\Delta\omega \approx 11$ and $r.m.s. \approx  0.036$ }
\end{figure}

\newpage \mbox{ }

\begin{table}
\begin{center}
\begin{tabular}{|c|c|c|c|} \hline
end-date & total \# minima & ``physical'' minima & $t_c$ of ``physical''
minima \\
 $85.00$ & $33$ & $1$  & $86.52$                           \\
 $85.14$ & $25$ & $4$  & $4$ in $\left[ 86.7:86.8 \right]$ \\
 $85.30$ & $26$ & $7$  & $5$ in $\left[ 86.5:87.0 \right]$, $2$ in $\left[
87.4:87.6 \right]$ \\
 $85.46$ & $29$ & $8$  & $7$ in $\left[ 86.6:86.9 \right]$ ,$1$ with
$87.22$  \\
 $85.62$ & $26$ & $13$ & $12$ in $\left[ 86.8:87.1\right]$ ,$1$ with
$87.65$  \\
 $85.78$ & $23$ & $7$  & $87.48$, $5$ in $\left[ 87.0-87.25\right]$,
$87.68$  \\
 $85.93$ & $17$ & $4$  & $87.25,87.01,87.34,86.80$           \\
 $86.09$ & $18$ & $4$  & $87.29,87.01,86,98,87.23$            \\
 $86.26$ & $28$ & $7$  & $5$ in $\left[ 87.2:87.4\right]$, $86.93,86.91,$  \\
 $86.41$ & $24$ & $4$  & $87.26,87.36,87.87,87.48$         \\
 $86.57$ & $20$ & $2$  & $87.67,87.34$  \\
 $86.73$ & $28$ & $7$  & $4$ in $\left[ 86.8:87.0\right]$,
$87.37,87.79,87.89$  \\
 $86.88$ & $22$ & $1$  & $87.79$                        \\
 $87.04$ & $18$ & $2$  & $87.68,88.35$               \\
 $87.20$ & $15$ & $2$  & $87.79,88.03$               \\
 $87.36$ & $15$ & $2$  & $88.19,88.30$               \\
 $87.52$ & $14$ & $3$  & $88.49,87.92,88.10$      \\
 $87.65$ & $15$ & $3$  & $87.81,88.08,88.04$      \\
\hline
\end{tabular}
\end{center}
\caption{\protect\label{tabel87a} Number of minima obtained by fitting
different truncated versions of the S\& P500 time series shown in figure
\ref{crashfit87}, using the procedure described in the text. }
\end{table}

\begin{table}
\begin{center}
\begin{tabular}{|c|c|c|c|c|} \hline
end-date & $t_c$ & $\alpha$ & $\omega$ & $\tau$ \\
 $86.88$ & $87.79$  & $0.66$ & $5.4$ & $7.8$    \\
 $87.04$ & $87.68,88.35$ & $0.61,0.77$ & $4.1,13.6$ & $12.3,10.2$  \\
 $87.20$ & $87.79,88.03$ & $0.76,0.77$ & $9.4,11.0$ & $10.0,9.6$   \\
 $87.36$ & $88.19,88.30$ & $0.66,0.79$ & $7.3,12.2$ & $7.9,8.1$    \\
 $87.52$ & $88.49,87.92,88.10$ & $0.51,0.71,0.65$ & $12.3,9.6,10.3$ &
$10.2,9.8,9.8$   \\
 $87.65$ & $87.81,88.08,88.04$ & $0.68,0.69,0.67$ & $8.9,10.4,10.1$ &
$10.8,9.7,10.2$ \\
\hline
\end{tabular}
\end{center}
\caption{\protect\label{tabel87b} For the last
five time intervals shown in table 1, we show the corresponding parameter
values for the other three variables $\beta ,\omega , \tau$.}
\end{table}

\begin{table}
\begin{center}
\begin{tabular}{|c|c|c|c|c|} \hline
end-date & $t_c$ & $\alpha$ & $\omega$ & $\tau$ \\
 $86.88$ & $87.79$  & $0.66$ & $5.4$ & $7.8$  \\
 $87.04$ & $88.02$ & $0.69$ & $8.6$ & $11.3$  \\
 $87.20$ & $87.91$ & $0.77$ & $10.20$ & $9.8$ \\
 $87.36$ & $88.25$ & $0.73$ & $9.6$ & $8.0$   \\
 $87.52$ & $88.17$ & $0.62$ & $10.7$ & $9.9$  \\
 $87.65$ & $87.98$ & $0.68$ & $9.8$  & $10.2$ \\
\hline
\end{tabular}
\end{center}
\caption{\protect\label{tabel87c} The average of the values listed in
table \protect\ref{tabel87b}.}
\end{table}

\begin{table}
\begin{center}
\begin{tabular}{|c|c|c|c|} \hline
end-date & total \# minima & ``physical'' minima & $t_c$ of ``physical''
minima \\
 $27.37$ & $12$ & $1$ & $31.08$       \\
 $27.56$ & $14$ & $2$ & $30.44,30.85$   \\
 $27.75$ & $24$ & $1$ & $30.34$       \\
 $27.94$ & $21$ & $1$ & $31.37$   \\
 $28.13$ & $21$ & $4$ & $29.85,30.75,30.72,30.50$   \\
 $28.35$ & $23$ & $4$ & $30.29,30.47,30.50,36.50$   \\
 $28.52$ & $18$ & $1$ & $31.3$   \\
 $28.70$ & $18$ & $1$ & $31.02$   \\
 $28.90$ & $16$ & $4$ & $30.40,30.72,31.07,30.94$   \\
 $29.09$ & $19$ & $2$ & $30.52,30.35$   \\
 $29.28$ & $33$ & $1$ & $30.61$          \\
 $29.47$ & $24$ & $3$ & $29.91,30.1,29.82$   \\
 $29.67$ & $23$ & $1$ & $29.87$   \\
\hline
\end{tabular}
\end{center}
\caption{\protect\label{tabel29a} Same as table 1 for the Oct. 1929 crash}
\end{table}

\begin{table}
\begin{center}
\begin{tabular}{|c|c|c|c|c|} \hline
end-date & $t_c$ & $\alpha$ & $\omega$ & $\tau$ \\
 $28.90$ & $30.40,30.72,31.07,30.94$ & $0.60,0.70,0.70,0.53$ &
$7.0,7.6,10.2,13.7$ & $12.3,9.5,9.0,11.6$     \\
 $29.09$ & $30.52,30.35$ & $0.54,0.62$ & $11.0,7.8$ & $12.6,10.2$    \\
 $29.28$ & $30.61$ & $0.63$  & $9.5$ & $9.5$  \\
 $29.47$ & $29.91,30.1,29.82$ & $0.60,0.67,0.69$ & $5.8,6.2,4.5$ &
$15.9,11.0,10.9$   \\
 $29.67$ & $29.87$ & $0.61$ & $5.4$ & $15.0$           \\
\hline
\end{tabular}
\end{center}
\caption{\protect\label{tabel29b} Same as table 2 for the Oct. 1929 crash}
\end{table}

\begin{table}
\begin{center}
\begin{tabular}{|c|c|c|c|c|} \hline
end-date & $t_c$ & $\alpha$ & $\omega$ & $\tau$ \\
 $28.90$ & $30.78$ & $0.63$ & $9.6$ & $10.6$     \\
 $29.09$ & $30.44$ & $0.58$ & $9.4$ & $11.4$    \\
 $29.28$ & $30.61$ & $0.63$ & $9.5$ & $9.5$  \\
 $29.47$ & $29.94$ & $0.65$ & $5.5$ & $12.6$   \\
 $29.67$ & $29.87$ & $0.61$ & $5.4$ & $15.0$           \\
\hline
\end{tabular}
\end{center}
\caption{\protect\label{tabel29c} The average of the values listed in
table \protect\ref{tabel29b}.}
\end{table}

\end{document}

%% file: newestdd1.tex
\setlength{\unitlength}{0.240900pt}
\ifx\plotpoint\undefined\newsavebox{\plotpoint}\fi
\sbox{\plotpoint}{\rule[-0.200pt]{0.400pt}{0.400pt}}%
\begin{picture}(1500,900)(0,0)
\font\gnuplot=cmr10 at 10pt
\gnuplot
\sbox{\plotpoint}{\rule[-0.200pt]{0.400pt}{0.400pt}}%
\put(220.0,113.0){\rule[-0.200pt]{2.409pt}{0.400pt}}
\put(1426.0,113.0){\rule[-0.200pt]{2.409pt}{0.400pt}}
\put(220.0,125.0){\rule[-0.200pt]{2.409pt}{0.400pt}}
\put(1426.0,125.0){\rule[-0.200pt]{2.409pt}{0.400pt}}
\put(220.0,136.0){\rule[-0.200pt]{4.818pt}{0.400pt}}
\put(198,136){\makebox(0,0)[r]{$1$}}
\put(1416.0,136.0){\rule[-0.200pt]{4.818pt}{0.400pt}}
\put(220.0,207.0){\rule[-0.200pt]{2.409pt}{0.400pt}}
\put(1426.0,207.0){\rule[-0.200pt]{2.409pt}{0.400pt}}
\put(220.0,248.0){\rule[-0.200pt]{2.409pt}{0.400pt}}
\put(1426.0,248.0){\rule[-0.200pt]{2.409pt}{0.400pt}}
\put(220.0,278.0){\rule[-0.200pt]{2.409pt}{0.400pt}}
\put(1426.0,278.0){\rule[-0.200pt]{2.409pt}{0.400pt}}
\put(220.0,300.0){\rule[-0.200pt]{2.409pt}{0.400pt}}
\put(1426.0,300.0){\rule[-0.200pt]{2.409pt}{0.400pt}}
\put(220.0,319.0){\rule[-0.200pt]{2.409pt}{0.400pt}}
\put(1426.0,319.0){\rule[-0.200pt]{2.409pt}{0.400pt}}
\put(220.0,335.0){\rule[-0.200pt]{2.409pt}{0.400pt}}
\put(1426.0,335.0){\rule[-0.200pt]{2.409pt}{0.400pt}}
\put(220.0,349.0){\rule[-0.200pt]{2.409pt}{0.400pt}}
\put(1426.0,349.0){\rule[-0.200pt]{2.409pt}{0.400pt}}
\put(220.0,361.0){\rule[-0.200pt]{2.409pt}{0.400pt}}
\put(1426.0,361.0){\rule[-0.200pt]{2.409pt}{0.400pt}}
\put(220.0,371.0){\rule[-0.200pt]{4.818pt}{0.400pt}}
\put(198,371){\makebox(0,0)[r]{$10$}}
\put(1416.0,371.0){\rule[-0.200pt]{4.818pt}{0.400pt}}
\put(220.0,442.0){\rule[-0.200pt]{2.409pt}{0.400pt}}
\put(1426.0,442.0){\rule[-0.200pt]{2.409pt}{0.400pt}}
\put(220.0,484.0){\rule[-0.200pt]{2.409pt}{0.400pt}}
\put(1426.0,484.0){\rule[-0.200pt]{2.409pt}{0.400pt}}
\put(220.0,513.0){\rule[-0.200pt]{2.409pt}{0.400pt}}
\put(1426.0,513.0){\rule[-0.200pt]{2.409pt}{0.400pt}}
\put(220.0,536.0){\rule[-0.200pt]{2.409pt}{0.400pt}}
\put(1426.0,536.0){\rule[-0.200pt]{2.409pt}{0.400pt}}
\put(220.0,555.0){\rule[-0.200pt]{2.409pt}{0.400pt}}
\put(1426.0,555.0){\rule[-0.200pt]{2.409pt}{0.400pt}}
\put(220.0,571.0){\rule[-0.200pt]{2.409pt}{0.400pt}}
\put(1426.0,571.0){\rule[-0.200pt]{2.409pt}{0.400pt}}
\put(220.0,584.0){\rule[-0.200pt]{2.409pt}{0.400pt}}
\put(1426.0,584.0){\rule[-0.200pt]{2.409pt}{0.400pt}}
\put(220.0,596.0){\rule[-0.200pt]{2.409pt}{0.400pt}}
\put(1426.0,596.0){\rule[-0.200pt]{2.409pt}{0.400pt}}
\put(220.0,607.0){\rule[-0.200pt]{4.818pt}{0.400pt}}
\put(198,607){\makebox(0,0)[r]{$100$}}
\put(1416.0,607.0){\rule[-0.200pt]{4.818pt}{0.400pt}}
\put(220.0,678.0){\rule[-0.200pt]{2.409pt}{0.400pt}}
\put(1426.0,678.0){\rule[-0.200pt]{2.409pt}{0.400pt}}
\put(220.0,719.0){\rule[-0.200pt]{2.409pt}{0.400pt}}
\put(1426.0,719.0){\rule[-0.200pt]{2.409pt}{0.400pt}}
\put(220.0,749.0){\rule[-0.200pt]{2.409pt}{0.400pt}}
\put(1426.0,749.0){\rule[-0.200pt]{2.409pt}{0.400pt}}
\put(220.0,772.0){\rule[-0.200pt]{2.409pt}{0.400pt}}
\put(1426.0,772.0){\rule[-0.200pt]{2.409pt}{0.400pt}}
\put(220.0,790.0){\rule[-0.200pt]{2.409pt}{0.400pt}}
\put(1426.0,790.0){\rule[-0.200pt]{2.409pt}{0.400pt}}
\put(220.0,806.0){\rule[-0.200pt]{2.409pt}{0.400pt}}
\put(1426.0,806.0){\rule[-0.200pt]{2.409pt}{0.400pt}}
\put(220.0,820.0){\rule[-0.200pt]{2.409pt}{0.400pt}}
\put(1426.0,820.0){\rule[-0.200pt]{2.409pt}{0.400pt}}
\put(220.0,832.0){\rule[-0.200pt]{2.409pt}{0.400pt}}
\put(1426.0,832.0){\rule[-0.200pt]{2.409pt}{0.400pt}}
\put(220.0,843.0){\rule[-0.200pt]{4.818pt}{0.400pt}}
\put(198,843){\makebox(0,0)[r]{$1000$}}
\put(1416.0,843.0){\rule[-0.200pt]{4.818pt}{0.400pt}}
\put(220.0,113.0){\rule[-0.200pt]{0.400pt}{4.818pt}}
\put(220,68){\makebox(0,0){$-0.35$}}
\put(220.0,857.0){\rule[-0.200pt]{0.400pt}{4.818pt}}
\put(399.0,113.0){\rule[-0.200pt]{0.400pt}{4.818pt}}
\put(399,68){\makebox(0,0){$-0.3$}}
\put(399.0,857.0){\rule[-0.200pt]{0.400pt}{4.818pt}}
\put(578.0,113.0){\rule[-0.200pt]{0.400pt}{4.818pt}}
\put(578,68){\makebox(0,0){$-0.25$}}
\put(578.0,857.0){\rule[-0.200pt]{0.400pt}{4.818pt}}
\put(756.0,113.0){\rule[-0.200pt]{0.400pt}{4.818pt}}
\put(756,68){\makebox(0,0){$-0.2$}}
\put(756.0,857.0){\rule[-0.200pt]{0.400pt}{4.818pt}}
\put(935.0,113.0){\rule[-0.200pt]{0.400pt}{4.818pt}}
\put(935,68){\makebox(0,0){$-0.15$}}
\put(935.0,857.0){\rule[-0.200pt]{0.400pt}{4.818pt}}
\put(1114.0,113.0){\rule[-0.200pt]{0.400pt}{4.818pt}}
\put(1114,68){\makebox(0,0){$-0.1$}}
\put(1114.0,857.0){\rule[-0.200pt]{0.400pt}{4.818pt}}
\put(1293.0,113.0){\rule[-0.200pt]{0.400pt}{4.818pt}}
\put(1293,68){\makebox(0,0){$-0.05$}}
\put(1293.0,857.0){\rule[-0.200pt]{0.400pt}{4.818pt}}
\put(220.0,113.0){\rule[-0.200pt]{292.934pt}{0.400pt}}
\put(1436.0,113.0){\rule[-0.200pt]{0.400pt}{184.048pt}}
\put(220.0,877.0){\rule[-0.200pt]{292.934pt}{0.400pt}}
\put(210,930){\makebox(0,0){N(DD)}}
\put(828,0){\makebox(0,0){Draw Down (DD)}}
\put(220.0,113.0){\rule[-0.200pt]{0.400pt}{184.048pt}}
\multiput(946.58,113.00)(0.492,0.826){19}{\rule{0.118pt}{0.755pt}}
\multiput(945.17,113.00)(11.000,16.434){2}{\rule{0.400pt}{0.377pt}}
\multiput(957.58,131.00)(0.492,0.798){21}{\rule{0.119pt}{0.733pt}}
\multiput(956.17,131.00)(12.000,17.478){2}{\rule{0.400pt}{0.367pt}}
\multiput(969.58,150.00)(0.493,0.774){23}{\rule{0.119pt}{0.715pt}}
\multiput(968.17,150.00)(13.000,18.515){2}{\rule{0.400pt}{0.358pt}}
\multiput(982.58,170.00)(0.492,0.798){21}{\rule{0.119pt}{0.733pt}}
\multiput(981.17,170.00)(12.000,17.478){2}{\rule{0.400pt}{0.367pt}}
\multiput(994.58,189.00)(0.492,0.841){21}{\rule{0.119pt}{0.767pt}}
\multiput(993.17,189.00)(12.000,18.409){2}{\rule{0.400pt}{0.383pt}}
\multiput(1006.58,209.00)(0.492,0.798){21}{\rule{0.119pt}{0.733pt}}
\multiput(1005.17,209.00)(12.000,17.478){2}{\rule{0.400pt}{0.367pt}}
\multiput(1018.58,228.00)(0.493,0.774){23}{\rule{0.119pt}{0.715pt}}
\multiput(1017.17,228.00)(13.000,18.515){2}{\rule{0.400pt}{0.358pt}}
\multiput(1031.58,248.00)(0.492,0.798){21}{\rule{0.119pt}{0.733pt}}
\multiput(1030.17,248.00)(12.000,17.478){2}{\rule{0.400pt}{0.367pt}}
\multiput(1043.58,267.00)(0.492,0.841){21}{\rule{0.119pt}{0.767pt}}
\multiput(1042.17,267.00)(12.000,18.409){2}{\rule{0.400pt}{0.383pt}}
\multiput(1055.58,287.00)(0.493,0.734){23}{\rule{0.119pt}{0.685pt}}
\multiput(1054.17,287.00)(13.000,17.579){2}{\rule{0.400pt}{0.342pt}}
\multiput(1068.58,306.00)(0.492,0.841){21}{\rule{0.119pt}{0.767pt}}
\multiput(1067.17,306.00)(12.000,18.409){2}{\rule{0.400pt}{0.383pt}}
\multiput(1080.58,326.00)(0.492,0.798){21}{\rule{0.119pt}{0.733pt}}
\multiput(1079.17,326.00)(12.000,17.478){2}{\rule{0.400pt}{0.367pt}}
\multiput(1092.58,345.00)(0.492,0.841){21}{\rule{0.119pt}{0.767pt}}
\multiput(1091.17,345.00)(12.000,18.409){2}{\rule{0.400pt}{0.383pt}}
\multiput(1104.58,365.00)(0.493,0.734){23}{\rule{0.119pt}{0.685pt}}
\multiput(1103.17,365.00)(13.000,17.579){2}{\rule{0.400pt}{0.342pt}}
\multiput(1117.58,384.00)(0.492,0.841){21}{\rule{0.119pt}{0.767pt}}
\multiput(1116.17,384.00)(12.000,18.409){2}{\rule{0.400pt}{0.383pt}}
\multiput(1129.58,404.00)(0.492,0.798){21}{\rule{0.119pt}{0.733pt}}
\multiput(1128.17,404.00)(12.000,17.478){2}{\rule{0.400pt}{0.367pt}}
\multiput(1141.58,423.00)(0.492,0.841){21}{\rule{0.119pt}{0.767pt}}
\multiput(1140.17,423.00)(12.000,18.409){2}{\rule{0.400pt}{0.383pt}}
\multiput(1153.58,443.00)(0.493,0.734){23}{\rule{0.119pt}{0.685pt}}
\multiput(1152.17,443.00)(13.000,17.579){2}{\rule{0.400pt}{0.342pt}}
\multiput(1166.58,462.00)(0.492,0.798){21}{\rule{0.119pt}{0.733pt}}
\multiput(1165.17,462.00)(12.000,17.478){2}{\rule{0.400pt}{0.367pt}}
\multiput(1178.58,481.00)(0.492,0.841){21}{\rule{0.119pt}{0.767pt}}
\multiput(1177.17,481.00)(12.000,18.409){2}{\rule{0.400pt}{0.383pt}}
\multiput(1190.58,501.00)(0.493,0.734){23}{\rule{0.119pt}{0.685pt}}
\multiput(1189.17,501.00)(13.000,17.579){2}{\rule{0.400pt}{0.342pt}}
\multiput(1203.58,520.00)(0.492,0.841){21}{\rule{0.119pt}{0.767pt}}
\multiput(1202.17,520.00)(12.000,18.409){2}{\rule{0.400pt}{0.383pt}}
\multiput(1215.58,540.00)(0.492,0.798){21}{\rule{0.119pt}{0.733pt}}
\multiput(1214.17,540.00)(12.000,17.478){2}{\rule{0.400pt}{0.367pt}}
\multiput(1227.58,559.00)(0.492,0.841){21}{\rule{0.119pt}{0.767pt}}
\multiput(1226.17,559.00)(12.000,18.409){2}{\rule{0.400pt}{0.383pt}}
\multiput(1239.58,579.00)(0.493,0.734){23}{\rule{0.119pt}{0.685pt}}
\multiput(1238.17,579.00)(13.000,17.579){2}{\rule{0.400pt}{0.342pt}}
\multiput(1252.58,598.00)(0.492,0.841){21}{\rule{0.119pt}{0.767pt}}
\multiput(1251.17,598.00)(12.000,18.409){2}{\rule{0.400pt}{0.383pt}}
\multiput(1264.58,618.00)(0.492,0.798){21}{\rule{0.119pt}{0.733pt}}
\multiput(1263.17,618.00)(12.000,17.478){2}{\rule{0.400pt}{0.367pt}}
\multiput(1276.58,637.00)(0.493,0.774){23}{\rule{0.119pt}{0.715pt}}
\multiput(1275.17,637.00)(13.000,18.515){2}{\rule{0.400pt}{0.358pt}}
\multiput(1289.58,657.00)(0.492,0.798){21}{\rule{0.119pt}{0.733pt}}
\multiput(1288.17,657.00)(12.000,17.478){2}{\rule{0.400pt}{0.367pt}}
\multiput(1301.58,676.00)(0.492,0.841){21}{\rule{0.119pt}{0.767pt}}
\multiput(1300.17,676.00)(12.000,18.409){2}{\rule{0.400pt}{0.383pt}}
\multiput(1313.58,696.00)(0.492,0.798){21}{\rule{0.119pt}{0.733pt}}
\multiput(1312.17,696.00)(12.000,17.478){2}{\rule{0.400pt}{0.367pt}}
\multiput(1325.58,715.00)(0.493,0.774){23}{\rule{0.119pt}{0.715pt}}
\multiput(1324.17,715.00)(13.000,18.515){2}{\rule{0.400pt}{0.358pt}}
\multiput(1338.58,735.00)(0.492,0.798){21}{\rule{0.119pt}{0.733pt}}
\multiput(1337.17,735.00)(12.000,17.478){2}{\rule{0.400pt}{0.367pt}}
\multiput(1350.58,754.00)(0.492,0.841){21}{\rule{0.119pt}{0.767pt}}
\multiput(1349.17,754.00)(12.000,18.409){2}{\rule{0.400pt}{0.383pt}}
\multiput(1362.58,774.00)(0.493,0.734){23}{\rule{0.119pt}{0.685pt}}
\multiput(1361.17,774.00)(13.000,17.579){2}{\rule{0.400pt}{0.342pt}}
\multiput(1375.58,793.00)(0.492,0.841){21}{\rule{0.119pt}{0.767pt}}
\multiput(1374.17,793.00)(12.000,18.409){2}{\rule{0.400pt}{0.383pt}}
\multiput(1387.58,813.00)(0.492,0.798){21}{\rule{0.119pt}{0.733pt}}
\multiput(1386.17,813.00)(12.000,17.478){2}{\rule{0.400pt}{0.367pt}}
\multiput(1399.58,832.00)(0.492,0.841){21}{\rule{0.119pt}{0.767pt}}
\multiput(1398.17,832.00)(12.000,18.409){2}{\rule{0.400pt}{0.383pt}}
\multiput(1411.58,852.00)(0.493,0.734){23}{\rule{0.119pt}{0.685pt}}
\multiput(1410.17,852.00)(13.000,17.579){2}{\rule{0.400pt}{0.342pt}}
\multiput(1424.61,871.00)(0.447,1.132){3}{\rule{0.108pt}{0.900pt}}
\multiput(1423.17,871.00)(3.000,4.132){2}{\rule{0.400pt}{0.450pt}}
\put(381,136){\raisebox{-.8pt}{\makebox(0,0){$\circ$}}}
\put(452,136){\raisebox{-.8pt}{\makebox(0,0){$\circ$}}}
\put(631,136){\raisebox{-.8pt}{\makebox(0,0){$\circ$}}}
\put(810,207){\raisebox{-.8pt}{\makebox(0,0){$\circ$}}}
\put(882,136){\raisebox{-.8pt}{\makebox(0,0){$\circ$}}}
\put(917,136){\raisebox{-.8pt}{\makebox(0,0){$\circ$}}}
\put(953,207){\raisebox{-.8pt}{\makebox(0,0){$\circ$}}}
\put(989,207){\raisebox{-.8pt}{\makebox(0,0){$\circ$}}}
\put(1025,207){\raisebox{-.8pt}{\makebox(0,0){$\circ$}}}
\put(1060,349){\raisebox{-.8pt}{\makebox(0,0){$\circ$}}}
\put(1096,361){\raisebox{-.8pt}{\makebox(0,0){$\circ$}}}
\put(1132,413){\raisebox{-.8pt}{\makebox(0,0){$\circ$}}}
\put(1168,452){\raisebox{-.8pt}{\makebox(0,0){$\circ$}}}
\put(1204,528){\raisebox{-.8pt}{\makebox(0,0){$\circ$}}}
\put(1239,563){\raisebox{-.8pt}{\makebox(0,0){$\circ$}}}
\put(1275,603){\raisebox{-.8pt}{\makebox(0,0){$\circ$}}}
\put(1311,662){\raisebox{-.8pt}{\makebox(0,0){$\circ$}}}
\put(1347,730){\raisebox{-.8pt}{\makebox(0,0){$\circ$}}}
\put(1382,793){\raisebox{-.8pt}{\makebox(0,0){$\circ$}}}
\put(1418,877){\raisebox{-.8pt}{\makebox(0,0){$\circ$}}}
\end{picture}

%% file: crash62.tex
\setlength{\unitlength}{0.240900pt}
\ifx\plotpoint\undefined\newsavebox{\plotpoint}\fi
\sbox{\plotpoint}{\rule[-0.200pt]{0.400pt}{0.400pt}}%
\begin{picture}(1500,900)(0,0)
\font\gnuplot=cmr10 at 10pt
\gnuplot
\sbox{\plotpoint}{\rule[-0.200pt]{0.400pt}{0.400pt}}%
\put(220.0,113.0){\rule[-0.200pt]{4.818pt}{0.400pt}}
\put(198,113){\makebox(0,0)[r]{$5.6$}}
\put(1416.0,113.0){\rule[-0.200pt]{4.818pt}{0.400pt}}
\put(220.0,216.0){\rule[-0.200pt]{4.818pt}{0.400pt}}
\put(198,216){\makebox(0,0)[r]{$5.8$}}
\put(1416.0,216.0){\rule[-0.200pt]{4.818pt}{0.400pt}}
\put(220.0,318.0){\rule[-0.200pt]{4.818pt}{0.400pt}}
\put(198,318){\makebox(0,0)[r]{$6$}}
\put(1416.0,318.0){\rule[-0.200pt]{4.818pt}{0.400pt}}
\put(220.0,421.0){\rule[-0.200pt]{4.818pt}{0.400pt}}
\put(198,421){\makebox(0,0)[r]{$6.2$}}
\put(1416.0,421.0){\rule[-0.200pt]{4.818pt}{0.400pt}}
\put(220.0,524.0){\rule[-0.200pt]{4.818pt}{0.400pt}}
\put(198,524){\makebox(0,0)[r]{$6.4$}}
\put(1416.0,524.0){\rule[-0.200pt]{4.818pt}{0.400pt}}
\put(220.0,627.0){\rule[-0.200pt]{4.818pt}{0.400pt}}
\put(198,627){\makebox(0,0)[r]{$6.6$}}
\put(1416.0,627.0){\rule[-0.200pt]{4.818pt}{0.400pt}}
\put(220.0,729.0){\rule[-0.200pt]{4.818pt}{0.400pt}}
\put(198,729){\makebox(0,0)[r]{$6.8$}}
\put(1416.0,729.0){\rule[-0.200pt]{4.818pt}{0.400pt}}
\put(220.0,832.0){\rule[-0.200pt]{4.818pt}{0.400pt}}
\put(198,832){\makebox(0,0)[r]{$7$}}
\put(1416.0,832.0){\rule[-0.200pt]{4.818pt}{0.400pt}}
\put(341.0,113.0){\rule[-0.200pt]{0.400pt}{4.818pt}}
\put(341,68){\makebox(0,0){$55$}}
\put(341.0,812.0){\rule[-0.200pt]{0.400pt}{4.818pt}}
\put(477.0,113.0){\rule[-0.200pt]{0.400pt}{4.818pt}}
\put(477,68){\makebox(0,0){$56$}}
\put(477.0,812.0){\rule[-0.200pt]{0.400pt}{4.818pt}}
\put(614.0,113.0){\rule[-0.200pt]{0.400pt}{4.818pt}}
\put(614,68){\makebox(0,0){$57$}}
\put(614.0,812.0){\rule[-0.200pt]{0.400pt}{4.818pt}}
\put(751.0,113.0){\rule[-0.200pt]{0.400pt}{4.818pt}}
\put(751,68){\makebox(0,0){$58$}}
\put(751.0,812.0){\rule[-0.200pt]{0.400pt}{4.818pt}}
\put(888.0,113.0){\rule[-0.200pt]{0.400pt}{4.818pt}}
\put(888,68){\makebox(0,0){$59$}}
\put(888.0,812.0){\rule[-0.200pt]{0.400pt}{4.818pt}}
\put(1025.0,113.0){\rule[-0.200pt]{0.400pt}{4.818pt}}
\put(1025,68){\makebox(0,0){$60$}}
\put(1025.0,812.0){\rule[-0.200pt]{0.400pt}{4.818pt}}
\put(1162.0,113.0){\rule[-0.200pt]{0.400pt}{4.818pt}}
\put(1162,68){\makebox(0,0){$61$}}
\put(1162.0,812.0){\rule[-0.200pt]{0.400pt}{4.818pt}}
\put(1299.0,113.0){\rule[-0.200pt]{0.400pt}{4.818pt}}
\put(1299,68){\makebox(0,0){$62$}}
\put(1299.0,812.0){\rule[-0.200pt]{0.400pt}{4.818pt}}
\put(1436.0,113.0){\rule[-0.200pt]{0.400pt}{4.818pt}}
\put(1436,68){\makebox(0,0){$63$}}
\put(1436.0,812.0){\rule[-0.200pt]{0.400pt}{4.818pt}}
\put(220.0,113.0){\rule[-0.200pt]{292.934pt}{0.400pt}}
\put(1436.0,113.0){\rule[-0.200pt]{0.400pt}{173.207pt}}
\put(220.0,832.0){\rule[-0.200pt]{292.934pt}{0.400pt}}
\put(90,900){\makebox(0,0){log(DJA)}}
\put(828,0){\makebox(0,0){Date}}
\put(220.0,113.0){\rule[-0.200pt]{0.400pt}{173.207pt}}
\put(222,151){\raisebox{-.8pt}{\makebox(0,0){$\cdot$}}}
\put(225,157){\raisebox{-.8pt}{\makebox(0,0){$\cdot$}}}
\put(228,165){\raisebox{-.8pt}{\makebox(0,0){$\cdot$}}}
\put(231,166){\raisebox{-.8pt}{\makebox(0,0){$\cdot$}}}
\put(234,169){\raisebox{-.8pt}{\makebox(0,0){$\cdot$}}}
\put(236,165){\raisebox{-.8pt}{\makebox(0,0){$\cdot$}}}
\put(239,178){\raisebox{-.8pt}{\makebox(0,0){$\cdot$}}}
\put(241,182){\raisebox{-.8pt}{\makebox(0,0){$\cdot$}}}
\put(244,189){\raisebox{-.8pt}{\makebox(0,0){$\cdot$}}}
\put(247,189){\raisebox{-.8pt}{\makebox(0,0){$\cdot$}}}
\put(249,198){\raisebox{-.8pt}{\makebox(0,0){$\cdot$}}}
\put(252,202){\raisebox{-.8pt}{\makebox(0,0){$\cdot$}}}
\put(255,203){\raisebox{-.8pt}{\makebox(0,0){$\cdot$}}}
\put(257,209){\raisebox{-.8pt}{\makebox(0,0){$\cdot$}}}
\put(260,211){\raisebox{-.8pt}{\makebox(0,0){$\cdot$}}}
\put(262,212){\raisebox{-.8pt}{\makebox(0,0){$\cdot$}}}
\put(265,203){\raisebox{-.8pt}{\makebox(0,0){$\cdot$}}}
\put(268,212){\raisebox{-.8pt}{\makebox(0,0){$\cdot$}}}
\put(270,219){\raisebox{-.8pt}{\makebox(0,0){$\cdot$}}}
\put(273,227){\raisebox{-.8pt}{\makebox(0,0){$\cdot$}}}
\put(276,232){\raisebox{-.8pt}{\makebox(0,0){$\cdot$}}}
\put(278,231){\raisebox{-.8pt}{\makebox(0,0){$\cdot$}}}
\put(281,236){\raisebox{-.8pt}{\makebox(0,0){$\cdot$}}}
\put(283,242){\raisebox{-.8pt}{\makebox(0,0){$\cdot$}}}
\put(286,235){\raisebox{-.8pt}{\makebox(0,0){$\cdot$}}}
\put(289,241){\raisebox{-.8pt}{\makebox(0,0){$\cdot$}}}
\put(291,246){\raisebox{-.8pt}{\makebox(0,0){$\cdot$}}}
\put(294,237){\raisebox{-.8pt}{\makebox(0,0){$\cdot$}}}
\put(296,235){\raisebox{-.8pt}{\makebox(0,0){$\cdot$}}}
\put(299,242){\raisebox{-.8pt}{\makebox(0,0){$\cdot$}}}
\put(301,253){\raisebox{-.8pt}{\makebox(0,0){$\cdot$}}}
\put(304,262){\raisebox{-.8pt}{\makebox(0,0){$\cdot$}}}
\put(307,260){\raisebox{-.8pt}{\makebox(0,0){$\cdot$}}}
\put(310,265){\raisebox{-.8pt}{\makebox(0,0){$\cdot$}}}
\put(312,250){\raisebox{-.8pt}{\makebox(0,0){$\cdot$}}}
\put(315,258){\raisebox{-.8pt}{\makebox(0,0){$\cdot$}}}
\put(317,249){\raisebox{-.8pt}{\makebox(0,0){$\cdot$}}}
\put(320,268){\raisebox{-.8pt}{\makebox(0,0){$\cdot$}}}
\put(322,284){\raisebox{-.8pt}{\makebox(0,0){$\cdot$}}}
\put(325,285){\raisebox{-.8pt}{\makebox(0,0){$\cdot$}}}
\put(328,298){\raisebox{-.8pt}{\makebox(0,0){$\cdot$}}}
\put(331,301){\raisebox{-.8pt}{\makebox(0,0){$\cdot$}}}
\put(333,301){\raisebox{-.8pt}{\makebox(0,0){$\cdot$}}}
\put(336,308){\raisebox{-.8pt}{\makebox(0,0){$\cdot$}}}
\put(338,310){\raisebox{-.8pt}{\makebox(0,0){$\cdot$}}}
\put(341,320){\raisebox{-.8pt}{\makebox(0,0){$\cdot$}}}
\put(343,308){\raisebox{-.8pt}{\makebox(0,0){$\cdot$}}}
\put(346,310){\raisebox{-.8pt}{\makebox(0,0){$\cdot$}}}
\put(348,309){\raisebox{-.8pt}{\makebox(0,0){$\cdot$}}}
\put(351,320){\raisebox{-.8pt}{\makebox(0,0){$\cdot$}}}
\put(353,326){\raisebox{-.8pt}{\makebox(0,0){$\cdot$}}}
\put(356,332){\raisebox{-.8pt}{\makebox(0,0){$\cdot$}}}
\put(359,329){\raisebox{-.8pt}{\makebox(0,0){$\cdot$}}}
\put(361,326){\raisebox{-.8pt}{\makebox(0,0){$\cdot$}}}
\put(365,339){\raisebox{-.8pt}{\makebox(0,0){$\cdot$}}}
\put(368,315){\raisebox{-.8pt}{\makebox(0,0){$\cdot$}}}
\put(370,320){\raisebox{-.8pt}{\makebox(0,0){$\cdot$}}}
\put(373,333){\raisebox{-.8pt}{\makebox(0,0){$\cdot$}}}
\put(375,332){\raisebox{-.8pt}{\makebox(0,0){$\cdot$}}}
\put(377,337){\raisebox{-.8pt}{\makebox(0,0){$\cdot$}}}
\put(380,346){\raisebox{-.8pt}{\makebox(0,0){$\cdot$}}}
\put(383,346){\raisebox{-.8pt}{\makebox(0,0){$\cdot$}}}
\put(386,346){\raisebox{-.8pt}{\makebox(0,0){$\cdot$}}}
\put(389,344){\raisebox{-.8pt}{\makebox(0,0){$\cdot$}}}
\put(391,339){\raisebox{-.8pt}{\makebox(0,0){$\cdot$}}}
\put(394,343){\raisebox{-.8pt}{\makebox(0,0){$\cdot$}}}
\put(396,346){\raisebox{-.8pt}{\makebox(0,0){$\cdot$}}}
\put(399,349){\raisebox{-.8pt}{\makebox(0,0){$\cdot$}}}
\put(401,360){\raisebox{-.8pt}{\makebox(0,0){$\cdot$}}}
\put(404,368){\raisebox{-.8pt}{\makebox(0,0){$\cdot$}}}
\put(407,373){\raisebox{-.8pt}{\makebox(0,0){$\cdot$}}}
\put(410,379){\raisebox{-.8pt}{\makebox(0,0){$\cdot$}}}
\put(412,387){\raisebox{-.8pt}{\makebox(0,0){$\cdot$}}}
\put(415,386){\raisebox{-.8pt}{\makebox(0,0){$\cdot$}}}
\put(417,391){\raisebox{-.8pt}{\makebox(0,0){$\cdot$}}}
\put(420,392){\raisebox{-.8pt}{\makebox(0,0){$\cdot$}}}
\put(422,382){\raisebox{-.8pt}{\makebox(0,0){$\cdot$}}}
\put(425,382){\raisebox{-.8pt}{\makebox(0,0){$\cdot$}}}
\put(428,379){\raisebox{-.8pt}{\makebox(0,0){$\cdot$}}}
\put(430,390){\raisebox{-.8pt}{\makebox(0,0){$\cdot$}}}
\put(433,400){\raisebox{-.8pt}{\makebox(0,0){$\cdot$}}}
\put(435,402){\raisebox{-.8pt}{\makebox(0,0){$\cdot$}}}
\put(438,412){\raisebox{-.8pt}{\makebox(0,0){$\cdot$}}}
\put(441,416){\raisebox{-.8pt}{\makebox(0,0){$\cdot$}}}
\put(443,393){\raisebox{-.8pt}{\makebox(0,0){$\cdot$}}}
\put(446,380){\raisebox{-.8pt}{\makebox(0,0){$\cdot$}}}
\put(449,368){\raisebox{-.8pt}{\makebox(0,0){$\cdot$}}}
\put(451,384){\raisebox{-.8pt}{\makebox(0,0){$\cdot$}}}
\put(454,380){\raisebox{-.8pt}{\makebox(0,0){$\cdot$}}}
\put(456,394){\raisebox{-.8pt}{\makebox(0,0){$\cdot$}}}
\put(459,404){\raisebox{-.8pt}{\makebox(0,0){$\cdot$}}}
\put(462,411){\raisebox{-.8pt}{\makebox(0,0){$\cdot$}}}
\put(464,411){\raisebox{-.8pt}{\makebox(0,0){$\cdot$}}}
\put(467,411){\raisebox{-.8pt}{\makebox(0,0){$\cdot$}}}
\put(470,416){\raisebox{-.8pt}{\makebox(0,0){$\cdot$}}}
\put(472,410){\raisebox{-.8pt}{\makebox(0,0){$\cdot$}}}
\put(475,415){\raisebox{-.8pt}{\makebox(0,0){$\cdot$}}}
\put(478,417){\raisebox{-.8pt}{\makebox(0,0){$\cdot$}}}
\put(480,414){\raisebox{-.8pt}{\makebox(0,0){$\cdot$}}}
\put(482,410){\raisebox{-.8pt}{\makebox(0,0){$\cdot$}}}
\put(485,391){\raisebox{-.8pt}{\makebox(0,0){$\cdot$}}}
\put(488,393){\raisebox{-.8pt}{\makebox(0,0){$\cdot$}}}
\put(490,405){\raisebox{-.8pt}{\makebox(0,0){$\cdot$}}}
\put(493,394){\raisebox{-.8pt}{\makebox(0,0){$\cdot$}}}
\put(495,405){\raisebox{-.8pt}{\makebox(0,0){$\cdot$}}}
\put(498,414){\raisebox{-.8pt}{\makebox(0,0){$\cdot$}}}
\put(501,417){\raisebox{-.8pt}{\makebox(0,0){$\cdot$}}}
\put(504,426){\raisebox{-.8pt}{\makebox(0,0){$\cdot$}}}
\put(506,436){\raisebox{-.8pt}{\makebox(0,0){$\cdot$}}}
\put(509,442){\raisebox{-.8pt}{\makebox(0,0){$\cdot$}}}
\put(511,441){\raisebox{-.8pt}{\makebox(0,0){$\cdot$}}}
\put(514,450){\raisebox{-.8pt}{\makebox(0,0){$\cdot$}}}
\put(517,439){\raisebox{-.8pt}{\makebox(0,0){$\cdot$}}}
\put(519,436){\raisebox{-.8pt}{\makebox(0,0){$\cdot$}}}
\put(522,441){\raisebox{-.8pt}{\makebox(0,0){$\cdot$}}}
\put(525,445){\raisebox{-.8pt}{\makebox(0,0){$\cdot$}}}
\put(527,430){\raisebox{-.8pt}{\makebox(0,0){$\cdot$}}}
\put(530,425){\raisebox{-.8pt}{\makebox(0,0){$\cdot$}}}
\put(533,400){\raisebox{-.8pt}{\makebox(0,0){$\cdot$}}}
\put(535,408){\raisebox{-.8pt}{\makebox(0,0){$\cdot$}}}
\put(538,403){\raisebox{-.8pt}{\makebox(0,0){$\cdot$}}}
\put(540,414){\raisebox{-.8pt}{\makebox(0,0){$\cdot$}}}
\put(543,416){\raisebox{-.8pt}{\makebox(0,0){$\cdot$}}}
\put(546,421){\raisebox{-.8pt}{\makebox(0,0){$\cdot$}}}
\put(548,433){\raisebox{-.8pt}{\makebox(0,0){$\cdot$}}}
\put(551,440){\raisebox{-.8pt}{\makebox(0,0){$\cdot$}}}
\put(554,443){\raisebox{-.8pt}{\makebox(0,0){$\cdot$}}}
\put(556,441){\raisebox{-.8pt}{\makebox(0,0){$\cdot$}}}
\put(559,449){\raisebox{-.8pt}{\makebox(0,0){$\cdot$}}}
\put(561,446){\raisebox{-.8pt}{\makebox(0,0){$\cdot$}}}
\put(564,445){\raisebox{-.8pt}{\makebox(0,0){$\cdot$}}}
\put(567,437){\raisebox{-.8pt}{\makebox(0,0){$\cdot$}}}
\put(569,431){\raisebox{-.8pt}{\makebox(0,0){$\cdot$}}}
\put(572,436){\raisebox{-.8pt}{\makebox(0,0){$\cdot$}}}
\put(574,429){\raisebox{-.8pt}{\makebox(0,0){$\cdot$}}}
\put(577,419){\raisebox{-.8pt}{\makebox(0,0){$\cdot$}}}
\put(579,403){\raisebox{-.8pt}{\makebox(0,0){$\cdot$}}}
\put(582,410){\raisebox{-.8pt}{\makebox(0,0){$\cdot$}}}
\put(585,418){\raisebox{-.8pt}{\makebox(0,0){$\cdot$}}}
\put(588,414){\raisebox{-.8pt}{\makebox(0,0){$\cdot$}}}
\put(590,414){\raisebox{-.8pt}{\makebox(0,0){$\cdot$}}}
\put(593,419){\raisebox{-.8pt}{\makebox(0,0){$\cdot$}}}
\put(595,413){\raisebox{-.8pt}{\makebox(0,0){$\cdot$}}}
\put(598,408){\raisebox{-.8pt}{\makebox(0,0){$\cdot$}}}
\put(600,400){\raisebox{-.8pt}{\makebox(0,0){$\cdot$}}}
\put(603,400){\raisebox{-.8pt}{\makebox(0,0){$\cdot$}}}
\put(606,423){\raisebox{-.8pt}{\makebox(0,0){$\cdot$}}}
\put(609,420){\raisebox{-.8pt}{\makebox(0,0){$\cdot$}}}
\put(611,423){\raisebox{-.8pt}{\makebox(0,0){$\cdot$}}}
\put(614,425){\raisebox{-.8pt}{\makebox(0,0){$\cdot$}}}
\put(616,427){\raisebox{-.8pt}{\makebox(0,0){$\cdot$}}}
\put(619,422){\raisebox{-.8pt}{\makebox(0,0){$\cdot$}}}
\put(621,405){\raisebox{-.8pt}{\makebox(0,0){$\cdot$}}}
\put(624,406){\raisebox{-.8pt}{\makebox(0,0){$\cdot$}}}
\put(626,405){\raisebox{-.8pt}{\makebox(0,0){$\cdot$}}}
\put(629,393){\raisebox{-.8pt}{\makebox(0,0){$\cdot$}}}
\put(631,395){\raisebox{-.8pt}{\makebox(0,0){$\cdot$}}}
\put(634,394){\raisebox{-.8pt}{\makebox(0,0){$\cdot$}}}
\put(638,396){\raisebox{-.8pt}{\makebox(0,0){$\cdot$}}}
\put(640,399){\raisebox{-.8pt}{\makebox(0,0){$\cdot$}}}
\put(643,402){\raisebox{-.8pt}{\makebox(0,0){$\cdot$}}}
\put(646,400){\raisebox{-.8pt}{\makebox(0,0){$\cdot$}}}
\put(648,402){\raisebox{-.8pt}{\makebox(0,0){$\cdot$}}}
\put(651,405){\raisebox{-.8pt}{\makebox(0,0){$\cdot$}}}
\put(653,415){\raisebox{-.8pt}{\makebox(0,0){$\cdot$}}}
\put(655,416){\raisebox{-.8pt}{\makebox(0,0){$\cdot$}}}
\put(658,420){\raisebox{-.8pt}{\makebox(0,0){$\cdot$}}}
\put(661,426){\raisebox{-.8pt}{\makebox(0,0){$\cdot$}}}
\put(664,427){\raisebox{-.8pt}{\makebox(0,0){$\cdot$}}}
\put(667,434){\raisebox{-.8pt}{\makebox(0,0){$\cdot$}}}
\put(669,433){\raisebox{-.8pt}{\makebox(0,0){$\cdot$}}}
\put(672,434){\raisebox{-.8pt}{\makebox(0,0){$\cdot$}}}
\put(674,434){\raisebox{-.8pt}{\makebox(0,0){$\cdot$}}}
\put(677,441){\raisebox{-.8pt}{\makebox(0,0){$\cdot$}}}
\put(679,429){\raisebox{-.8pt}{\makebox(0,0){$\cdot$}}}
\put(682,432){\raisebox{-.8pt}{\makebox(0,0){$\cdot$}}}
\put(685,446){\raisebox{-.8pt}{\makebox(0,0){$\cdot$}}}
\put(688,450){\raisebox{-.8pt}{\makebox(0,0){$\cdot$}}}
\put(690,445){\raisebox{-.8pt}{\makebox(0,0){$\cdot$}}}
\put(693,443){\raisebox{-.8pt}{\makebox(0,0){$\cdot$}}}
\put(695,434){\raisebox{-.8pt}{\makebox(0,0){$\cdot$}}}
\put(698,425){\raisebox{-.8pt}{\makebox(0,0){$\cdot$}}}
\put(700,416){\raisebox{-.8pt}{\makebox(0,0){$\cdot$}}}
\put(703,403){\raisebox{-.8pt}{\makebox(0,0){$\cdot$}}}
\put(706,412){\raisebox{-.8pt}{\makebox(0,0){$\cdot$}}}
\put(708,406){\raisebox{-.8pt}{\makebox(0,0){$\cdot$}}}
\put(711,409){\raisebox{-.8pt}{\makebox(0,0){$\cdot$}}}
\put(713,395){\raisebox{-.8pt}{\makebox(0,0){$\cdot$}}}
\put(716,382){\raisebox{-.8pt}{\makebox(0,0){$\cdot$}}}
\put(719,388){\raisebox{-.8pt}{\makebox(0,0){$\cdot$}}}
\put(721,364){\raisebox{-.8pt}{\makebox(0,0){$\cdot$}}}
\put(724,356){\raisebox{-.8pt}{\makebox(0,0){$\cdot$}}}
\put(727,357){\raisebox{-.8pt}{\makebox(0,0){$\cdot$}}}
\put(729,357){\raisebox{-.8pt}{\makebox(0,0){$\cdot$}}}
\put(732,356){\raisebox{-.8pt}{\makebox(0,0){$\cdot$}}}
\put(734,362){\raisebox{-.8pt}{\makebox(0,0){$\cdot$}}}
\put(737,366){\raisebox{-.8pt}{\makebox(0,0){$\cdot$}}}
\put(740,374){\raisebox{-.8pt}{\makebox(0,0){$\cdot$}}}
\put(742,371){\raisebox{-.8pt}{\makebox(0,0){$\cdot$}}}
\put(745,364){\raisebox{-.8pt}{\makebox(0,0){$\cdot$}}}
\put(748,348){\raisebox{-.8pt}{\makebox(0,0){$\cdot$}}}
\put(750,355){\raisebox{-.8pt}{\makebox(0,0){$\cdot$}}}
\put(752,368){\raisebox{-.8pt}{\makebox(0,0){$\cdot$}}}
\put(755,361){\raisebox{-.8pt}{\makebox(0,0){$\cdot$}}}
\put(758,368){\raisebox{-.8pt}{\makebox(0,0){$\cdot$}}}
\put(760,375){\raisebox{-.8pt}{\makebox(0,0){$\cdot$}}}
\put(763,375){\raisebox{-.8pt}{\makebox(0,0){$\cdot$}}}
\put(765,373){\raisebox{-.8pt}{\makebox(0,0){$\cdot$}}}
\put(768,368){\raisebox{-.8pt}{\makebox(0,0){$\cdot$}}}
\put(771,363){\raisebox{-.8pt}{\makebox(0,0){$\cdot$}}}
\put(773,363){\raisebox{-.8pt}{\makebox(0,0){$\cdot$}}}
\put(777,376){\raisebox{-.8pt}{\makebox(0,0){$\cdot$}}}
\put(779,378){\raisebox{-.8pt}{\makebox(0,0){$\cdot$}}}
\put(782,377){\raisebox{-.8pt}{\makebox(0,0){$\cdot$}}}
\put(785,373){\raisebox{-.8pt}{\makebox(0,0){$\cdot$}}}
\put(787,364){\raisebox{-.8pt}{\makebox(0,0){$\cdot$}}}
\put(790,364){\raisebox{-.8pt}{\makebox(0,0){$\cdot$}}}
\put(792,374){\raisebox{-.8pt}{\makebox(0,0){$\cdot$}}}
\put(795,380){\raisebox{-.8pt}{\makebox(0,0){$\cdot$}}}
\put(798,385){\raisebox{-.8pt}{\makebox(0,0){$\cdot$}}}
\put(800,389){\raisebox{-.8pt}{\makebox(0,0){$\cdot$}}}
\put(803,383){\raisebox{-.8pt}{\makebox(0,0){$\cdot$}}}
\put(806,387){\raisebox{-.8pt}{\makebox(0,0){$\cdot$}}}
\put(808,389){\raisebox{-.8pt}{\makebox(0,0){$\cdot$}}}
\put(811,396){\raisebox{-.8pt}{\makebox(0,0){$\cdot$}}}
\put(813,402){\raisebox{-.8pt}{\makebox(0,0){$\cdot$}}}
\put(816,401){\raisebox{-.8pt}{\makebox(0,0){$\cdot$}}}
\put(819,403){\raisebox{-.8pt}{\makebox(0,0){$\cdot$}}}
\put(821,408){\raisebox{-.8pt}{\makebox(0,0){$\cdot$}}}
\put(824,411){\raisebox{-.8pt}{\makebox(0,0){$\cdot$}}}
\put(827,415){\raisebox{-.8pt}{\makebox(0,0){$\cdot$}}}
\put(829,430){\raisebox{-.8pt}{\makebox(0,0){$\cdot$}}}
\put(832,434){\raisebox{-.8pt}{\makebox(0,0){$\cdot$}}}
\put(834,439){\raisebox{-.8pt}{\makebox(0,0){$\cdot$}}}
\put(837,435){\raisebox{-.8pt}{\makebox(0,0){$\cdot$}}}
\put(840,437){\raisebox{-.8pt}{\makebox(0,0){$\cdot$}}}
\put(842,437){\raisebox{-.8pt}{\makebox(0,0){$\cdot$}}}
\put(845,442){\raisebox{-.8pt}{\makebox(0,0){$\cdot$}}}
\put(847,448){\raisebox{-.8pt}{\makebox(0,0){$\cdot$}}}
\put(850,455){\raisebox{-.8pt}{\makebox(0,0){$\cdot$}}}
\put(853,455){\raisebox{-.8pt}{\makebox(0,0){$\cdot$}}}
\put(855,462){\raisebox{-.8pt}{\makebox(0,0){$\cdot$}}}
\put(858,471){\raisebox{-.8pt}{\makebox(0,0){$\cdot$}}}
\put(861,474){\raisebox{-.8pt}{\makebox(0,0){$\cdot$}}}
\put(863,468){\raisebox{-.8pt}{\makebox(0,0){$\cdot$}}}
\put(866,471){\raisebox{-.8pt}{\makebox(0,0){$\cdot$}}}
\put(868,482){\raisebox{-.8pt}{\makebox(0,0){$\cdot$}}}
\put(871,491){\raisebox{-.8pt}{\makebox(0,0){$\cdot$}}}
\put(874,486){\raisebox{-.8pt}{\makebox(0,0){$\cdot$}}}
\put(876,485){\raisebox{-.8pt}{\makebox(0,0){$\cdot$}}}
\put(879,484){\raisebox{-.8pt}{\makebox(0,0){$\cdot$}}}
\put(882,489){\raisebox{-.8pt}{\makebox(0,0){$\cdot$}}}
\put(884,499){\raisebox{-.8pt}{\makebox(0,0){$\cdot$}}}
\put(886,498){\raisebox{-.8pt}{\makebox(0,0){$\cdot$}}}
\put(889,512){\raisebox{-.8pt}{\makebox(0,0){$\cdot$}}}
\put(892,516){\raisebox{-.8pt}{\makebox(0,0){$\cdot$}}}
\put(894,519){\raisebox{-.8pt}{\makebox(0,0){$\cdot$}}}
\put(897,519){\raisebox{-.8pt}{\makebox(0,0){$\cdot$}}}
\put(900,517){\raisebox{-.8pt}{\makebox(0,0){$\cdot$}}}
\put(902,507){\raisebox{-.8pt}{\makebox(0,0){$\cdot$}}}
\put(905,512){\raisebox{-.8pt}{\makebox(0,0){$\cdot$}}}
\put(907,524){\raisebox{-.8pt}{\makebox(0,0){$\cdot$}}}
\put(910,525){\raisebox{-.8pt}{\makebox(0,0){$\cdot$}}}
\put(913,530){\raisebox{-.8pt}{\makebox(0,0){$\cdot$}}}
\put(916,535){\raisebox{-.8pt}{\makebox(0,0){$\cdot$}}}
\put(919,531){\raisebox{-.8pt}{\makebox(0,0){$\cdot$}}}
\put(921,528){\raisebox{-.8pt}{\makebox(0,0){$\cdot$}}}
\put(924,532){\raisebox{-.8pt}{\makebox(0,0){$\cdot$}}}
\put(926,527){\raisebox{-.8pt}{\makebox(0,0){$\cdot$}}}
\put(929,542){\raisebox{-.8pt}{\makebox(0,0){$\cdot$}}}
\put(932,545){\raisebox{-.8pt}{\makebox(0,0){$\cdot$}}}
\put(934,543){\raisebox{-.8pt}{\makebox(0,0){$\cdot$}}}
\put(937,540){\raisebox{-.8pt}{\makebox(0,0){$\cdot$}}}
\put(940,551){\raisebox{-.8pt}{\makebox(0,0){$\cdot$}}}
\put(942,551){\raisebox{-.8pt}{\makebox(0,0){$\cdot$}}}
\put(945,558){\raisebox{-.8pt}{\makebox(0,0){$\cdot$}}}
\put(947,547){\raisebox{-.8pt}{\makebox(0,0){$\cdot$}}}
\put(950,545){\raisebox{-.8pt}{\makebox(0,0){$\cdot$}}}
\put(953,547){\raisebox{-.8pt}{\makebox(0,0){$\cdot$}}}
\put(955,555){\raisebox{-.8pt}{\makebox(0,0){$\cdot$}}}
\put(958,567){\raisebox{-.8pt}{\makebox(0,0){$\cdot$}}}
\put(961,574){\raisebox{-.8pt}{\makebox(0,0){$\cdot$}}}
\put(963,569){\raisebox{-.8pt}{\makebox(0,0){$\cdot$}}}
\put(966,574){\raisebox{-.8pt}{\makebox(0,0){$\cdot$}}}
\put(969,583){\raisebox{-.8pt}{\makebox(0,0){$\cdot$}}}
\put(971,578){\raisebox{-.8pt}{\makebox(0,0){$\cdot$}}}
\put(974,570){\raisebox{-.8pt}{\makebox(0,0){$\cdot$}}}
\put(976,568){\raisebox{-.8pt}{\makebox(0,0){$\cdot$}}}
\put(979,574){\raisebox{-.8pt}{\makebox(0,0){$\cdot$}}}
\put(981,565){\raisebox{-.8pt}{\makebox(0,0){$\cdot$}}}
\put(984,553){\raisebox{-.8pt}{\makebox(0,0){$\cdot$}}}
\put(987,544){\raisebox{-.8pt}{\makebox(0,0){$\cdot$}}}
\put(989,549){\raisebox{-.8pt}{\makebox(0,0){$\cdot$}}}
\put(992,553){\raisebox{-.8pt}{\makebox(0,0){$\cdot$}}}
\put(995,553){\raisebox{-.8pt}{\makebox(0,0){$\cdot$}}}
\put(997,558){\raisebox{-.8pt}{\makebox(0,0){$\cdot$}}}
\put(1000,550){\raisebox{-.8pt}{\makebox(0,0){$\cdot$}}}
\put(1002,561){\raisebox{-.8pt}{\makebox(0,0){$\cdot$}}}
\put(1005,564){\raisebox{-.8pt}{\makebox(0,0){$\cdot$}}}
\put(1008,557){\raisebox{-.8pt}{\makebox(0,0){$\cdot$}}}
\put(1010,560){\raisebox{-.8pt}{\makebox(0,0){$\cdot$}}}
\put(1013,565){\raisebox{-.8pt}{\makebox(0,0){$\cdot$}}}
\put(1016,574){\raisebox{-.8pt}{\makebox(0,0){$\cdot$}}}
\put(1018,579){\raisebox{-.8pt}{\makebox(0,0){$\cdot$}}}
\put(1021,584){\raisebox{-.8pt}{\makebox(0,0){$\cdot$}}}
\put(1023,579){\raisebox{-.8pt}{\makebox(0,0){$\cdot$}}}
\put(1026,586){\raisebox{-.8pt}{\makebox(0,0){$\cdot$}}}
\put(1028,583){\raisebox{-.8pt}{\makebox(0,0){$\cdot$}}}
\put(1031,571){\raisebox{-.8pt}{\makebox(0,0){$\cdot$}}}
\put(1033,560){\raisebox{-.8pt}{\makebox(0,0){$\cdot$}}}
\put(1036,541){\raisebox{-.8pt}{\makebox(0,0){$\cdot$}}}
\put(1039,545){\raisebox{-.8pt}{\makebox(0,0){$\cdot$}}}
\put(1041,541){\raisebox{-.8pt}{\makebox(0,0){$\cdot$}}}
\put(1044,546){\raisebox{-.8pt}{\makebox(0,0){$\cdot$}}}
\put(1046,549){\raisebox{-.8pt}{\makebox(0,0){$\cdot$}}}
\put(1050,531){\raisebox{-.8pt}{\makebox(0,0){$\cdot$}}}
\put(1052,527){\raisebox{-.8pt}{\makebox(0,0){$\cdot$}}}
\put(1055,536){\raisebox{-.8pt}{\makebox(0,0){$\cdot$}}}
\put(1057,541){\raisebox{-.8pt}{\makebox(0,0){$\cdot$}}}
\put(1060,536){\raisebox{-.8pt}{\makebox(0,0){$\cdot$}}}
\put(1063,546){\raisebox{-.8pt}{\makebox(0,0){$\cdot$}}}
\put(1065,547){\raisebox{-.8pt}{\makebox(0,0){$\cdot$}}}
\put(1068,536){\raisebox{-.8pt}{\makebox(0,0){$\cdot$}}}
\put(1070,524){\raisebox{-.8pt}{\makebox(0,0){$\cdot$}}}
\put(1073,529){\raisebox{-.8pt}{\makebox(0,0){$\cdot$}}}
\put(1076,536){\raisebox{-.8pt}{\makebox(0,0){$\cdot$}}}
\put(1078,543){\raisebox{-.8pt}{\makebox(0,0){$\cdot$}}}
\put(1081,543){\raisebox{-.8pt}{\makebox(0,0){$\cdot$}}}
\put(1084,547){\raisebox{-.8pt}{\makebox(0,0){$\cdot$}}}
\put(1086,567){\raisebox{-.8pt}{\makebox(0,0){$\cdot$}}}
\put(1089,564){\raisebox{-.8pt}{\makebox(0,0){$\cdot$}}}
\put(1091,561){\raisebox{-.8pt}{\makebox(0,0){$\cdot$}}}
\put(1094,556){\raisebox{-.8pt}{\makebox(0,0){$\cdot$}}}
\put(1097,561){\raisebox{-.8pt}{\makebox(0,0){$\cdot$}}}
\put(1099,548){\raisebox{-.8pt}{\makebox(0,0){$\cdot$}}}
\put(1102,531){\raisebox{-.8pt}{\makebox(0,0){$\cdot$}}}
\put(1105,536){\raisebox{-.8pt}{\makebox(0,0){$\cdot$}}}
\put(1107,534){\raisebox{-.8pt}{\makebox(0,0){$\cdot$}}}
\put(1110,544){\raisebox{-.8pt}{\makebox(0,0){$\cdot$}}}
\put(1112,547){\raisebox{-.8pt}{\makebox(0,0){$\cdot$}}}
\put(1115,552){\raisebox{-.8pt}{\makebox(0,0){$\cdot$}}}
\put(1117,543){\raisebox{-.8pt}{\makebox(0,0){$\cdot$}}}
\put(1120,534){\raisebox{-.8pt}{\makebox(0,0){$\cdot$}}}
\put(1123,524){\raisebox{-.8pt}{\makebox(0,0){$\cdot$}}}
\put(1125,509){\raisebox{-.8pt}{\makebox(0,0){$\cdot$}}}
\put(1128,505){\raisebox{-.8pt}{\makebox(0,0){$\cdot$}}}
\put(1131,511){\raisebox{-.8pt}{\makebox(0,0){$\cdot$}}}
\put(1133,519){\raisebox{-.8pt}{\makebox(0,0){$\cdot$}}}
\put(1136,503){\raisebox{-.8pt}{\makebox(0,0){$\cdot$}}}
\put(1139,503){\raisebox{-.8pt}{\makebox(0,0){$\cdot$}}}
\put(1141,519){\raisebox{-.8pt}{\makebox(0,0){$\cdot$}}}
\put(1144,530){\raisebox{-.8pt}{\makebox(0,0){$\cdot$}}}
\put(1146,525){\raisebox{-.8pt}{\makebox(0,0){$\cdot$}}}
\put(1149,528){\raisebox{-.8pt}{\makebox(0,0){$\cdot$}}}
\put(1152,519){\raisebox{-.8pt}{\makebox(0,0){$\cdot$}}}
\put(1154,532){\raisebox{-.8pt}{\makebox(0,0){$\cdot$}}}
\put(1157,537){\raisebox{-.8pt}{\makebox(0,0){$\cdot$}}}
\put(1160,533){\raisebox{-.8pt}{\makebox(0,0){$\cdot$}}}
\put(1162,536){\raisebox{-.8pt}{\makebox(0,0){$\cdot$}}}
\put(1164,540){\raisebox{-.8pt}{\makebox(0,0){$\cdot$}}}
\put(1167,550){\raisebox{-.8pt}{\makebox(0,0){$\cdot$}}}
\put(1170,551){\raisebox{-.8pt}{\makebox(0,0){$\cdot$}}}
\put(1172,558){\raisebox{-.8pt}{\makebox(0,0){$\cdot$}}}
\put(1175,566){\raisebox{-.8pt}{\makebox(0,0){$\cdot$}}}
\put(1177,555){\raisebox{-.8pt}{\makebox(0,0){$\cdot$}}}
\put(1180,565){\raisebox{-.8pt}{\makebox(0,0){$\cdot$}}}
\put(1183,568){\raisebox{-.8pt}{\makebox(0,0){$\cdot$}}}
\put(1186,580){\raisebox{-.8pt}{\makebox(0,0){$\cdot$}}}
\put(1189,574){\raisebox{-.8pt}{\makebox(0,0){$\cdot$}}}
\put(1191,584){\raisebox{-.8pt}{\makebox(0,0){$\cdot$}}}
\put(1194,581){\raisebox{-.8pt}{\makebox(0,0){$\cdot$}}}
\put(1196,584){\raisebox{-.8pt}{\makebox(0,0){$\cdot$}}}
\put(1199,589){\raisebox{-.8pt}{\makebox(0,0){$\cdot$}}}
\put(1202,597){\raisebox{-.8pt}{\makebox(0,0){$\cdot$}}}
\put(1204,591){\raisebox{-.8pt}{\makebox(0,0){$\cdot$}}}
\put(1207,586){\raisebox{-.8pt}{\makebox(0,0){$\cdot$}}}
\put(1210,595){\raisebox{-.8pt}{\makebox(0,0){$\cdot$}}}
\put(1212,593){\raisebox{-.8pt}{\makebox(0,0){$\cdot$}}}
\put(1215,606){\raisebox{-.8pt}{\makebox(0,0){$\cdot$}}}
\put(1218,599){\raisebox{-.8pt}{\makebox(0,0){$\cdot$}}}
\put(1220,600){\raisebox{-.8pt}{\makebox(0,0){$\cdot$}}}
\put(1223,602){\raisebox{-.8pt}{\makebox(0,0){$\cdot$}}}
\put(1225,591){\raisebox{-.8pt}{\makebox(0,0){$\cdot$}}}
\put(1228,593){\raisebox{-.8pt}{\makebox(0,0){$\cdot$}}}
\put(1231,590){\raisebox{-.8pt}{\makebox(0,0){$\cdot$}}}
\put(1233,596){\raisebox{-.8pt}{\makebox(0,0){$\cdot$}}}
\put(1236,595){\raisebox{-.8pt}{\makebox(0,0){$\cdot$}}}
\put(1239,589){\raisebox{-.8pt}{\makebox(0,0){$\cdot$}}}
\put(1241,605){\raisebox{-.8pt}{\makebox(0,0){$\cdot$}}}
\put(1244,616){\raisebox{-.8pt}{\makebox(0,0){$\cdot$}}}
\put(1246,618){\raisebox{-.8pt}{\makebox(0,0){$\cdot$}}}
\put(1249,618){\raisebox{-.8pt}{\makebox(0,0){$\cdot$}}}
\put(1252,614){\raisebox{-.8pt}{\makebox(0,0){$\cdot$}}}
\put(1254,617){\raisebox{-.8pt}{\makebox(0,0){$\cdot$}}}
\put(1257,617){\raisebox{-.8pt}{\makebox(0,0){$\cdot$}}}
\put(1259,613){\raisebox{-.8pt}{\makebox(0,0){$\cdot$}}}
\put(1262,603){\raisebox{-.8pt}{\makebox(0,0){$\cdot$}}}
\put(1265,602){\raisebox{-.8pt}{\makebox(0,0){$\cdot$}}}
\put(1267,607){\raisebox{-.8pt}{\makebox(0,0){$\cdot$}}}
\put(1270,604){\raisebox{-.8pt}{\makebox(0,0){$\cdot$}}}
\put(1273,606){\raisebox{-.8pt}{\makebox(0,0){$\cdot$}}}
\put(1275,601){\raisebox{-.8pt}{\makebox(0,0){$\cdot$}}}
\put(1278,608){\raisebox{-.8pt}{\makebox(0,0){$\cdot$}}}
\put(1280,619){\raisebox{-.8pt}{\makebox(0,0){$\cdot$}}}
\put(1283,623){\raisebox{-.8pt}{\makebox(0,0){$\cdot$}}}
\put(1286,625){\raisebox{-.8pt}{\makebox(0,0){$\cdot$}}}
\put(1288,622){\raisebox{-.8pt}{\makebox(0,0){$\cdot$}}}
\put(1291,622){\raisebox{-.8pt}{\makebox(0,0){$\cdot$}}}
\put(1294,623){\raisebox{-.8pt}{\makebox(0,0){$\cdot$}}}
\put(1296,617){\raisebox{-.8pt}{\makebox(0,0){$\cdot$}}}
\put(1299,624){\raisebox{-.8pt}{\makebox(0,0){$\cdot$}}}
\put(1301,612){\raisebox{-.8pt}{\makebox(0,0){$\cdot$}}}
\put(1304,610){\raisebox{-.8pt}{\makebox(0,0){$\cdot$}}}
\put(1306,602){\raisebox{-.8pt}{\makebox(0,0){$\cdot$}}}
\put(1309,596){\raisebox{-.8pt}{\makebox(0,0){$\cdot$}}}
\put(1311,606){\raisebox{-.8pt}{\makebox(0,0){$\cdot$}}}
\put(1314,612){\raisebox{-.8pt}{\makebox(0,0){$\cdot$}}}
\put(1317,613){\raisebox{-.8pt}{\makebox(0,0){$\cdot$}}}
\put(1319,608){\raisebox{-.8pt}{\makebox(0,0){$\cdot$}}}
\put(1323,609){\raisebox{-.8pt}{\makebox(0,0){$\cdot$}}}
\put(1325,612){\raisebox{-.8pt}{\makebox(0,0){$\cdot$}}}
\put(1328,618){\raisebox{-.8pt}{\makebox(0,0){$\cdot$}}}
\put(1331,613){\raisebox{-.8pt}{\makebox(0,0){$\cdot$}}}
\put(1333,607){\raisebox{-.8pt}{\makebox(0,0){$\cdot$}}}
\put(1336,601){\raisebox{-.8pt}{\makebox(0,0){$\cdot$}}}
\put(1338,592){\raisebox{-.8pt}{\makebox(0,0){$\cdot$}}}
\put(1341,597){\raisebox{-.8pt}{\makebox(0,0){$\cdot$}}}
\put(1344,581){\raisebox{-.8pt}{\makebox(0,0){$\cdot$}}}
\put(1346,580){\raisebox{-.8pt}{\makebox(0,0){$\cdot$}}}
\put(1349,556){\raisebox{-.8pt}{\makebox(0,0){$\cdot$}}}
\put(1352,564){\raisebox{-.8pt}{\makebox(0,0){$\cdot$}}}
\put(1354,532){\raisebox{-.8pt}{\makebox(0,0){$\cdot$}}}
\put(1357,532){\raisebox{-.8pt}{\makebox(0,0){$\cdot$}}}
\put(1359,524){\raisebox{-.8pt}{\makebox(0,0){$\cdot$}}}
\put(1362,503){\raisebox{-.8pt}{\makebox(0,0){$\cdot$}}}
\put(1365,467){\raisebox{-.8pt}{\makebox(0,0){$\cdot$}}}
\put(1367,488){\raisebox{-.8pt}{\makebox(0,0){$\cdot$}}}
\put(1370,501){\raisebox{-.8pt}{\makebox(0,0){$\cdot$}}}
\put(1373,514){\raisebox{-.8pt}{\makebox(0,0){$\cdot$}}}
\put(1375,502){\raisebox{-.8pt}{\makebox(0,0){$\cdot$}}}
\put(1378,509){\raisebox{-.8pt}{\makebox(0,0){$\cdot$}}}
\put(1380,519){\raisebox{-.8pt}{\makebox(0,0){$\cdot$}}}
\put(1383,516){\raisebox{-.8pt}{\makebox(0,0){$\cdot$}}}
\put(1386,531){\raisebox{-.8pt}{\makebox(0,0){$\cdot$}}}
\put(1388,534){\raisebox{-.8pt}{\makebox(0,0){$\cdot$}}}
\put(1391,530){\raisebox{-.8pt}{\makebox(0,0){$\cdot$}}}
\put(1393,523){\raisebox{-.8pt}{\makebox(0,0){$\cdot$}}}
\put(1396,527){\raisebox{-.8pt}{\makebox(0,0){$\cdot$}}}
\put(1398,515){\raisebox{-.8pt}{\makebox(0,0){$\cdot$}}}
\put(1401,504){\raisebox{-.8pt}{\makebox(0,0){$\cdot$}}}
\put(1404,511){\raisebox{-.8pt}{\makebox(0,0){$\cdot$}}}
\put(1407,511){\raisebox{-.8pt}{\makebox(0,0){$\cdot$}}}
\put(1409,499){\raisebox{-.8pt}{\makebox(0,0){$\cdot$}}}
\put(1412,495){\raisebox{-.8pt}{\makebox(0,0){$\cdot$}}}
\put(1414,526){\raisebox{-.8pt}{\makebox(0,0){$\cdot$}}}
\put(1417,536){\raisebox{-.8pt}{\makebox(0,0){$\cdot$}}}
\put(1419,548){\raisebox{-.8pt}{\makebox(0,0){$\cdot$}}}
\put(1422,559){\raisebox{-.8pt}{\makebox(0,0){$\cdot$}}}
\put(1425,563){\raisebox{-.8pt}{\makebox(0,0){$\cdot$}}}
\put(1428,565){\raisebox{-.8pt}{\makebox(0,0){$\cdot$}}}
\put(1430,562){\raisebox{-.8pt}{\makebox(0,0){$\cdot$}}}
\put(1433,561){\raisebox{-.8pt}{\makebox(0,0){$\cdot$}}}
\put(1435,565){\raisebox{-.8pt}{\makebox(0,0){$\cdot$}}}
\put(220,146){\usebox{\plotpoint}}
\multiput(220.00,146.58)(0.496,0.492){21}{\rule{0.500pt}{0.119pt}}
\multiput(220.00,145.17)(10.962,12.000){2}{\rule{0.250pt}{0.400pt}}
\multiput(232.58,158.00)(0.493,0.536){23}{\rule{0.119pt}{0.531pt}}
\multiput(231.17,158.00)(13.000,12.898){2}{\rule{0.400pt}{0.265pt}}
\multiput(245.58,172.00)(0.492,0.625){21}{\rule{0.119pt}{0.600pt}}
\multiput(244.17,172.00)(12.000,13.755){2}{\rule{0.400pt}{0.300pt}}
\multiput(257.58,187.00)(0.492,0.712){21}{\rule{0.119pt}{0.667pt}}
\multiput(256.17,187.00)(12.000,15.616){2}{\rule{0.400pt}{0.333pt}}
\multiput(269.58,204.00)(0.492,0.755){21}{\rule{0.119pt}{0.700pt}}
\multiput(268.17,204.00)(12.000,16.547){2}{\rule{0.400pt}{0.350pt}}
\multiput(281.58,222.00)(0.493,0.695){23}{\rule{0.119pt}{0.654pt}}
\multiput(280.17,222.00)(13.000,16.643){2}{\rule{0.400pt}{0.327pt}}
\multiput(294.58,240.00)(0.492,0.798){21}{\rule{0.119pt}{0.733pt}}
\multiput(293.17,240.00)(12.000,17.478){2}{\rule{0.400pt}{0.367pt}}
\multiput(306.58,259.00)(0.492,0.755){21}{\rule{0.119pt}{0.700pt}}
\multiput(305.17,259.00)(12.000,16.547){2}{\rule{0.400pt}{0.350pt}}
\multiput(318.58,277.00)(0.493,0.695){23}{\rule{0.119pt}{0.654pt}}
\multiput(317.17,277.00)(13.000,16.643){2}{\rule{0.400pt}{0.327pt}}
\multiput(331.58,295.00)(0.492,0.755){21}{\rule{0.119pt}{0.700pt}}
\multiput(330.17,295.00)(12.000,16.547){2}{\rule{0.400pt}{0.350pt}}
\multiput(343.58,313.00)(0.492,0.669){21}{\rule{0.119pt}{0.633pt}}
\multiput(342.17,313.00)(12.000,14.685){2}{\rule{0.400pt}{0.317pt}}
\multiput(355.58,329.00)(0.492,0.625){21}{\rule{0.119pt}{0.600pt}}
\multiput(354.17,329.00)(12.000,13.755){2}{\rule{0.400pt}{0.300pt}}
\multiput(367.58,344.00)(0.493,0.536){23}{\rule{0.119pt}{0.531pt}}
\multiput(366.17,344.00)(13.000,12.898){2}{\rule{0.400pt}{0.265pt}}
\multiput(380.00,358.58)(0.496,0.492){21}{\rule{0.500pt}{0.119pt}}
\multiput(380.00,357.17)(10.962,12.000){2}{\rule{0.250pt}{0.400pt}}
\multiput(392.00,370.58)(0.543,0.492){19}{\rule{0.536pt}{0.118pt}}
\multiput(392.00,369.17)(10.887,11.000){2}{\rule{0.268pt}{0.400pt}}
\multiput(404.00,381.59)(0.728,0.489){15}{\rule{0.678pt}{0.118pt}}
\multiput(404.00,380.17)(11.593,9.000){2}{\rule{0.339pt}{0.400pt}}
\multiput(417.00,390.59)(0.758,0.488){13}{\rule{0.700pt}{0.117pt}}
\multiput(417.00,389.17)(10.547,8.000){2}{\rule{0.350pt}{0.400pt}}
\multiput(429.00,398.59)(1.033,0.482){9}{\rule{0.900pt}{0.116pt}}
\multiput(429.00,397.17)(10.132,6.000){2}{\rule{0.450pt}{0.400pt}}
\multiput(441.00,404.59)(1.267,0.477){7}{\rule{1.060pt}{0.115pt}}
\multiput(441.00,403.17)(9.800,5.000){2}{\rule{0.530pt}{0.400pt}}
\multiput(453.00,409.60)(1.797,0.468){5}{\rule{1.400pt}{0.113pt}}
\multiput(453.00,408.17)(10.094,4.000){2}{\rule{0.700pt}{0.400pt}}
\multiput(466.00,413.61)(2.472,0.447){3}{\rule{1.700pt}{0.108pt}}
\multiput(466.00,412.17)(8.472,3.000){2}{\rule{0.850pt}{0.400pt}}
\put(478,415.67){\rule{2.891pt}{0.400pt}}
\multiput(478.00,415.17)(6.000,1.000){2}{\rule{1.445pt}{0.400pt}}
\put(490,416.67){\rule{3.132pt}{0.400pt}}
\multiput(490.00,416.17)(6.500,1.000){2}{\rule{1.566pt}{0.400pt}}
\put(515,416.67){\rule{2.891pt}{0.400pt}}
\multiput(515.00,417.17)(6.000,-1.000){2}{\rule{1.445pt}{0.400pt}}
\put(527,415.67){\rule{2.891pt}{0.400pt}}
\multiput(527.00,416.17)(6.000,-1.000){2}{\rule{1.445pt}{0.400pt}}
\put(539,414.17){\rule{2.700pt}{0.400pt}}
\multiput(539.00,415.17)(7.396,-2.000){2}{\rule{1.350pt}{0.400pt}}
\put(552,412.17){\rule{2.500pt}{0.400pt}}
\multiput(552.00,413.17)(6.811,-2.000){2}{\rule{1.250pt}{0.400pt}}
\put(564,410.17){\rule{2.500pt}{0.400pt}}
\multiput(564.00,411.17)(6.811,-2.000){2}{\rule{1.250pt}{0.400pt}}
\put(576,408.17){\rule{2.500pt}{0.400pt}}
\multiput(576.00,409.17)(6.811,-2.000){2}{\rule{1.250pt}{0.400pt}}
\put(588,406.17){\rule{2.700pt}{0.400pt}}
\multiput(588.00,407.17)(7.396,-2.000){2}{\rule{1.350pt}{0.400pt}}
\put(601,404.17){\rule{2.500pt}{0.400pt}}
\multiput(601.00,405.17)(6.811,-2.000){2}{\rule{1.250pt}{0.400pt}}
\put(613,402.67){\rule{2.891pt}{0.400pt}}
\multiput(613.00,403.17)(6.000,-1.000){2}{\rule{1.445pt}{0.400pt}}
\put(625,401.67){\rule{3.132pt}{0.400pt}}
\multiput(625.00,402.17)(6.500,-1.000){2}{\rule{1.566pt}{0.400pt}}
\put(638,400.67){\rule{2.891pt}{0.400pt}}
\multiput(638.00,401.17)(6.000,-1.000){2}{\rule{1.445pt}{0.400pt}}
\put(650,399.67){\rule{2.891pt}{0.400pt}}
\multiput(650.00,400.17)(6.000,-1.000){2}{\rule{1.445pt}{0.400pt}}
\put(503.0,418.0){\rule[-0.200pt]{2.891pt}{0.400pt}}
\put(674,399.67){\rule{3.132pt}{0.400pt}}
\multiput(674.00,399.17)(6.500,1.000){2}{\rule{1.566pt}{0.400pt}}
\put(662.0,400.0){\rule[-0.200pt]{2.891pt}{0.400pt}}
\put(699,401.17){\rule{2.500pt}{0.400pt}}
\multiput(699.00,400.17)(6.811,2.000){2}{\rule{1.250pt}{0.400pt}}
\put(711,403.17){\rule{2.700pt}{0.400pt}}
\multiput(711.00,402.17)(7.396,2.000){2}{\rule{1.350pt}{0.400pt}}
\put(724,405.17){\rule{2.500pt}{0.400pt}}
\multiput(724.00,404.17)(6.811,2.000){2}{\rule{1.250pt}{0.400pt}}
\multiput(736.00,407.61)(2.472,0.447){3}{\rule{1.700pt}{0.108pt}}
\multiput(736.00,406.17)(8.472,3.000){2}{\rule{0.850pt}{0.400pt}}
\multiput(748.00,410.61)(2.472,0.447){3}{\rule{1.700pt}{0.108pt}}
\multiput(748.00,409.17)(8.472,3.000){2}{\rule{0.850pt}{0.400pt}}
\multiput(760.00,413.60)(1.797,0.468){5}{\rule{1.400pt}{0.113pt}}
\multiput(760.00,412.17)(10.094,4.000){2}{\rule{0.700pt}{0.400pt}}
\multiput(773.00,417.59)(1.267,0.477){7}{\rule{1.060pt}{0.115pt}}
\multiput(773.00,416.17)(9.800,5.000){2}{\rule{0.530pt}{0.400pt}}
\multiput(785.00,422.59)(1.267,0.477){7}{\rule{1.060pt}{0.115pt}}
\multiput(785.00,421.17)(9.800,5.000){2}{\rule{0.530pt}{0.400pt}}
\multiput(797.00,427.59)(1.378,0.477){7}{\rule{1.140pt}{0.115pt}}
\multiput(797.00,426.17)(10.634,5.000){2}{\rule{0.570pt}{0.400pt}}
\multiput(810.00,432.59)(1.033,0.482){9}{\rule{0.900pt}{0.116pt}}
\multiput(810.00,431.17)(10.132,6.000){2}{\rule{0.450pt}{0.400pt}}
\multiput(822.00,438.59)(1.033,0.482){9}{\rule{0.900pt}{0.116pt}}
\multiput(822.00,437.17)(10.132,6.000){2}{\rule{0.450pt}{0.400pt}}
\multiput(834.00,444.59)(0.874,0.485){11}{\rule{0.786pt}{0.117pt}}
\multiput(834.00,443.17)(10.369,7.000){2}{\rule{0.393pt}{0.400pt}}
\multiput(846.00,451.59)(0.824,0.488){13}{\rule{0.750pt}{0.117pt}}
\multiput(846.00,450.17)(11.443,8.000){2}{\rule{0.375pt}{0.400pt}}
\multiput(859.00,459.59)(0.874,0.485){11}{\rule{0.786pt}{0.117pt}}
\multiput(859.00,458.17)(10.369,7.000){2}{\rule{0.393pt}{0.400pt}}
\multiput(871.00,466.59)(0.669,0.489){15}{\rule{0.633pt}{0.118pt}}
\multiput(871.00,465.17)(10.685,9.000){2}{\rule{0.317pt}{0.400pt}}
\multiput(883.00,475.59)(0.824,0.488){13}{\rule{0.750pt}{0.117pt}}
\multiput(883.00,474.17)(11.443,8.000){2}{\rule{0.375pt}{0.400pt}}
\multiput(896.00,483.59)(0.669,0.489){15}{\rule{0.633pt}{0.118pt}}
\multiput(896.00,482.17)(10.685,9.000){2}{\rule{0.317pt}{0.400pt}}
\multiput(908.00,492.59)(0.669,0.489){15}{\rule{0.633pt}{0.118pt}}
\multiput(908.00,491.17)(10.685,9.000){2}{\rule{0.317pt}{0.400pt}}
\multiput(920.00,501.59)(0.669,0.489){15}{\rule{0.633pt}{0.118pt}}
\multiput(920.00,500.17)(10.685,9.000){2}{\rule{0.317pt}{0.400pt}}
\multiput(932.00,510.59)(0.728,0.489){15}{\rule{0.678pt}{0.118pt}}
\multiput(932.00,509.17)(11.593,9.000){2}{\rule{0.339pt}{0.400pt}}
\multiput(945.00,519.59)(0.758,0.488){13}{\rule{0.700pt}{0.117pt}}
\multiput(945.00,518.17)(10.547,8.000){2}{\rule{0.350pt}{0.400pt}}
\multiput(957.00,527.59)(0.669,0.489){15}{\rule{0.633pt}{0.118pt}}
\multiput(957.00,526.17)(10.685,9.000){2}{\rule{0.317pt}{0.400pt}}
\multiput(969.00,536.59)(0.950,0.485){11}{\rule{0.843pt}{0.117pt}}
\multiput(969.00,535.17)(11.251,7.000){2}{\rule{0.421pt}{0.400pt}}
\multiput(982.00,543.59)(1.033,0.482){9}{\rule{0.900pt}{0.116pt}}
\multiput(982.00,542.17)(10.132,6.000){2}{\rule{0.450pt}{0.400pt}}
\multiput(994.00,549.59)(1.033,0.482){9}{\rule{0.900pt}{0.116pt}}
\multiput(994.00,548.17)(10.132,6.000){2}{\rule{0.450pt}{0.400pt}}
\multiput(1006.00,555.61)(2.472,0.447){3}{\rule{1.700pt}{0.108pt}}
\multiput(1006.00,554.17)(8.472,3.000){2}{\rule{0.850pt}{0.400pt}}
\put(1018,558.17){\rule{2.700pt}{0.400pt}}
\multiput(1018.00,557.17)(7.396,2.000){2}{\rule{1.350pt}{0.400pt}}
\put(1031,558.67){\rule{2.891pt}{0.400pt}}
\multiput(1031.00,559.17)(6.000,-1.000){2}{\rule{1.445pt}{0.400pt}}
\multiput(1043.00,557.95)(2.472,-0.447){3}{\rule{1.700pt}{0.108pt}}
\multiput(1043.00,558.17)(8.472,-3.000){2}{\rule{0.850pt}{0.400pt}}
\multiput(1055.00,554.93)(1.378,-0.477){7}{\rule{1.140pt}{0.115pt}}
\multiput(1055.00,555.17)(10.634,-5.000){2}{\rule{0.570pt}{0.400pt}}
\multiput(1068.00,549.93)(1.033,-0.482){9}{\rule{0.900pt}{0.116pt}}
\multiput(1068.00,550.17)(10.132,-6.000){2}{\rule{0.450pt}{0.400pt}}
\multiput(1080.00,543.93)(0.758,-0.488){13}{\rule{0.700pt}{0.117pt}}
\multiput(1080.00,544.17)(10.547,-8.000){2}{\rule{0.350pt}{0.400pt}}
\multiput(1092.00,535.93)(0.758,-0.488){13}{\rule{0.700pt}{0.117pt}}
\multiput(1092.00,536.17)(10.547,-8.000){2}{\rule{0.350pt}{0.400pt}}
\multiput(1104.00,527.93)(0.950,-0.485){11}{\rule{0.843pt}{0.117pt}}
\multiput(1104.00,528.17)(11.251,-7.000){2}{\rule{0.421pt}{0.400pt}}
\multiput(1117.00,520.93)(1.267,-0.477){7}{\rule{1.060pt}{0.115pt}}
\multiput(1117.00,521.17)(9.800,-5.000){2}{\rule{0.530pt}{0.400pt}}
\put(687.0,401.0){\rule[-0.200pt]{2.891pt}{0.400pt}}
\multiput(1141.00,517.59)(1.033,0.482){9}{\rule{0.900pt}{0.116pt}}
\multiput(1141.00,516.17)(10.132,6.000){2}{\rule{0.450pt}{0.400pt}}
\multiput(1153.00,523.58)(0.497,0.493){23}{\rule{0.500pt}{0.119pt}}
\multiput(1153.00,522.17)(11.962,13.000){2}{\rule{0.250pt}{0.400pt}}
\multiput(1166.58,536.00)(0.492,0.841){21}{\rule{0.119pt}{0.767pt}}
\multiput(1165.17,536.00)(12.000,18.409){2}{\rule{0.400pt}{0.383pt}}
\multiput(1178.58,556.00)(0.492,0.970){21}{\rule{0.119pt}{0.867pt}}
\multiput(1177.17,556.00)(12.000,21.201){2}{\rule{0.400pt}{0.433pt}}
\multiput(1190.58,579.00)(0.493,0.972){23}{\rule{0.119pt}{0.869pt}}
\multiput(1189.17,579.00)(13.000,23.196){2}{\rule{0.400pt}{0.435pt}}
\multiput(1203.58,604.00)(0.492,0.798){21}{\rule{0.119pt}{0.733pt}}
\multiput(1202.17,604.00)(12.000,17.478){2}{\rule{0.400pt}{0.367pt}}
\multiput(1215.00,623.58)(0.600,0.491){17}{\rule{0.580pt}{0.118pt}}
\multiput(1215.00,622.17)(10.796,10.000){2}{\rule{0.290pt}{0.400pt}}
\multiput(1227.00,631.95)(2.472,-0.447){3}{\rule{1.700pt}{0.108pt}}
\multiput(1227.00,632.17)(8.472,-3.000){2}{\rule{0.850pt}{0.400pt}}
\multiput(1239.00,628.92)(0.497,-0.493){23}{\rule{0.500pt}{0.119pt}}
\multiput(1239.00,629.17)(11.962,-13.000){2}{\rule{0.250pt}{0.400pt}}
\multiput(1252.58,614.65)(0.492,-0.582){21}{\rule{0.119pt}{0.567pt}}
\multiput(1251.17,615.82)(12.000,-12.824){2}{\rule{0.400pt}{0.283pt}}
\put(1264,601.17){\rule{2.500pt}{0.400pt}}
\multiput(1264.00,602.17)(6.811,-2.000){2}{\rule{1.250pt}{0.400pt}}
\multiput(1276.58,601.00)(0.493,0.734){23}{\rule{0.119pt}{0.685pt}}
\multiput(1275.17,601.00)(13.000,17.579){2}{\rule{0.400pt}{0.342pt}}
\multiput(1289.58,620.00)(0.492,1.315){21}{\rule{0.119pt}{1.133pt}}
\multiput(1288.17,620.00)(12.000,28.648){2}{\rule{0.400pt}{0.567pt}}
\multiput(1301.58,651.00)(0.492,0.798){21}{\rule{0.119pt}{0.733pt}}
\multiput(1300.17,651.00)(12.000,17.478){2}{\rule{0.400pt}{0.367pt}}
\multiput(1313.00,668.92)(0.600,-0.491){17}{\rule{0.580pt}{0.118pt}}
\multiput(1313.00,669.17)(10.796,-10.000){2}{\rule{0.290pt}{0.400pt}}
\multiput(1325.00,658.92)(0.497,-0.493){23}{\rule{0.500pt}{0.119pt}}
\multiput(1325.00,659.17)(11.962,-13.000){2}{\rule{0.250pt}{0.400pt}}
\multiput(1338.58,647.00)(0.492,1.013){21}{\rule{0.119pt}{0.900pt}}
\multiput(1337.17,647.00)(12.000,22.132){2}{\rule{0.400pt}{0.450pt}}
\multiput(1350.58,671.00)(0.492,0.884){21}{\rule{0.119pt}{0.800pt}}
\multiput(1349.17,671.00)(12.000,19.340){2}{\rule{0.400pt}{0.400pt}}
\multiput(1362.58,689.67)(0.493,-0.576){23}{\rule{0.119pt}{0.562pt}}
\multiput(1361.17,690.83)(13.000,-13.834){2}{\rule{0.400pt}{0.281pt}}
\multiput(1375.58,677.00)(0.492,1.315){21}{\rule{0.119pt}{1.133pt}}
\multiput(1374.17,677.00)(12.000,28.648){2}{\rule{0.400pt}{0.567pt}}
\multiput(1387.00,706.95)(2.472,-0.447){3}{\rule{1.700pt}{0.108pt}}
\multiput(1387.00,707.17)(8.472,-3.000){2}{\rule{0.850pt}{0.400pt}}
\multiput(1399.58,705.00)(0.492,0.582){21}{\rule{0.119pt}{0.567pt}}
\multiput(1398.17,705.00)(12.000,12.824){2}{\rule{0.400pt}{0.283pt}}
\multiput(1411.00,719.58)(0.539,0.492){21}{\rule{0.533pt}{0.119pt}}
\multiput(1411.00,718.17)(11.893,12.000){2}{\rule{0.267pt}{0.400pt}}
\put(1129.0,517.0){\rule[-0.200pt]{2.891pt}{0.400pt}}
\end{picture}

%% file: goodsurrfit1.tex
\setlength{\unitlength}{0.240900pt}
\ifx\plotpoint\undefined\newsavebox{\plotpoint}\fi
\sbox{\plotpoint}{\rule[-0.200pt]{0.400pt}{0.400pt}}%
\begin{picture}(1500,900)(0,0)
\font\gnuplot=cmr10 at 10pt
\gnuplot
\sbox{\plotpoint}{\rule[-0.200pt]{0.400pt}{0.400pt}}%
\put(220.0,113.0){\rule[-0.200pt]{4.818pt}{0.400pt}}
\put(198,113){\makebox(0,0)[r]{$5.7$}}
\put(1416.0,113.0){\rule[-0.200pt]{4.818pt}{0.400pt}}
\put(220.0,198.0){\rule[-0.200pt]{4.818pt}{0.400pt}}
\put(198,198){\makebox(0,0)[r]{$5.8$}}
\put(1416.0,198.0){\rule[-0.200pt]{4.818pt}{0.400pt}}
\put(220.0,283.0){\rule[-0.200pt]{4.818pt}{0.400pt}}
\put(198,283){\makebox(0,0)[r]{$5.9$}}
\put(1416.0,283.0){\rule[-0.200pt]{4.818pt}{0.400pt}}
\put(220.0,368.0){\rule[-0.200pt]{4.818pt}{0.400pt}}
\put(198,368){\makebox(0,0)[r]{$6$}}
\put(1416.0,368.0){\rule[-0.200pt]{4.818pt}{0.400pt}}
\put(220.0,453.0){\rule[-0.200pt]{4.818pt}{0.400pt}}
\put(198,453){\makebox(0,0)[r]{$6.1$}}
\put(1416.0,453.0){\rule[-0.200pt]{4.818pt}{0.400pt}}
\put(220.0,537.0){\rule[-0.200pt]{4.818pt}{0.400pt}}
\put(198,537){\makebox(0,0)[r]{$6.2$}}
\put(1416.0,537.0){\rule[-0.200pt]{4.818pt}{0.400pt}}
\put(220.0,622.0){\rule[-0.200pt]{4.818pt}{0.400pt}}
\put(198,622){\makebox(0,0)[r]{$6.3$}}
\put(1416.0,622.0){\rule[-0.200pt]{4.818pt}{0.400pt}}
\put(220.0,707.0){\rule[-0.200pt]{4.818pt}{0.400pt}}
\put(198,707){\makebox(0,0)[r]{$6.4$}}
\put(1416.0,707.0){\rule[-0.200pt]{4.818pt}{0.400pt}}
\put(220.0,792.0){\rule[-0.200pt]{4.818pt}{0.400pt}}
\put(198,792){\makebox(0,0)[r]{$6.5$}}
\put(1416.0,792.0){\rule[-0.200pt]{4.818pt}{0.400pt}}
\put(220.0,877.0){\rule[-0.200pt]{4.818pt}{0.400pt}}
\put(198,877){\makebox(0,0)[r]{$6.6$}}
\put(1416.0,877.0){\rule[-0.200pt]{4.818pt}{0.400pt}}
\put(375.0,113.0){\rule[-0.200pt]{0.400pt}{4.818pt}}
\put(375,68){\makebox(0,0){$1951$}}
\put(375.0,857.0){\rule[-0.200pt]{0.400pt}{4.818pt}}
\put(534.0,113.0){\rule[-0.200pt]{0.400pt}{4.818pt}}
\put(534,68){\makebox(0,0){$1952$}}
\put(534.0,857.0){\rule[-0.200pt]{0.400pt}{4.818pt}}
\put(692.0,113.0){\rule[-0.200pt]{0.400pt}{4.818pt}}
\put(692,68){\makebox(0,0){$1953$}}
\put(692.0,857.0){\rule[-0.200pt]{0.400pt}{4.818pt}}
\put(851.0,113.0){\rule[-0.200pt]{0.400pt}{4.818pt}}
\put(851,68){\makebox(0,0){$1954$}}
\put(851.0,857.0){\rule[-0.200pt]{0.400pt}{4.818pt}}
\put(1009.0,113.0){\rule[-0.200pt]{0.400pt}{4.818pt}}
\put(1009,68){\makebox(0,0){$1955$}}
\put(1009.0,857.0){\rule[-0.200pt]{0.400pt}{4.818pt}}
\put(1168.0,113.0){\rule[-0.200pt]{0.400pt}{4.818pt}}
\put(1168,68){\makebox(0,0){$1956$}}
\put(1168.0,857.0){\rule[-0.200pt]{0.400pt}{4.818pt}}
\put(1326.0,113.0){\rule[-0.200pt]{0.400pt}{4.818pt}}
\put(1326,68){\makebox(0,0){$1957$}}
\put(1326.0,857.0){\rule[-0.200pt]{0.400pt}{4.818pt}}
\put(220.0,113.0){\rule[-0.200pt]{292.934pt}{0.400pt}}
\put(1436.0,113.0){\rule[-0.200pt]{0.400pt}{184.048pt}}
\put(220.0,877.0){\rule[-0.200pt]{292.934pt}{0.400pt}}
\put(100,945){\makebox(0,0){Artificial $\ln I\lp t\rp$}}
\put(828,0){\makebox(0,0){Artificial Time}}
\put(220.0,113.0){\rule[-0.200pt]{0.400pt}{184.048pt}}
\put(220,284){\raisebox{-.8pt}{\makebox(0,0){$\cdot$}}}
\put(223,271){\raisebox{-.8pt}{\makebox(0,0){$\cdot$}}}
\put(226,249){\raisebox{-.8pt}{\makebox(0,0){$\cdot$}}}
\put(229,242){\raisebox{-.8pt}{\makebox(0,0){$\cdot$}}}
\put(232,250){\raisebox{-.8pt}{\makebox(0,0){$\cdot$}}}
\put(235,287){\raisebox{-.8pt}{\makebox(0,0){$\cdot$}}}
\put(238,299){\raisebox{-.8pt}{\makebox(0,0){$\cdot$}}}
\put(241,270){\raisebox{-.8pt}{\makebox(0,0){$\cdot$}}}
\put(244,265){\raisebox{-.8pt}{\makebox(0,0){$\cdot$}}}
\put(247,271){\raisebox{-.8pt}{\makebox(0,0){$\cdot$}}}
\put(251,254){\raisebox{-.8pt}{\makebox(0,0){$\cdot$}}}
\put(254,276){\raisebox{-.8pt}{\makebox(0,0){$\cdot$}}}
\put(257,289){\raisebox{-.8pt}{\makebox(0,0){$\cdot$}}}
\put(260,274){\raisebox{-.8pt}{\makebox(0,0){$\cdot$}}}
\put(263,263){\raisebox{-.8pt}{\makebox(0,0){$\cdot$}}}
\put(266,248){\raisebox{-.8pt}{\makebox(0,0){$\cdot$}}}
\put(269,214){\raisebox{-.8pt}{\makebox(0,0){$\cdot$}}}
\put(272,215){\raisebox{-.8pt}{\makebox(0,0){$\cdot$}}}
\put(275,208){\raisebox{-.8pt}{\makebox(0,0){$\cdot$}}}
\put(278,212){\raisebox{-.8pt}{\makebox(0,0){$\cdot$}}}
\put(281,238){\raisebox{-.8pt}{\makebox(0,0){$\cdot$}}}
\put(284,237){\raisebox{-.8pt}{\makebox(0,0){$\cdot$}}}
\put(287,235){\raisebox{-.8pt}{\makebox(0,0){$\cdot$}}}
\put(290,186){\raisebox{-.8pt}{\makebox(0,0){$\cdot$}}}
\put(293,201){\raisebox{-.8pt}{\makebox(0,0){$\cdot$}}}
\put(296,186){\raisebox{-.8pt}{\makebox(0,0){$\cdot$}}}
\put(299,185){\raisebox{-.8pt}{\makebox(0,0){$\cdot$}}}
\put(302,220){\raisebox{-.8pt}{\makebox(0,0){$\cdot$}}}
\put(305,255){\raisebox{-.8pt}{\makebox(0,0){$\cdot$}}}
\put(308,253){\raisebox{-.8pt}{\makebox(0,0){$\cdot$}}}
\put(311,240){\raisebox{-.8pt}{\makebox(0,0){$\cdot$}}}
\put(315,238){\raisebox{-.8pt}{\makebox(0,0){$\cdot$}}}
\put(318,256){\raisebox{-.8pt}{\makebox(0,0){$\cdot$}}}
\put(321,255){\raisebox{-.8pt}{\makebox(0,0){$\cdot$}}}
\put(324,248){\raisebox{-.8pt}{\makebox(0,0){$\cdot$}}}
\put(327,248){\raisebox{-.8pt}{\makebox(0,0){$\cdot$}}}
\put(330,238){\raisebox{-.8pt}{\makebox(0,0){$\cdot$}}}
\put(333,239){\raisebox{-.8pt}{\makebox(0,0){$\cdot$}}}
\put(336,244){\raisebox{-.8pt}{\makebox(0,0){$\cdot$}}}
\put(339,253){\raisebox{-.8pt}{\makebox(0,0){$\cdot$}}}
\put(342,261){\raisebox{-.8pt}{\makebox(0,0){$\cdot$}}}
\put(345,251){\raisebox{-.8pt}{\makebox(0,0){$\cdot$}}}
\put(348,262){\raisebox{-.8pt}{\makebox(0,0){$\cdot$}}}
\put(351,273){\raisebox{-.8pt}{\makebox(0,0){$\cdot$}}}
\put(354,281){\raisebox{-.8pt}{\makebox(0,0){$\cdot$}}}
\put(357,264){\raisebox{-.8pt}{\makebox(0,0){$\cdot$}}}
\put(360,270){\raisebox{-.8pt}{\makebox(0,0){$\cdot$}}}
\put(363,254){\raisebox{-.8pt}{\makebox(0,0){$\cdot$}}}
\put(366,274){\raisebox{-.8pt}{\makebox(0,0){$\cdot$}}}
\put(369,294){\raisebox{-.8pt}{\makebox(0,0){$\cdot$}}}
\put(372,304){\raisebox{-.8pt}{\makebox(0,0){$\cdot$}}}
\put(375,332){\raisebox{-.8pt}{\makebox(0,0){$\cdot$}}}
\put(379,336){\raisebox{-.8pt}{\makebox(0,0){$\cdot$}}}
\put(382,310){\raisebox{-.8pt}{\makebox(0,0){$\cdot$}}}
\put(385,294){\raisebox{-.8pt}{\makebox(0,0){$\cdot$}}}
\put(388,279){\raisebox{-.8pt}{\makebox(0,0){$\cdot$}}}
\put(391,267){\raisebox{-.8pt}{\makebox(0,0){$\cdot$}}}
\put(394,269){\raisebox{-.8pt}{\makebox(0,0){$\cdot$}}}
\put(397,257){\raisebox{-.8pt}{\makebox(0,0){$\cdot$}}}
\put(400,268){\raisebox{-.8pt}{\makebox(0,0){$\cdot$}}}
\put(403,251){\raisebox{-.8pt}{\makebox(0,0){$\cdot$}}}
\put(406,259){\raisebox{-.8pt}{\makebox(0,0){$\cdot$}}}
\put(409,273){\raisebox{-.8pt}{\makebox(0,0){$\cdot$}}}
\put(412,269){\raisebox{-.8pt}{\makebox(0,0){$\cdot$}}}
\put(415,262){\raisebox{-.8pt}{\makebox(0,0){$\cdot$}}}
\put(418,254){\raisebox{-.8pt}{\makebox(0,0){$\cdot$}}}
\put(421,258){\raisebox{-.8pt}{\makebox(0,0){$\cdot$}}}
\put(424,258){\raisebox{-.8pt}{\makebox(0,0){$\cdot$}}}
\put(427,255){\raisebox{-.8pt}{\makebox(0,0){$\cdot$}}}
\put(430,256){\raisebox{-.8pt}{\makebox(0,0){$\cdot$}}}
\put(433,244){\raisebox{-.8pt}{\makebox(0,0){$\cdot$}}}
\put(436,253){\raisebox{-.8pt}{\makebox(0,0){$\cdot$}}}
\put(439,274){\raisebox{-.8pt}{\makebox(0,0){$\cdot$}}}
\put(443,283){\raisebox{-.8pt}{\makebox(0,0){$\cdot$}}}
\put(446,300){\raisebox{-.8pt}{\makebox(0,0){$\cdot$}}}
\put(449,319){\raisebox{-.8pt}{\makebox(0,0){$\cdot$}}}
\put(452,314){\raisebox{-.8pt}{\makebox(0,0){$\cdot$}}}
\put(455,321){\raisebox{-.8pt}{\makebox(0,0){$\cdot$}}}
\put(458,355){\raisebox{-.8pt}{\makebox(0,0){$\cdot$}}}
\put(461,350){\raisebox{-.8pt}{\makebox(0,0){$\cdot$}}}
\put(464,333){\raisebox{-.8pt}{\makebox(0,0){$\cdot$}}}
\put(467,335){\raisebox{-.8pt}{\makebox(0,0){$\cdot$}}}
\put(470,339){\raisebox{-.8pt}{\makebox(0,0){$\cdot$}}}
\put(473,339){\raisebox{-.8pt}{\makebox(0,0){$\cdot$}}}
\put(476,324){\raisebox{-.8pt}{\makebox(0,0){$\cdot$}}}
\put(479,308){\raisebox{-.8pt}{\makebox(0,0){$\cdot$}}}
\put(482,282){\raisebox{-.8pt}{\makebox(0,0){$\cdot$}}}
\put(485,281){\raisebox{-.8pt}{\makebox(0,0){$\cdot$}}}
\put(488,255){\raisebox{-.8pt}{\makebox(0,0){$\cdot$}}}
\put(491,252){\raisebox{-.8pt}{\makebox(0,0){$\cdot$}}}
\put(494,245){\raisebox{-.8pt}{\makebox(0,0){$\cdot$}}}
\put(497,244){\raisebox{-.8pt}{\makebox(0,0){$\cdot$}}}
\put(500,238){\raisebox{-.8pt}{\makebox(0,0){$\cdot$}}}
\put(503,246){\raisebox{-.8pt}{\makebox(0,0){$\cdot$}}}
\put(507,261){\raisebox{-.8pt}{\makebox(0,0){$\cdot$}}}
\put(510,268){\raisebox{-.8pt}{\makebox(0,0){$\cdot$}}}
\put(513,286){\raisebox{-.8pt}{\makebox(0,0){$\cdot$}}}
\put(516,277){\raisebox{-.8pt}{\makebox(0,0){$\cdot$}}}
\put(519,293){\raisebox{-.8pt}{\makebox(0,0){$\cdot$}}}
\put(522,280){\raisebox{-.8pt}{\makebox(0,0){$\cdot$}}}
\put(525,269){\raisebox{-.8pt}{\makebox(0,0){$\cdot$}}}
\put(528,281){\raisebox{-.8pt}{\makebox(0,0){$\cdot$}}}
\put(531,272){\raisebox{-.8pt}{\makebox(0,0){$\cdot$}}}
\put(534,266){\raisebox{-.8pt}{\makebox(0,0){$\cdot$}}}
\put(537,265){\raisebox{-.8pt}{\makebox(0,0){$\cdot$}}}
\put(540,274){\raisebox{-.8pt}{\makebox(0,0){$\cdot$}}}
\put(543,270){\raisebox{-.8pt}{\makebox(0,0){$\cdot$}}}
\put(546,267){\raisebox{-.8pt}{\makebox(0,0){$\cdot$}}}
\put(549,276){\raisebox{-.8pt}{\makebox(0,0){$\cdot$}}}
\put(552,284){\raisebox{-.8pt}{\makebox(0,0){$\cdot$}}}
\put(555,310){\raisebox{-.8pt}{\makebox(0,0){$\cdot$}}}
\put(558,273){\raisebox{-.8pt}{\makebox(0,0){$\cdot$}}}
\put(561,280){\raisebox{-.8pt}{\makebox(0,0){$\cdot$}}}
\put(564,293){\raisebox{-.8pt}{\makebox(0,0){$\cdot$}}}
\put(567,302){\raisebox{-.8pt}{\makebox(0,0){$\cdot$}}}
\put(571,297){\raisebox{-.8pt}{\makebox(0,0){$\cdot$}}}
\put(574,283){\raisebox{-.8pt}{\makebox(0,0){$\cdot$}}}
\put(577,275){\raisebox{-.8pt}{\makebox(0,0){$\cdot$}}}
\put(580,282){\raisebox{-.8pt}{\makebox(0,0){$\cdot$}}}
\put(583,282){\raisebox{-.8pt}{\makebox(0,0){$\cdot$}}}
\put(586,267){\raisebox{-.8pt}{\makebox(0,0){$\cdot$}}}
\put(589,291){\raisebox{-.8pt}{\makebox(0,0){$\cdot$}}}
\put(592,275){\raisebox{-.8pt}{\makebox(0,0){$\cdot$}}}
\put(595,278){\raisebox{-.8pt}{\makebox(0,0){$\cdot$}}}
\put(598,252){\raisebox{-.8pt}{\makebox(0,0){$\cdot$}}}
\put(601,245){\raisebox{-.8pt}{\makebox(0,0){$\cdot$}}}
\put(604,246){\raisebox{-.8pt}{\makebox(0,0){$\cdot$}}}
\put(607,251){\raisebox{-.8pt}{\makebox(0,0){$\cdot$}}}
\put(610,246){\raisebox{-.8pt}{\makebox(0,0){$\cdot$}}}
\put(613,209){\raisebox{-.8pt}{\makebox(0,0){$\cdot$}}}
\put(616,193){\raisebox{-.8pt}{\makebox(0,0){$\cdot$}}}
\put(619,201){\raisebox{-.8pt}{\makebox(0,0){$\cdot$}}}
\put(622,206){\raisebox{-.8pt}{\makebox(0,0){$\cdot$}}}
\put(625,207){\raisebox{-.8pt}{\makebox(0,0){$\cdot$}}}
\put(628,202){\raisebox{-.8pt}{\makebox(0,0){$\cdot$}}}
\put(631,165){\raisebox{-.8pt}{\makebox(0,0){$\cdot$}}}
\put(635,182){\raisebox{-.8pt}{\makebox(0,0){$\cdot$}}}
\put(638,160){\raisebox{-.8pt}{\makebox(0,0){$\cdot$}}}
\put(641,161){\raisebox{-.8pt}{\makebox(0,0){$\cdot$}}}
\put(644,168){\raisebox{-.8pt}{\makebox(0,0){$\cdot$}}}
\put(647,160){\raisebox{-.8pt}{\makebox(0,0){$\cdot$}}}
\put(650,163){\raisebox{-.8pt}{\makebox(0,0){$\cdot$}}}
\put(653,164){\raisebox{-.8pt}{\makebox(0,0){$\cdot$}}}
\put(656,138){\raisebox{-.8pt}{\makebox(0,0){$\cdot$}}}
\put(659,170){\raisebox{-.8pt}{\makebox(0,0){$\cdot$}}}
\put(662,182){\raisebox{-.8pt}{\makebox(0,0){$\cdot$}}}
\put(665,180){\raisebox{-.8pt}{\makebox(0,0){$\cdot$}}}
\put(668,138){\raisebox{-.8pt}{\makebox(0,0){$\cdot$}}}
\put(671,128){\raisebox{-.8pt}{\makebox(0,0){$\cdot$}}}
\put(674,120){\raisebox{-.8pt}{\makebox(0,0){$\cdot$}}}
\put(677,125){\raisebox{-.8pt}{\makebox(0,0){$\cdot$}}}
\put(680,135){\raisebox{-.8pt}{\makebox(0,0){$\cdot$}}}
\put(683,141){\raisebox{-.8pt}{\makebox(0,0){$\cdot$}}}
\put(686,128){\raisebox{-.8pt}{\makebox(0,0){$\cdot$}}}
\put(689,139){\raisebox{-.8pt}{\makebox(0,0){$\cdot$}}}
\put(692,136){\raisebox{-.8pt}{\makebox(0,0){$\cdot$}}}
\put(695,142){\raisebox{-.8pt}{\makebox(0,0){$\cdot$}}}
\put(699,142){\raisebox{-.8pt}{\makebox(0,0){$\cdot$}}}
\put(702,157){\raisebox{-.8pt}{\makebox(0,0){$\cdot$}}}
\put(705,184){\raisebox{-.8pt}{\makebox(0,0){$\cdot$}}}
\put(708,171){\raisebox{-.8pt}{\makebox(0,0){$\cdot$}}}
\put(711,165){\raisebox{-.8pt}{\makebox(0,0){$\cdot$}}}
\put(714,181){\raisebox{-.8pt}{\makebox(0,0){$\cdot$}}}
\put(717,178){\raisebox{-.8pt}{\makebox(0,0){$\cdot$}}}
\put(720,196){\raisebox{-.8pt}{\makebox(0,0){$\cdot$}}}
\put(723,193){\raisebox{-.8pt}{\makebox(0,0){$\cdot$}}}
\put(726,182){\raisebox{-.8pt}{\makebox(0,0){$\cdot$}}}
\put(729,194){\raisebox{-.8pt}{\makebox(0,0){$\cdot$}}}
\put(732,194){\raisebox{-.8pt}{\makebox(0,0){$\cdot$}}}
\put(735,201){\raisebox{-.8pt}{\makebox(0,0){$\cdot$}}}
\put(738,186){\raisebox{-.8pt}{\makebox(0,0){$\cdot$}}}
\put(741,179){\raisebox{-.8pt}{\makebox(0,0){$\cdot$}}}
\put(744,190){\raisebox{-.8pt}{\makebox(0,0){$\cdot$}}}
\put(747,192){\raisebox{-.8pt}{\makebox(0,0){$\cdot$}}}
\put(750,185){\raisebox{-.8pt}{\makebox(0,0){$\cdot$}}}
\put(753,193){\raisebox{-.8pt}{\makebox(0,0){$\cdot$}}}
\put(756,207){\raisebox{-.8pt}{\makebox(0,0){$\cdot$}}}
\put(759,185){\raisebox{-.8pt}{\makebox(0,0){$\cdot$}}}
\put(763,184){\raisebox{-.8pt}{\makebox(0,0){$\cdot$}}}
\put(766,182){\raisebox{-.8pt}{\makebox(0,0){$\cdot$}}}
\put(769,181){\raisebox{-.8pt}{\makebox(0,0){$\cdot$}}}
\put(772,196){\raisebox{-.8pt}{\makebox(0,0){$\cdot$}}}
\put(775,204){\raisebox{-.8pt}{\makebox(0,0){$\cdot$}}}
\put(778,221){\raisebox{-.8pt}{\makebox(0,0){$\cdot$}}}
\put(781,230){\raisebox{-.8pt}{\makebox(0,0){$\cdot$}}}
\put(784,261){\raisebox{-.8pt}{\makebox(0,0){$\cdot$}}}
\put(787,274){\raisebox{-.8pt}{\makebox(0,0){$\cdot$}}}
\put(790,213){\raisebox{-.8pt}{\makebox(0,0){$\cdot$}}}
\put(793,213){\raisebox{-.8pt}{\makebox(0,0){$\cdot$}}}
\put(796,200){\raisebox{-.8pt}{\makebox(0,0){$\cdot$}}}
\put(799,201){\raisebox{-.8pt}{\makebox(0,0){$\cdot$}}}
\put(802,219){\raisebox{-.8pt}{\makebox(0,0){$\cdot$}}}
\put(805,248){\raisebox{-.8pt}{\makebox(0,0){$\cdot$}}}
\put(808,286){\raisebox{-.8pt}{\makebox(0,0){$\cdot$}}}
\put(811,277){\raisebox{-.8pt}{\makebox(0,0){$\cdot$}}}
\put(814,292){\raisebox{-.8pt}{\makebox(0,0){$\cdot$}}}
\put(817,301){\raisebox{-.8pt}{\makebox(0,0){$\cdot$}}}
\put(820,314){\raisebox{-.8pt}{\makebox(0,0){$\cdot$}}}
\put(823,325){\raisebox{-.8pt}{\makebox(0,0){$\cdot$}}}
\put(827,337){\raisebox{-.8pt}{\makebox(0,0){$\cdot$}}}
\put(830,307){\raisebox{-.8pt}{\makebox(0,0){$\cdot$}}}
\put(833,334){\raisebox{-.8pt}{\makebox(0,0){$\cdot$}}}
\put(836,348){\raisebox{-.8pt}{\makebox(0,0){$\cdot$}}}
\put(839,344){\raisebox{-.8pt}{\makebox(0,0){$\cdot$}}}
\put(842,339){\raisebox{-.8pt}{\makebox(0,0){$\cdot$}}}
\put(845,352){\raisebox{-.8pt}{\makebox(0,0){$\cdot$}}}
\put(848,347){\raisebox{-.8pt}{\makebox(0,0){$\cdot$}}}
\put(851,356){\raisebox{-.8pt}{\makebox(0,0){$\cdot$}}}
\put(854,363){\raisebox{-.8pt}{\makebox(0,0){$\cdot$}}}
\put(857,357){\raisebox{-.8pt}{\makebox(0,0){$\cdot$}}}
\put(860,347){\raisebox{-.8pt}{\makebox(0,0){$\cdot$}}}
\put(863,352){\raisebox{-.8pt}{\makebox(0,0){$\cdot$}}}
\put(866,376){\raisebox{-.8pt}{\makebox(0,0){$\cdot$}}}
\put(869,361){\raisebox{-.8pt}{\makebox(0,0){$\cdot$}}}
\put(872,347){\raisebox{-.8pt}{\makebox(0,0){$\cdot$}}}
\put(875,330){\raisebox{-.8pt}{\makebox(0,0){$\cdot$}}}
\put(878,319){\raisebox{-.8pt}{\makebox(0,0){$\cdot$}}}
\put(881,318){\raisebox{-.8pt}{\makebox(0,0){$\cdot$}}}
\put(884,323){\raisebox{-.8pt}{\makebox(0,0){$\cdot$}}}
\put(887,350){\raisebox{-.8pt}{\makebox(0,0){$\cdot$}}}
\put(891,350){\raisebox{-.8pt}{\makebox(0,0){$\cdot$}}}
\put(894,340){\raisebox{-.8pt}{\makebox(0,0){$\cdot$}}}
\put(897,327){\raisebox{-.8pt}{\makebox(0,0){$\cdot$}}}
\put(900,334){\raisebox{-.8pt}{\makebox(0,0){$\cdot$}}}
\put(903,322){\raisebox{-.8pt}{\makebox(0,0){$\cdot$}}}
\put(906,317){\raisebox{-.8pt}{\makebox(0,0){$\cdot$}}}
\put(909,323){\raisebox{-.8pt}{\makebox(0,0){$\cdot$}}}
\put(912,334){\raisebox{-.8pt}{\makebox(0,0){$\cdot$}}}
\put(915,341){\raisebox{-.8pt}{\makebox(0,0){$\cdot$}}}
\put(918,374){\raisebox{-.8pt}{\makebox(0,0){$\cdot$}}}
\put(921,388){\raisebox{-.8pt}{\makebox(0,0){$\cdot$}}}
\put(924,400){\raisebox{-.8pt}{\makebox(0,0){$\cdot$}}}
\put(927,384){\raisebox{-.8pt}{\makebox(0,0){$\cdot$}}}
\put(930,398){\raisebox{-.8pt}{\makebox(0,0){$\cdot$}}}
\put(933,388){\raisebox{-.8pt}{\makebox(0,0){$\cdot$}}}
\put(936,409){\raisebox{-.8pt}{\makebox(0,0){$\cdot$}}}
\put(939,423){\raisebox{-.8pt}{\makebox(0,0){$\cdot$}}}
\put(942,415){\raisebox{-.8pt}{\makebox(0,0){$\cdot$}}}
\put(945,441){\raisebox{-.8pt}{\makebox(0,0){$\cdot$}}}
\put(948,440){\raisebox{-.8pt}{\makebox(0,0){$\cdot$}}}
\put(951,433){\raisebox{-.8pt}{\makebox(0,0){$\cdot$}}}
\put(955,434){\raisebox{-.8pt}{\makebox(0,0){$\cdot$}}}
\put(958,450){\raisebox{-.8pt}{\makebox(0,0){$\cdot$}}}
\put(961,437){\raisebox{-.8pt}{\makebox(0,0){$\cdot$}}}
\put(964,439){\raisebox{-.8pt}{\makebox(0,0){$\cdot$}}}
\put(967,436){\raisebox{-.8pt}{\makebox(0,0){$\cdot$}}}
\put(970,471){\raisebox{-.8pt}{\makebox(0,0){$\cdot$}}}
\put(973,486){\raisebox{-.8pt}{\makebox(0,0){$\cdot$}}}
\put(976,471){\raisebox{-.8pt}{\makebox(0,0){$\cdot$}}}
\put(979,458){\raisebox{-.8pt}{\makebox(0,0){$\cdot$}}}
\put(982,447){\raisebox{-.8pt}{\makebox(0,0){$\cdot$}}}
\put(985,406){\raisebox{-.8pt}{\makebox(0,0){$\cdot$}}}
\put(988,406){\raisebox{-.8pt}{\makebox(0,0){$\cdot$}}}
\put(991,395){\raisebox{-.8pt}{\makebox(0,0){$\cdot$}}}
\put(994,402){\raisebox{-.8pt}{\makebox(0,0){$\cdot$}}}
\put(997,401){\raisebox{-.8pt}{\makebox(0,0){$\cdot$}}}
\put(1000,407){\raisebox{-.8pt}{\makebox(0,0){$\cdot$}}}
\put(1003,434){\raisebox{-.8pt}{\makebox(0,0){$\cdot$}}}
\put(1006,450){\raisebox{-.8pt}{\makebox(0,0){$\cdot$}}}
\put(1009,438){\raisebox{-.8pt}{\makebox(0,0){$\cdot$}}}
\put(1012,399){\raisebox{-.8pt}{\makebox(0,0){$\cdot$}}}
\put(1015,416){\raisebox{-.8pt}{\makebox(0,0){$\cdot$}}}
\put(1019,417){\raisebox{-.8pt}{\makebox(0,0){$\cdot$}}}
\put(1022,461){\raisebox{-.8pt}{\makebox(0,0){$\cdot$}}}
\put(1025,462){\raisebox{-.8pt}{\makebox(0,0){$\cdot$}}}
\put(1028,464){\raisebox{-.8pt}{\makebox(0,0){$\cdot$}}}
\put(1031,447){\raisebox{-.8pt}{\makebox(0,0){$\cdot$}}}
\put(1034,440){\raisebox{-.8pt}{\makebox(0,0){$\cdot$}}}
\put(1037,420){\raisebox{-.8pt}{\makebox(0,0){$\cdot$}}}
\put(1040,411){\raisebox{-.8pt}{\makebox(0,0){$\cdot$}}}
\put(1043,410){\raisebox{-.8pt}{\makebox(0,0){$\cdot$}}}
\put(1046,399){\raisebox{-.8pt}{\makebox(0,0){$\cdot$}}}
\put(1049,410){\raisebox{-.8pt}{\makebox(0,0){$\cdot$}}}
\put(1052,425){\raisebox{-.8pt}{\makebox(0,0){$\cdot$}}}
\put(1055,427){\raisebox{-.8pt}{\makebox(0,0){$\cdot$}}}
\put(1058,432){\raisebox{-.8pt}{\makebox(0,0){$\cdot$}}}
\put(1061,441){\raisebox{-.8pt}{\makebox(0,0){$\cdot$}}}
\put(1064,459){\raisebox{-.8pt}{\makebox(0,0){$\cdot$}}}
\put(1067,394){\raisebox{-.8pt}{\makebox(0,0){$\cdot$}}}
\put(1070,376){\raisebox{-.8pt}{\makebox(0,0){$\cdot$}}}
\put(1073,365){\raisebox{-.8pt}{\makebox(0,0){$\cdot$}}}
\put(1076,356){\raisebox{-.8pt}{\makebox(0,0){$\cdot$}}}
\put(1079,360){\raisebox{-.8pt}{\makebox(0,0){$\cdot$}}}
\put(1083,359){\raisebox{-.8pt}{\makebox(0,0){$\cdot$}}}
\put(1086,377){\raisebox{-.8pt}{\makebox(0,0){$\cdot$}}}
\put(1089,360){\raisebox{-.8pt}{\makebox(0,0){$\cdot$}}}
\put(1092,359){\raisebox{-.8pt}{\makebox(0,0){$\cdot$}}}
\put(1095,348){\raisebox{-.8pt}{\makebox(0,0){$\cdot$}}}
\put(1098,371){\raisebox{-.8pt}{\makebox(0,0){$\cdot$}}}
\put(1101,370){\raisebox{-.8pt}{\makebox(0,0){$\cdot$}}}
\put(1104,370){\raisebox{-.8pt}{\makebox(0,0){$\cdot$}}}
\put(1107,378){\raisebox{-.8pt}{\makebox(0,0){$\cdot$}}}
\put(1110,386){\raisebox{-.8pt}{\makebox(0,0){$\cdot$}}}
\put(1113,387){\raisebox{-.8pt}{\makebox(0,0){$\cdot$}}}
\put(1116,399){\raisebox{-.8pt}{\makebox(0,0){$\cdot$}}}
\put(1119,413){\raisebox{-.8pt}{\makebox(0,0){$\cdot$}}}
\put(1122,396){\raisebox{-.8pt}{\makebox(0,0){$\cdot$}}}
\put(1125,393){\raisebox{-.8pt}{\makebox(0,0){$\cdot$}}}
\put(1128,401){\raisebox{-.8pt}{\makebox(0,0){$\cdot$}}}
\put(1131,399){\raisebox{-.8pt}{\makebox(0,0){$\cdot$}}}
\put(1134,413){\raisebox{-.8pt}{\makebox(0,0){$\cdot$}}}
\put(1137,404){\raisebox{-.8pt}{\makebox(0,0){$\cdot$}}}
\put(1140,425){\raisebox{-.8pt}{\makebox(0,0){$\cdot$}}}
\put(1143,423){\raisebox{-.8pt}{\makebox(0,0){$\cdot$}}}
\put(1147,439){\raisebox{-.8pt}{\makebox(0,0){$\cdot$}}}
\put(1150,438){\raisebox{-.8pt}{\makebox(0,0){$\cdot$}}}
\put(1153,443){\raisebox{-.8pt}{\makebox(0,0){$\cdot$}}}
\put(1156,465){\raisebox{-.8pt}{\makebox(0,0){$\cdot$}}}
\put(1159,486){\raisebox{-.8pt}{\makebox(0,0){$\cdot$}}}
\put(1162,484){\raisebox{-.8pt}{\makebox(0,0){$\cdot$}}}
\put(1165,473){\raisebox{-.8pt}{\makebox(0,0){$\cdot$}}}
\put(1168,467){\raisebox{-.8pt}{\makebox(0,0){$\cdot$}}}
\put(1171,477){\raisebox{-.8pt}{\makebox(0,0){$\cdot$}}}
\put(1174,483){\raisebox{-.8pt}{\makebox(0,0){$\cdot$}}}
\put(1177,497){\raisebox{-.8pt}{\makebox(0,0){$\cdot$}}}
\put(1180,504){\raisebox{-.8pt}{\makebox(0,0){$\cdot$}}}
\put(1183,520){\raisebox{-.8pt}{\makebox(0,0){$\cdot$}}}
\put(1186,459){\raisebox{-.8pt}{\makebox(0,0){$\cdot$}}}
\put(1189,494){\raisebox{-.8pt}{\makebox(0,0){$\cdot$}}}
\put(1192,505){\raisebox{-.8pt}{\makebox(0,0){$\cdot$}}}
\put(1195,543){\raisebox{-.8pt}{\makebox(0,0){$\cdot$}}}
\put(1198,534){\raisebox{-.8pt}{\makebox(0,0){$\cdot$}}}
\put(1201,516){\raisebox{-.8pt}{\makebox(0,0){$\cdot$}}}
\put(1204,532){\raisebox{-.8pt}{\makebox(0,0){$\cdot$}}}
\put(1207,568){\raisebox{-.8pt}{\makebox(0,0){$\cdot$}}}
\put(1211,567){\raisebox{-.8pt}{\makebox(0,0){$\cdot$}}}
\put(1214,556){\raisebox{-.8pt}{\makebox(0,0){$\cdot$}}}
\put(1217,568){\raisebox{-.8pt}{\makebox(0,0){$\cdot$}}}
\put(1220,560){\raisebox{-.8pt}{\makebox(0,0){$\cdot$}}}
\put(1223,571){\raisebox{-.8pt}{\makebox(0,0){$\cdot$}}}
\put(1226,616){\raisebox{-.8pt}{\makebox(0,0){$\cdot$}}}
\put(1229,599){\raisebox{-.8pt}{\makebox(0,0){$\cdot$}}}
\put(1232,620){\raisebox{-.8pt}{\makebox(0,0){$\cdot$}}}
\put(1235,628){\raisebox{-.8pt}{\makebox(0,0){$\cdot$}}}
\put(1238,590){\raisebox{-.8pt}{\makebox(0,0){$\cdot$}}}
\put(1241,584){\raisebox{-.8pt}{\makebox(0,0){$\cdot$}}}
\put(1244,518){\raisebox{-.8pt}{\makebox(0,0){$\cdot$}}}
\put(1247,510){\raisebox{-.8pt}{\makebox(0,0){$\cdot$}}}
\put(1250,503){\raisebox{-.8pt}{\makebox(0,0){$\cdot$}}}
\put(1253,535){\raisebox{-.8pt}{\makebox(0,0){$\cdot$}}}
\put(1256,536){\raisebox{-.8pt}{\makebox(0,0){$\cdot$}}}
\put(1259,534){\raisebox{-.8pt}{\makebox(0,0){$\cdot$}}}
\put(1262,511){\raisebox{-.8pt}{\makebox(0,0){$\cdot$}}}
\put(1265,512){\raisebox{-.8pt}{\makebox(0,0){$\cdot$}}}
\put(1268,517){\raisebox{-.8pt}{\makebox(0,0){$\cdot$}}}
\put(1271,540){\raisebox{-.8pt}{\makebox(0,0){$\cdot$}}}
\put(1275,531){\raisebox{-.8pt}{\makebox(0,0){$\cdot$}}}
\put(1278,541){\raisebox{-.8pt}{\makebox(0,0){$\cdot$}}}
\put(1281,562){\raisebox{-.8pt}{\makebox(0,0){$\cdot$}}}
\put(1284,578){\raisebox{-.8pt}{\makebox(0,0){$\cdot$}}}
\put(1287,579){\raisebox{-.8pt}{\makebox(0,0){$\cdot$}}}
\put(1290,589){\raisebox{-.8pt}{\makebox(0,0){$\cdot$}}}
\put(1293,592){\raisebox{-.8pt}{\makebox(0,0){$\cdot$}}}
\put(1296,584){\raisebox{-.8pt}{\makebox(0,0){$\cdot$}}}
\put(1299,588){\raisebox{-.8pt}{\makebox(0,0){$\cdot$}}}
\put(1302,648){\raisebox{-.8pt}{\makebox(0,0){$\cdot$}}}
\put(1305,655){\raisebox{-.8pt}{\makebox(0,0){$\cdot$}}}
\put(1308,662){\raisebox{-.8pt}{\makebox(0,0){$\cdot$}}}
\put(1311,654){\raisebox{-.8pt}{\makebox(0,0){$\cdot$}}}
\put(1314,660){\raisebox{-.8pt}{\makebox(0,0){$\cdot$}}}
\put(1317,669){\raisebox{-.8pt}{\makebox(0,0){$\cdot$}}}
\put(1320,690){\raisebox{-.8pt}{\makebox(0,0){$\cdot$}}}
\put(1323,668){\raisebox{-.8pt}{\makebox(0,0){$\cdot$}}}
\put(1326,678){\raisebox{-.8pt}{\makebox(0,0){$\cdot$}}}
\put(1329,702){\raisebox{-.8pt}{\makebox(0,0){$\cdot$}}}
\put(1332,694){\raisebox{-.8pt}{\makebox(0,0){$\cdot$}}}
\put(1335,691){\raisebox{-.8pt}{\makebox(0,0){$\cdot$}}}
\put(1339,692){\raisebox{-.8pt}{\makebox(0,0){$\cdot$}}}
\put(1342,705){\raisebox{-.8pt}{\makebox(0,0){$\cdot$}}}
\put(1345,697){\raisebox{-.8pt}{\makebox(0,0){$\cdot$}}}
\put(1348,691){\raisebox{-.8pt}{\makebox(0,0){$\cdot$}}}
\put(1351,693){\raisebox{-.8pt}{\makebox(0,0){$\cdot$}}}
\put(1354,685){\raisebox{-.8pt}{\makebox(0,0){$\cdot$}}}
\put(1357,697){\raisebox{-.8pt}{\makebox(0,0){$\cdot$}}}
\put(1360,710){\raisebox{-.8pt}{\makebox(0,0){$\cdot$}}}
\put(1363,716){\raisebox{-.8pt}{\makebox(0,0){$\cdot$}}}
\put(1366,729){\raisebox{-.8pt}{\makebox(0,0){$\cdot$}}}
\put(1369,745){\raisebox{-.8pt}{\makebox(0,0){$\cdot$}}}
\put(1372,755){\raisebox{-.8pt}{\makebox(0,0){$\cdot$}}}
\put(1375,735){\raisebox{-.8pt}{\makebox(0,0){$\cdot$}}}
\put(1378,735){\raisebox{-.8pt}{\makebox(0,0){$\cdot$}}}
\put(1381,750){\raisebox{-.8pt}{\makebox(0,0){$\cdot$}}}
\put(1384,744){\raisebox{-.8pt}{\makebox(0,0){$\cdot$}}}
\put(1387,755){\raisebox{-.8pt}{\makebox(0,0){$\cdot$}}}
\put(1390,749){\raisebox{-.8pt}{\makebox(0,0){$\cdot$}}}
\put(1393,745){\raisebox{-.8pt}{\makebox(0,0){$\cdot$}}}
\put(1396,745){\raisebox{-.8pt}{\makebox(0,0){$\cdot$}}}
\put(1399,747){\raisebox{-.8pt}{\makebox(0,0){$\cdot$}}}
\put(1403,759){\raisebox{-.8pt}{\makebox(0,0){$\cdot$}}}
\put(1406,772){\raisebox{-.8pt}{\makebox(0,0){$\cdot$}}}
\put(1409,779){\raisebox{-.8pt}{\makebox(0,0){$\cdot$}}}
\put(1412,785){\raisebox{-.8pt}{\makebox(0,0){$\cdot$}}}
\put(1415,803){\raisebox{-.8pt}{\makebox(0,0){$\cdot$}}}
\put(1418,810){\raisebox{-.8pt}{\makebox(0,0){$\cdot$}}}
\put(1421,821){\raisebox{-.8pt}{\makebox(0,0){$\cdot$}}}
\put(1424,812){\raisebox{-.8pt}{\makebox(0,0){$\cdot$}}}
\put(1427,812){\raisebox{-.8pt}{\makebox(0,0){$\cdot$}}}
\put(1430,809){\raisebox{-.8pt}{\makebox(0,0){$\cdot$}}}
\put(1433,815){\raisebox{-.8pt}{\makebox(0,0){$\cdot$}}}
\put(220,214){\usebox{\plotpoint}}
\multiput(220.00,214.59)(0.758,0.488){13}{\rule{0.700pt}{0.117pt}}
\multiput(220.00,213.17)(10.547,8.000){2}{\rule{0.350pt}{0.400pt}}
\multiput(232.00,222.59)(0.824,0.488){13}{\rule{0.750pt}{0.117pt}}
\multiput(232.00,221.17)(11.443,8.000){2}{\rule{0.375pt}{0.400pt}}
\multiput(245.00,230.59)(0.874,0.485){11}{\rule{0.786pt}{0.117pt}}
\multiput(245.00,229.17)(10.369,7.000){2}{\rule{0.393pt}{0.400pt}}
\multiput(257.00,237.59)(0.874,0.485){11}{\rule{0.786pt}{0.117pt}}
\multiput(257.00,236.17)(10.369,7.000){2}{\rule{0.393pt}{0.400pt}}
\multiput(269.00,244.59)(0.874,0.485){11}{\rule{0.786pt}{0.117pt}}
\multiput(269.00,243.17)(10.369,7.000){2}{\rule{0.393pt}{0.400pt}}
\multiput(281.00,251.59)(0.950,0.485){11}{\rule{0.843pt}{0.117pt}}
\multiput(281.00,250.17)(11.251,7.000){2}{\rule{0.421pt}{0.400pt}}
\multiput(294.00,258.59)(1.033,0.482){9}{\rule{0.900pt}{0.116pt}}
\multiput(294.00,257.17)(10.132,6.000){2}{\rule{0.450pt}{0.400pt}}
\multiput(306.00,264.59)(1.267,0.477){7}{\rule{1.060pt}{0.115pt}}
\multiput(306.00,263.17)(9.800,5.000){2}{\rule{0.530pt}{0.400pt}}
\multiput(318.00,269.59)(1.378,0.477){7}{\rule{1.140pt}{0.115pt}}
\multiput(318.00,268.17)(10.634,5.000){2}{\rule{0.570pt}{0.400pt}}
\multiput(331.00,274.59)(1.267,0.477){7}{\rule{1.060pt}{0.115pt}}
\multiput(331.00,273.17)(9.800,5.000){2}{\rule{0.530pt}{0.400pt}}
\multiput(343.00,279.60)(1.651,0.468){5}{\rule{1.300pt}{0.113pt}}
\multiput(343.00,278.17)(9.302,4.000){2}{\rule{0.650pt}{0.400pt}}
\multiput(355.00,283.61)(2.472,0.447){3}{\rule{1.700pt}{0.108pt}}
\multiput(355.00,282.17)(8.472,3.000){2}{\rule{0.850pt}{0.400pt}}
\put(367,286.17){\rule{2.700pt}{0.400pt}}
\multiput(367.00,285.17)(7.396,2.000){2}{\rule{1.350pt}{0.400pt}}
\put(380,288.17){\rule{2.500pt}{0.400pt}}
\multiput(380.00,287.17)(6.811,2.000){2}{\rule{1.250pt}{0.400pt}}
\put(392,289.67){\rule{2.891pt}{0.400pt}}
\multiput(392.00,289.17)(6.000,1.000){2}{\rule{1.445pt}{0.400pt}}
\put(404,290.67){\rule{3.132pt}{0.400pt}}
\multiput(404.00,290.17)(6.500,1.000){2}{\rule{1.566pt}{0.400pt}}
\put(429,290.67){\rule{2.891pt}{0.400pt}}
\multiput(429.00,291.17)(6.000,-1.000){2}{\rule{1.445pt}{0.400pt}}
\put(441,289.17){\rule{2.500pt}{0.400pt}}
\multiput(441.00,290.17)(6.811,-2.000){2}{\rule{1.250pt}{0.400pt}}
\multiput(453.00,287.95)(2.695,-0.447){3}{\rule{1.833pt}{0.108pt}}
\multiput(453.00,288.17)(9.195,-3.000){2}{\rule{0.917pt}{0.400pt}}
\multiput(466.00,284.95)(2.472,-0.447){3}{\rule{1.700pt}{0.108pt}}
\multiput(466.00,285.17)(8.472,-3.000){2}{\rule{0.850pt}{0.400pt}}
\multiput(478.00,281.94)(1.651,-0.468){5}{\rule{1.300pt}{0.113pt}}
\multiput(478.00,282.17)(9.302,-4.000){2}{\rule{0.650pt}{0.400pt}}
\multiput(490.00,277.94)(1.797,-0.468){5}{\rule{1.400pt}{0.113pt}}
\multiput(490.00,278.17)(10.094,-4.000){2}{\rule{0.700pt}{0.400pt}}
\multiput(503.00,273.93)(1.033,-0.482){9}{\rule{0.900pt}{0.116pt}}
\multiput(503.00,274.17)(10.132,-6.000){2}{\rule{0.450pt}{0.400pt}}
\multiput(515.00,267.93)(1.267,-0.477){7}{\rule{1.060pt}{0.115pt}}
\multiput(515.00,268.17)(9.800,-5.000){2}{\rule{0.530pt}{0.400pt}}
\multiput(527.00,262.93)(1.033,-0.482){9}{\rule{0.900pt}{0.116pt}}
\multiput(527.00,263.17)(10.132,-6.000){2}{\rule{0.450pt}{0.400pt}}
\multiput(539.00,256.93)(0.950,-0.485){11}{\rule{0.843pt}{0.117pt}}
\multiput(539.00,257.17)(11.251,-7.000){2}{\rule{0.421pt}{0.400pt}}
\multiput(552.00,249.93)(0.874,-0.485){11}{\rule{0.786pt}{0.117pt}}
\multiput(552.00,250.17)(10.369,-7.000){2}{\rule{0.393pt}{0.400pt}}
\multiput(564.00,242.93)(0.874,-0.485){11}{\rule{0.786pt}{0.117pt}}
\multiput(564.00,243.17)(10.369,-7.000){2}{\rule{0.393pt}{0.400pt}}
\multiput(576.00,235.93)(0.874,-0.485){11}{\rule{0.786pt}{0.117pt}}
\multiput(576.00,236.17)(10.369,-7.000){2}{\rule{0.393pt}{0.400pt}}
\multiput(588.00,228.93)(0.950,-0.485){11}{\rule{0.843pt}{0.117pt}}
\multiput(588.00,229.17)(11.251,-7.000){2}{\rule{0.421pt}{0.400pt}}
\multiput(601.00,221.93)(0.874,-0.485){11}{\rule{0.786pt}{0.117pt}}
\multiput(601.00,222.17)(10.369,-7.000){2}{\rule{0.393pt}{0.400pt}}
\multiput(613.00,214.93)(1.033,-0.482){9}{\rule{0.900pt}{0.116pt}}
\multiput(613.00,215.17)(10.132,-6.000){2}{\rule{0.450pt}{0.400pt}}
\multiput(625.00,208.93)(1.123,-0.482){9}{\rule{0.967pt}{0.116pt}}
\multiput(625.00,209.17)(10.994,-6.000){2}{\rule{0.483pt}{0.400pt}}
\multiput(638.00,202.93)(1.033,-0.482){9}{\rule{0.900pt}{0.116pt}}
\multiput(638.00,203.17)(10.132,-6.000){2}{\rule{0.450pt}{0.400pt}}
\multiput(650.00,196.93)(1.267,-0.477){7}{\rule{1.060pt}{0.115pt}}
\multiput(650.00,197.17)(9.800,-5.000){2}{\rule{0.530pt}{0.400pt}}
\multiput(662.00,191.95)(2.472,-0.447){3}{\rule{1.700pt}{0.108pt}}
\multiput(662.00,192.17)(8.472,-3.000){2}{\rule{0.850pt}{0.400pt}}
\multiput(674.00,188.95)(2.695,-0.447){3}{\rule{1.833pt}{0.108pt}}
\multiput(674.00,189.17)(9.195,-3.000){2}{\rule{0.917pt}{0.400pt}}
\put(687,185.67){\rule{2.891pt}{0.400pt}}
\multiput(687.00,186.17)(6.000,-1.000){2}{\rule{1.445pt}{0.400pt}}
\put(417.0,292.0){\rule[-0.200pt]{2.891pt}{0.400pt}}
\put(711,185.67){\rule{3.132pt}{0.400pt}}
\multiput(711.00,185.17)(6.500,1.000){2}{\rule{1.566pt}{0.400pt}}
\multiput(724.00,187.61)(2.472,0.447){3}{\rule{1.700pt}{0.108pt}}
\multiput(724.00,186.17)(8.472,3.000){2}{\rule{0.850pt}{0.400pt}}
\multiput(736.00,190.59)(1.267,0.477){7}{\rule{1.060pt}{0.115pt}}
\multiput(736.00,189.17)(9.800,5.000){2}{\rule{0.530pt}{0.400pt}}
\multiput(748.00,195.59)(0.874,0.485){11}{\rule{0.786pt}{0.117pt}}
\multiput(748.00,194.17)(10.369,7.000){2}{\rule{0.393pt}{0.400pt}}
\multiput(760.00,202.59)(0.824,0.488){13}{\rule{0.750pt}{0.117pt}}
\multiput(760.00,201.17)(11.443,8.000){2}{\rule{0.375pt}{0.400pt}}
\multiput(773.00,210.58)(0.600,0.491){17}{\rule{0.580pt}{0.118pt}}
\multiput(773.00,209.17)(10.796,10.000){2}{\rule{0.290pt}{0.400pt}}
\multiput(785.00,220.58)(0.496,0.492){21}{\rule{0.500pt}{0.119pt}}
\multiput(785.00,219.17)(10.962,12.000){2}{\rule{0.250pt}{0.400pt}}
\multiput(797.00,232.58)(0.497,0.493){23}{\rule{0.500pt}{0.119pt}}
\multiput(797.00,231.17)(11.962,13.000){2}{\rule{0.250pt}{0.400pt}}
\multiput(810.58,245.00)(0.492,0.625){21}{\rule{0.119pt}{0.600pt}}
\multiput(809.17,245.00)(12.000,13.755){2}{\rule{0.400pt}{0.300pt}}
\multiput(822.58,260.00)(0.492,0.669){21}{\rule{0.119pt}{0.633pt}}
\multiput(821.17,260.00)(12.000,14.685){2}{\rule{0.400pt}{0.317pt}}
\multiput(834.58,276.00)(0.492,0.712){21}{\rule{0.119pt}{0.667pt}}
\multiput(833.17,276.00)(12.000,15.616){2}{\rule{0.400pt}{0.333pt}}
\multiput(846.58,293.00)(0.493,0.655){23}{\rule{0.119pt}{0.623pt}}
\multiput(845.17,293.00)(13.000,15.707){2}{\rule{0.400pt}{0.312pt}}
\multiput(859.58,310.00)(0.492,0.755){21}{\rule{0.119pt}{0.700pt}}
\multiput(858.17,310.00)(12.000,16.547){2}{\rule{0.400pt}{0.350pt}}
\multiput(871.58,328.00)(0.492,0.755){21}{\rule{0.119pt}{0.700pt}}
\multiput(870.17,328.00)(12.000,16.547){2}{\rule{0.400pt}{0.350pt}}
\multiput(883.58,346.00)(0.493,0.655){23}{\rule{0.119pt}{0.623pt}}
\multiput(882.17,346.00)(13.000,15.707){2}{\rule{0.400pt}{0.312pt}}
\multiput(896.58,363.00)(0.492,0.712){21}{\rule{0.119pt}{0.667pt}}
\multiput(895.17,363.00)(12.000,15.616){2}{\rule{0.400pt}{0.333pt}}
\multiput(908.58,380.00)(0.492,0.582){21}{\rule{0.119pt}{0.567pt}}
\multiput(907.17,380.00)(12.000,12.824){2}{\rule{0.400pt}{0.283pt}}
\multiput(920.58,394.00)(0.492,0.539){21}{\rule{0.119pt}{0.533pt}}
\multiput(919.17,394.00)(12.000,11.893){2}{\rule{0.400pt}{0.267pt}}
\multiput(932.00,407.58)(0.590,0.492){19}{\rule{0.573pt}{0.118pt}}
\multiput(932.00,406.17)(11.811,11.000){2}{\rule{0.286pt}{0.400pt}}
\multiput(945.00,418.59)(0.758,0.488){13}{\rule{0.700pt}{0.117pt}}
\multiput(945.00,417.17)(10.547,8.000){2}{\rule{0.350pt}{0.400pt}}
\multiput(957.00,426.59)(1.267,0.477){7}{\rule{1.060pt}{0.115pt}}
\multiput(957.00,425.17)(9.800,5.000){2}{\rule{0.530pt}{0.400pt}}
\put(969,431.17){\rule{2.700pt}{0.400pt}}
\multiput(969.00,430.17)(7.396,2.000){2}{\rule{1.350pt}{0.400pt}}
\put(982,431.67){\rule{2.891pt}{0.400pt}}
\multiput(982.00,432.17)(6.000,-1.000){2}{\rule{1.445pt}{0.400pt}}
\multiput(994.00,430.94)(1.651,-0.468){5}{\rule{1.300pt}{0.113pt}}
\multiput(994.00,431.17)(9.302,-4.000){2}{\rule{0.650pt}{0.400pt}}
\multiput(1006.00,426.93)(0.874,-0.485){11}{\rule{0.786pt}{0.117pt}}
\multiput(1006.00,427.17)(10.369,-7.000){2}{\rule{0.393pt}{0.400pt}}
\multiput(1018.00,419.93)(0.824,-0.488){13}{\rule{0.750pt}{0.117pt}}
\multiput(1018.00,420.17)(11.443,-8.000){2}{\rule{0.375pt}{0.400pt}}
\multiput(1031.00,411.92)(0.600,-0.491){17}{\rule{0.580pt}{0.118pt}}
\multiput(1031.00,412.17)(10.796,-10.000){2}{\rule{0.290pt}{0.400pt}}
\multiput(1043.00,401.93)(0.669,-0.489){15}{\rule{0.633pt}{0.118pt}}
\multiput(1043.00,402.17)(10.685,-9.000){2}{\rule{0.317pt}{0.400pt}}
\multiput(1055.00,392.93)(0.824,-0.488){13}{\rule{0.750pt}{0.117pt}}
\multiput(1055.00,393.17)(11.443,-8.000){2}{\rule{0.375pt}{0.400pt}}
\multiput(1068.00,384.93)(1.267,-0.477){7}{\rule{1.060pt}{0.115pt}}
\multiput(1068.00,385.17)(9.800,-5.000){2}{\rule{0.530pt}{0.400pt}}
\put(1080,379.17){\rule{2.500pt}{0.400pt}}
\multiput(1080.00,380.17)(6.811,-2.000){2}{\rule{1.250pt}{0.400pt}}
\multiput(1092.00,379.60)(1.651,0.468){5}{\rule{1.300pt}{0.113pt}}
\multiput(1092.00,378.17)(9.302,4.000){2}{\rule{0.650pt}{0.400pt}}
\multiput(1104.00,383.59)(0.728,0.489){15}{\rule{0.678pt}{0.118pt}}
\multiput(1104.00,382.17)(11.593,9.000){2}{\rule{0.339pt}{0.400pt}}
\multiput(1117.58,392.00)(0.492,0.669){21}{\rule{0.119pt}{0.633pt}}
\multiput(1116.17,392.00)(12.000,14.685){2}{\rule{0.400pt}{0.317pt}}
\multiput(1129.58,408.00)(0.492,0.884){21}{\rule{0.119pt}{0.800pt}}
\multiput(1128.17,408.00)(12.000,19.340){2}{\rule{0.400pt}{0.400pt}}
\multiput(1141.58,429.00)(0.492,1.142){21}{\rule{0.119pt}{1.000pt}}
\multiput(1140.17,429.00)(12.000,24.924){2}{\rule{0.400pt}{0.500pt}}
\multiput(1153.58,456.00)(0.493,1.171){23}{\rule{0.119pt}{1.023pt}}
\multiput(1152.17,456.00)(13.000,27.877){2}{\rule{0.400pt}{0.512pt}}
\multiput(1166.58,486.00)(0.492,1.272){21}{\rule{0.119pt}{1.100pt}}
\multiput(1165.17,486.00)(12.000,27.717){2}{\rule{0.400pt}{0.550pt}}
\multiput(1178.58,516.00)(0.492,1.099){21}{\rule{0.119pt}{0.967pt}}
\multiput(1177.17,516.00)(12.000,23.994){2}{\rule{0.400pt}{0.483pt}}
\multiput(1190.58,542.00)(0.493,0.814){23}{\rule{0.119pt}{0.746pt}}
\multiput(1189.17,542.00)(13.000,19.451){2}{\rule{0.400pt}{0.373pt}}
\multiput(1203.00,563.58)(0.496,0.492){21}{\rule{0.500pt}{0.119pt}}
\multiput(1203.00,562.17)(10.962,12.000){2}{\rule{0.250pt}{0.400pt}}
\put(1215,575.17){\rule{2.500pt}{0.400pt}}
\multiput(1215.00,574.17)(6.811,2.000){2}{\rule{1.250pt}{0.400pt}}
\multiput(1227.00,575.93)(0.874,-0.485){11}{\rule{0.786pt}{0.117pt}}
\multiput(1227.00,576.17)(10.369,-7.000){2}{\rule{0.393pt}{0.400pt}}
\multiput(1239.00,568.92)(0.590,-0.492){19}{\rule{0.573pt}{0.118pt}}
\multiput(1239.00,569.17)(11.811,-11.000){2}{\rule{0.286pt}{0.400pt}}
\multiput(1252.00,557.93)(0.758,-0.488){13}{\rule{0.700pt}{0.117pt}}
\multiput(1252.00,558.17)(10.547,-8.000){2}{\rule{0.350pt}{0.400pt}}
\multiput(1264.00,551.61)(2.472,0.447){3}{\rule{1.700pt}{0.108pt}}
\multiput(1264.00,550.17)(8.472,3.000){2}{\rule{0.850pt}{0.400pt}}
\multiput(1276.58,554.00)(0.493,0.734){23}{\rule{0.119pt}{0.685pt}}
\multiput(1275.17,554.00)(13.000,17.579){2}{\rule{0.400pt}{0.342pt}}
\multiput(1289.58,573.00)(0.492,1.530){21}{\rule{0.119pt}{1.300pt}}
\multiput(1288.17,573.00)(12.000,33.302){2}{\rule{0.400pt}{0.650pt}}
\multiput(1301.58,609.00)(0.492,1.746){21}{\rule{0.119pt}{1.467pt}}
\multiput(1300.17,609.00)(12.000,37.956){2}{\rule{0.400pt}{0.733pt}}
\multiput(1313.58,650.00)(0.492,1.272){21}{\rule{0.119pt}{1.100pt}}
\multiput(1312.17,650.00)(12.000,27.717){2}{\rule{0.400pt}{0.550pt}}
\multiput(1325.00,680.60)(1.797,0.468){5}{\rule{1.400pt}{0.113pt}}
\multiput(1325.00,679.17)(10.094,4.000){2}{\rule{0.700pt}{0.400pt}}
\multiput(1338.00,682.92)(0.496,-0.492){21}{\rule{0.500pt}{0.119pt}}
\multiput(1338.00,683.17)(10.962,-12.000){2}{\rule{0.250pt}{0.400pt}}
\multiput(1350.00,672.59)(0.758,0.488){13}{\rule{0.700pt}{0.117pt}}
\multiput(1350.00,671.17)(10.547,8.000){2}{\rule{0.350pt}{0.400pt}}
\multiput(1362.58,680.00)(0.493,1.845){23}{\rule{0.119pt}{1.546pt}}
\multiput(1361.17,680.00)(13.000,43.791){2}{\rule{0.400pt}{0.773pt}}
\multiput(1375.58,727.00)(0.492,1.229){21}{\rule{0.119pt}{1.067pt}}
\multiput(1374.17,727.00)(12.000,26.786){2}{\rule{0.400pt}{0.533pt}}
\multiput(1387.00,754.93)(1.033,-0.482){9}{\rule{0.900pt}{0.116pt}}
\multiput(1387.00,755.17)(10.132,-6.000){2}{\rule{0.450pt}{0.400pt}}
\multiput(1399.58,750.00)(0.492,2.090){21}{\rule{0.119pt}{1.733pt}}
\multiput(1398.17,750.00)(12.000,45.402){2}{\rule{0.400pt}{0.867pt}}
\multiput(1411.58,799.00)(0.493,0.853){23}{\rule{0.119pt}{0.777pt}}
\multiput(1410.17,799.00)(13.000,20.387){2}{\rule{0.400pt}{0.388pt}}
\multiput(1424.58,821.00)(0.492,1.272){21}{\rule{0.119pt}{1.100pt}}
\multiput(1423.17,821.00)(12.000,27.717){2}{\rule{0.400pt}{0.550pt}}
\put(699.0,186.0){\rule[-0.200pt]{2.891pt}{0.400pt}}
\end{picture}

%% file: goodsurrfit2.tex
\setlength{\unitlength}{0.240900pt}
\ifx\plotpoint\undefined\newsavebox{\plotpoint}\fi
\begin{picture}(1500,900)(0,0)
\font\gnuplot=cmr10 at 10pt
\gnuplot
\sbox{\plotpoint}{\rule[-0.200pt]{0.400pt}{0.400pt}}%
\put(220.0,113.0){\rule[-0.200pt]{4.818pt}{0.400pt}}
\put(198,113){\makebox(0,0)[r]{$3.8$}}
\put(1416.0,113.0){\rule[-0.200pt]{4.818pt}{0.400pt}}
\put(220.0,222.0){\rule[-0.200pt]{4.818pt}{0.400pt}}
\put(198,222){\makebox(0,0)[r]{$4$}}
\put(1416.0,222.0){\rule[-0.200pt]{4.818pt}{0.400pt}}
\put(220.0,331.0){\rule[-0.200pt]{4.818pt}{0.400pt}}
\put(198,331){\makebox(0,0)[r]{$4.2$}}
\put(1416.0,331.0){\rule[-0.200pt]{4.818pt}{0.400pt}}
\put(220.0,440.0){\rule[-0.200pt]{4.818pt}{0.400pt}}
\put(198,440){\makebox(0,0)[r]{$4.4$}}
\put(1416.0,440.0){\rule[-0.200pt]{4.818pt}{0.400pt}}
\put(220.0,550.0){\rule[-0.200pt]{4.818pt}{0.400pt}}
\put(198,550){\makebox(0,0)[r]{$4.6$}}
\put(1416.0,550.0){\rule[-0.200pt]{4.818pt}{0.400pt}}
\put(220.0,659.0){\rule[-0.200pt]{4.818pt}{0.400pt}}
\put(198,659){\makebox(0,0)[r]{$4.8$}}
\put(1416.0,659.0){\rule[-0.200pt]{4.818pt}{0.400pt}}
\put(220.0,768.0){\rule[-0.200pt]{4.818pt}{0.400pt}}
\put(198,768){\makebox(0,0)[r]{$5$}}
\put(1416.0,768.0){\rule[-0.200pt]{4.818pt}{0.400pt}}
\put(220.0,877.0){\rule[-0.200pt]{4.818pt}{0.400pt}}
\put(198,877){\makebox(0,0)[r]{$5.2$}}
\put(1416.0,877.0){\rule[-0.200pt]{4.818pt}{0.400pt}}
\put(375.0,113.0){\rule[-0.200pt]{0.400pt}{4.818pt}}
\put(375,68){\makebox(0,0){$1951$}}
\put(375.0,857.0){\rule[-0.200pt]{0.400pt}{4.818pt}}
\put(534.0,113.0){\rule[-0.200pt]{0.400pt}{4.818pt}}
\put(534,68){\makebox(0,0){$1952$}}
\put(534.0,857.0){\rule[-0.200pt]{0.400pt}{4.818pt}}
\put(692.0,113.0){\rule[-0.200pt]{0.400pt}{4.818pt}}
\put(692,68){\makebox(0,0){$1953$}}
\put(692.0,857.0){\rule[-0.200pt]{0.400pt}{4.818pt}}
\put(851.0,113.0){\rule[-0.200pt]{0.400pt}{4.818pt}}
\put(851,68){\makebox(0,0){$1954$}}
\put(851.0,857.0){\rule[-0.200pt]{0.400pt}{4.818pt}}
\put(1009.0,113.0){\rule[-0.200pt]{0.400pt}{4.818pt}}
\put(1009,68){\makebox(0,0){$1955$}}
\put(1009.0,857.0){\rule[-0.200pt]{0.400pt}{4.818pt}}
\put(1168.0,113.0){\rule[-0.200pt]{0.400pt}{4.818pt}}
\put(1168,68){\makebox(0,0){$1956$}}
\put(1168.0,857.0){\rule[-0.200pt]{0.400pt}{4.818pt}}
\put(1326.0,113.0){\rule[-0.200pt]{0.400pt}{4.818pt}}
\put(1326,68){\makebox(0,0){$1957$}}
\put(1326.0,857.0){\rule[-0.200pt]{0.400pt}{4.818pt}}
\put(220.0,113.0){\rule[-0.200pt]{292.934pt}{0.400pt}}
\put(1436.0,113.0){\rule[-0.200pt]{0.400pt}{184.048pt}}
\put(220.0,877.0){\rule[-0.200pt]{292.934pt}{0.400pt}}
\put(100,945){\makebox(0,0){Artificial $\ln I\lp t\rp$}}
\put(828,0){\makebox(0,0){Artificial Time}}
\put(220.0,113.0){\rule[-0.200pt]{0.400pt}{184.048pt}}
\put(220,240){\raisebox{-.8pt}{\makebox(0,0){$\cdot$}}}
\put(223,245){\raisebox{-.8pt}{\makebox(0,0){$\cdot$}}}
\put(226,247){\raisebox{-.8pt}{\makebox(0,0){$\cdot$}}}
\put(229,234){\raisebox{-.8pt}{\makebox(0,0){$\cdot$}}}
\put(232,225){\raisebox{-.8pt}{\makebox(0,0){$\cdot$}}}
\put(235,227){\raisebox{-.8pt}{\makebox(0,0){$\cdot$}}}
\put(238,232){\raisebox{-.8pt}{\makebox(0,0){$\cdot$}}}
\put(241,243){\raisebox{-.8pt}{\makebox(0,0){$\cdot$}}}
\put(244,255){\raisebox{-.8pt}{\makebox(0,0){$\cdot$}}}
\put(247,252){\raisebox{-.8pt}{\makebox(0,0){$\cdot$}}}
\put(251,251){\raisebox{-.8pt}{\makebox(0,0){$\cdot$}}}
\put(254,252){\raisebox{-.8pt}{\makebox(0,0){$\cdot$}}}
\put(257,245){\raisebox{-.8pt}{\makebox(0,0){$\cdot$}}}
\put(260,241){\raisebox{-.8pt}{\makebox(0,0){$\cdot$}}}
\put(263,224){\raisebox{-.8pt}{\makebox(0,0){$\cdot$}}}
\put(266,230){\raisebox{-.8pt}{\makebox(0,0){$\cdot$}}}
\put(269,233){\raisebox{-.8pt}{\makebox(0,0){$\cdot$}}}
\put(272,229){\raisebox{-.8pt}{\makebox(0,0){$\cdot$}}}
\put(275,220){\raisebox{-.8pt}{\makebox(0,0){$\cdot$}}}
\put(278,246){\raisebox{-.8pt}{\makebox(0,0){$\cdot$}}}
\put(281,246){\raisebox{-.8pt}{\makebox(0,0){$\cdot$}}}
\put(284,220){\raisebox{-.8pt}{\makebox(0,0){$\cdot$}}}
\put(287,214){\raisebox{-.8pt}{\makebox(0,0){$\cdot$}}}
\put(290,245){\raisebox{-.8pt}{\makebox(0,0){$\cdot$}}}
\put(293,249){\raisebox{-.8pt}{\makebox(0,0){$\cdot$}}}
\put(296,246){\raisebox{-.8pt}{\makebox(0,0){$\cdot$}}}
\put(299,241){\raisebox{-.8pt}{\makebox(0,0){$\cdot$}}}
\put(302,225){\raisebox{-.8pt}{\makebox(0,0){$\cdot$}}}
\put(305,223){\raisebox{-.8pt}{\makebox(0,0){$\cdot$}}}
\put(308,217){\raisebox{-.8pt}{\makebox(0,0){$\cdot$}}}
\put(311,213){\raisebox{-.8pt}{\makebox(0,0){$\cdot$}}}
\put(315,208){\raisebox{-.8pt}{\makebox(0,0){$\cdot$}}}
\put(318,206){\raisebox{-.8pt}{\makebox(0,0){$\cdot$}}}
\put(321,205){\raisebox{-.8pt}{\makebox(0,0){$\cdot$}}}
\put(324,214){\raisebox{-.8pt}{\makebox(0,0){$\cdot$}}}
\put(327,211){\raisebox{-.8pt}{\makebox(0,0){$\cdot$}}}
\put(330,214){\raisebox{-.8pt}{\makebox(0,0){$\cdot$}}}
\put(333,198){\raisebox{-.8pt}{\makebox(0,0){$\cdot$}}}
\put(336,203){\raisebox{-.8pt}{\makebox(0,0){$\cdot$}}}
\put(339,203){\raisebox{-.8pt}{\makebox(0,0){$\cdot$}}}
\put(342,208){\raisebox{-.8pt}{\makebox(0,0){$\cdot$}}}
\put(345,201){\raisebox{-.8pt}{\makebox(0,0){$\cdot$}}}
\put(348,232){\raisebox{-.8pt}{\makebox(0,0){$\cdot$}}}
\put(351,249){\raisebox{-.8pt}{\makebox(0,0){$\cdot$}}}
\put(354,254){\raisebox{-.8pt}{\makebox(0,0){$\cdot$}}}
\put(357,253){\raisebox{-.8pt}{\makebox(0,0){$\cdot$}}}
\put(360,249){\raisebox{-.8pt}{\makebox(0,0){$\cdot$}}}
\put(363,254){\raisebox{-.8pt}{\makebox(0,0){$\cdot$}}}
\put(366,257){\raisebox{-.8pt}{\makebox(0,0){$\cdot$}}}
\put(369,260){\raisebox{-.8pt}{\makebox(0,0){$\cdot$}}}
\put(372,260){\raisebox{-.8pt}{\makebox(0,0){$\cdot$}}}
\put(375,274){\raisebox{-.8pt}{\makebox(0,0){$\cdot$}}}
\put(379,279){\raisebox{-.8pt}{\makebox(0,0){$\cdot$}}}
\put(382,277){\raisebox{-.8pt}{\makebox(0,0){$\cdot$}}}
\put(385,251){\raisebox{-.8pt}{\makebox(0,0){$\cdot$}}}
\put(388,225){\raisebox{-.8pt}{\makebox(0,0){$\cdot$}}}
\put(391,234){\raisebox{-.8pt}{\makebox(0,0){$\cdot$}}}
\put(394,227){\raisebox{-.8pt}{\makebox(0,0){$\cdot$}}}
\put(397,223){\raisebox{-.8pt}{\makebox(0,0){$\cdot$}}}
\put(400,257){\raisebox{-.8pt}{\makebox(0,0){$\cdot$}}}
\put(403,259){\raisebox{-.8pt}{\makebox(0,0){$\cdot$}}}
\put(406,284){\raisebox{-.8pt}{\makebox(0,0){$\cdot$}}}
\put(409,292){\raisebox{-.8pt}{\makebox(0,0){$\cdot$}}}
\put(412,315){\raisebox{-.8pt}{\makebox(0,0){$\cdot$}}}
\put(415,320){\raisebox{-.8pt}{\makebox(0,0){$\cdot$}}}
\put(418,327){\raisebox{-.8pt}{\makebox(0,0){$\cdot$}}}
\put(421,322){\raisebox{-.8pt}{\makebox(0,0){$\cdot$}}}
\put(424,323){\raisebox{-.8pt}{\makebox(0,0){$\cdot$}}}
\put(427,338){\raisebox{-.8pt}{\makebox(0,0){$\cdot$}}}
\put(430,346){\raisebox{-.8pt}{\makebox(0,0){$\cdot$}}}
\put(433,337){\raisebox{-.8pt}{\makebox(0,0){$\cdot$}}}
\put(436,318){\raisebox{-.8pt}{\makebox(0,0){$\cdot$}}}
\put(439,319){\raisebox{-.8pt}{\makebox(0,0){$\cdot$}}}
\put(443,344){\raisebox{-.8pt}{\makebox(0,0){$\cdot$}}}
\put(446,340){\raisebox{-.8pt}{\makebox(0,0){$\cdot$}}}
\put(449,339){\raisebox{-.8pt}{\makebox(0,0){$\cdot$}}}
\put(452,369){\raisebox{-.8pt}{\makebox(0,0){$\cdot$}}}
\put(455,362){\raisebox{-.8pt}{\makebox(0,0){$\cdot$}}}
\put(458,358){\raisebox{-.8pt}{\makebox(0,0){$\cdot$}}}
\put(461,356){\raisebox{-.8pt}{\makebox(0,0){$\cdot$}}}
\put(464,367){\raisebox{-.8pt}{\makebox(0,0){$\cdot$}}}
\put(467,370){\raisebox{-.8pt}{\makebox(0,0){$\cdot$}}}
\put(470,370){\raisebox{-.8pt}{\makebox(0,0){$\cdot$}}}
\put(473,382){\raisebox{-.8pt}{\makebox(0,0){$\cdot$}}}
\put(476,393){\raisebox{-.8pt}{\makebox(0,0){$\cdot$}}}
\put(479,381){\raisebox{-.8pt}{\makebox(0,0){$\cdot$}}}
\put(482,360){\raisebox{-.8pt}{\makebox(0,0){$\cdot$}}}
\put(485,352){\raisebox{-.8pt}{\makebox(0,0){$\cdot$}}}
\put(488,345){\raisebox{-.8pt}{\makebox(0,0){$\cdot$}}}
\put(491,359){\raisebox{-.8pt}{\makebox(0,0){$\cdot$}}}
\put(494,360){\raisebox{-.8pt}{\makebox(0,0){$\cdot$}}}
\put(497,372){\raisebox{-.8pt}{\makebox(0,0){$\cdot$}}}
\put(500,372){\raisebox{-.8pt}{\makebox(0,0){$\cdot$}}}
\put(503,385){\raisebox{-.8pt}{\makebox(0,0){$\cdot$}}}
\put(507,400){\raisebox{-.8pt}{\makebox(0,0){$\cdot$}}}
\put(510,386){\raisebox{-.8pt}{\makebox(0,0){$\cdot$}}}
\put(513,383){\raisebox{-.8pt}{\makebox(0,0){$\cdot$}}}
\put(516,381){\raisebox{-.8pt}{\makebox(0,0){$\cdot$}}}
\put(519,396){\raisebox{-.8pt}{\makebox(0,0){$\cdot$}}}
\put(522,398){\raisebox{-.8pt}{\makebox(0,0){$\cdot$}}}
\put(525,398){\raisebox{-.8pt}{\makebox(0,0){$\cdot$}}}
\put(528,393){\raisebox{-.8pt}{\makebox(0,0){$\cdot$}}}
\put(531,390){\raisebox{-.8pt}{\makebox(0,0){$\cdot$}}}
\put(534,393){\raisebox{-.8pt}{\makebox(0,0){$\cdot$}}}
\put(537,387){\raisebox{-.8pt}{\makebox(0,0){$\cdot$}}}
\put(540,391){\raisebox{-.8pt}{\makebox(0,0){$\cdot$}}}
\put(543,401){\raisebox{-.8pt}{\makebox(0,0){$\cdot$}}}
\put(546,397){\raisebox{-.8pt}{\makebox(0,0){$\cdot$}}}
\put(549,414){\raisebox{-.8pt}{\makebox(0,0){$\cdot$}}}
\put(552,416){\raisebox{-.8pt}{\makebox(0,0){$\cdot$}}}
\put(555,416){\raisebox{-.8pt}{\makebox(0,0){$\cdot$}}}
\put(558,456){\raisebox{-.8pt}{\makebox(0,0){$\cdot$}}}
\put(561,442){\raisebox{-.8pt}{\makebox(0,0){$\cdot$}}}
\put(564,459){\raisebox{-.8pt}{\makebox(0,0){$\cdot$}}}
\put(567,461){\raisebox{-.8pt}{\makebox(0,0){$\cdot$}}}
\put(571,450){\raisebox{-.8pt}{\makebox(0,0){$\cdot$}}}
\put(574,449){\raisebox{-.8pt}{\makebox(0,0){$\cdot$}}}
\put(577,449){\raisebox{-.8pt}{\makebox(0,0){$\cdot$}}}
\put(580,449){\raisebox{-.8pt}{\makebox(0,0){$\cdot$}}}
\put(583,445){\raisebox{-.8pt}{\makebox(0,0){$\cdot$}}}
\put(586,449){\raisebox{-.8pt}{\makebox(0,0){$\cdot$}}}
\put(589,449){\raisebox{-.8pt}{\makebox(0,0){$\cdot$}}}
\put(592,443){\raisebox{-.8pt}{\makebox(0,0){$\cdot$}}}
\put(595,438){\raisebox{-.8pt}{\makebox(0,0){$\cdot$}}}
\put(598,410){\raisebox{-.8pt}{\makebox(0,0){$\cdot$}}}
\put(601,393){\raisebox{-.8pt}{\makebox(0,0){$\cdot$}}}
\put(604,404){\raisebox{-.8pt}{\makebox(0,0){$\cdot$}}}
\put(607,402){\raisebox{-.8pt}{\makebox(0,0){$\cdot$}}}
\put(610,394){\raisebox{-.8pt}{\makebox(0,0){$\cdot$}}}
\put(613,401){\raisebox{-.8pt}{\makebox(0,0){$\cdot$}}}
\put(616,398){\raisebox{-.8pt}{\makebox(0,0){$\cdot$}}}
\put(619,390){\raisebox{-.8pt}{\makebox(0,0){$\cdot$}}}
\put(622,402){\raisebox{-.8pt}{\makebox(0,0){$\cdot$}}}
\put(625,413){\raisebox{-.8pt}{\makebox(0,0){$\cdot$}}}
\put(628,420){\raisebox{-.8pt}{\makebox(0,0){$\cdot$}}}
\put(631,423){\raisebox{-.8pt}{\makebox(0,0){$\cdot$}}}
\put(635,425){\raisebox{-.8pt}{\makebox(0,0){$\cdot$}}}
\put(638,427){\raisebox{-.8pt}{\makebox(0,0){$\cdot$}}}
\put(641,444){\raisebox{-.8pt}{\makebox(0,0){$\cdot$}}}
\put(644,459){\raisebox{-.8pt}{\makebox(0,0){$\cdot$}}}
\put(647,457){\raisebox{-.8pt}{\makebox(0,0){$\cdot$}}}
\put(650,459){\raisebox{-.8pt}{\makebox(0,0){$\cdot$}}}
\put(653,455){\raisebox{-.8pt}{\makebox(0,0){$\cdot$}}}
\put(656,452){\raisebox{-.8pt}{\makebox(0,0){$\cdot$}}}
\put(659,449){\raisebox{-.8pt}{\makebox(0,0){$\cdot$}}}
\put(662,445){\raisebox{-.8pt}{\makebox(0,0){$\cdot$}}}
\put(665,448){\raisebox{-.8pt}{\makebox(0,0){$\cdot$}}}
\put(668,450){\raisebox{-.8pt}{\makebox(0,0){$\cdot$}}}
\put(671,447){\raisebox{-.8pt}{\makebox(0,0){$\cdot$}}}
\put(674,452){\raisebox{-.8pt}{\makebox(0,0){$\cdot$}}}
\put(677,447){\raisebox{-.8pt}{\makebox(0,0){$\cdot$}}}
\put(680,445){\raisebox{-.8pt}{\makebox(0,0){$\cdot$}}}
\put(683,437){\raisebox{-.8pt}{\makebox(0,0){$\cdot$}}}
\put(686,448){\raisebox{-.8pt}{\makebox(0,0){$\cdot$}}}
\put(689,453){\raisebox{-.8pt}{\makebox(0,0){$\cdot$}}}
\put(692,451){\raisebox{-.8pt}{\makebox(0,0){$\cdot$}}}
\put(695,460){\raisebox{-.8pt}{\makebox(0,0){$\cdot$}}}
\put(699,469){\raisebox{-.8pt}{\makebox(0,0){$\cdot$}}}
\put(702,478){\raisebox{-.8pt}{\makebox(0,0){$\cdot$}}}
\put(705,466){\raisebox{-.8pt}{\makebox(0,0){$\cdot$}}}
\put(708,476){\raisebox{-.8pt}{\makebox(0,0){$\cdot$}}}
\put(711,474){\raisebox{-.8pt}{\makebox(0,0){$\cdot$}}}
\put(714,477){\raisebox{-.8pt}{\makebox(0,0){$\cdot$}}}
\put(717,471){\raisebox{-.8pt}{\makebox(0,0){$\cdot$}}}
\put(720,477){\raisebox{-.8pt}{\makebox(0,0){$\cdot$}}}
\put(723,478){\raisebox{-.8pt}{\makebox(0,0){$\cdot$}}}
\put(726,483){\raisebox{-.8pt}{\makebox(0,0){$\cdot$}}}
\put(729,484){\raisebox{-.8pt}{\makebox(0,0){$\cdot$}}}
\put(732,469){\raisebox{-.8pt}{\makebox(0,0){$\cdot$}}}
\put(735,461){\raisebox{-.8pt}{\makebox(0,0){$\cdot$}}}
\put(738,462){\raisebox{-.8pt}{\makebox(0,0){$\cdot$}}}
\put(741,456){\raisebox{-.8pt}{\makebox(0,0){$\cdot$}}}
\put(744,440){\raisebox{-.8pt}{\makebox(0,0){$\cdot$}}}
\put(747,447){\raisebox{-.8pt}{\makebox(0,0){$\cdot$}}}
\put(750,442){\raisebox{-.8pt}{\makebox(0,0){$\cdot$}}}
\put(753,438){\raisebox{-.8pt}{\makebox(0,0){$\cdot$}}}
\put(756,456){\raisebox{-.8pt}{\makebox(0,0){$\cdot$}}}
\put(759,452){\raisebox{-.8pt}{\makebox(0,0){$\cdot$}}}
\put(763,449){\raisebox{-.8pt}{\makebox(0,0){$\cdot$}}}
\put(766,436){\raisebox{-.8pt}{\makebox(0,0){$\cdot$}}}
\put(769,422){\raisebox{-.8pt}{\makebox(0,0){$\cdot$}}}
\put(772,417){\raisebox{-.8pt}{\makebox(0,0){$\cdot$}}}
\put(775,410){\raisebox{-.8pt}{\makebox(0,0){$\cdot$}}}
\put(778,407){\raisebox{-.8pt}{\makebox(0,0){$\cdot$}}}
\put(781,412){\raisebox{-.8pt}{\makebox(0,0){$\cdot$}}}
\put(784,406){\raisebox{-.8pt}{\makebox(0,0){$\cdot$}}}
\put(787,410){\raisebox{-.8pt}{\makebox(0,0){$\cdot$}}}
\put(790,417){\raisebox{-.8pt}{\makebox(0,0){$\cdot$}}}
\put(793,413){\raisebox{-.8pt}{\makebox(0,0){$\cdot$}}}
\put(796,405){\raisebox{-.8pt}{\makebox(0,0){$\cdot$}}}
\put(799,408){\raisebox{-.8pt}{\makebox(0,0){$\cdot$}}}
\put(802,404){\raisebox{-.8pt}{\makebox(0,0){$\cdot$}}}
\put(805,408){\raisebox{-.8pt}{\makebox(0,0){$\cdot$}}}
\put(808,400){\raisebox{-.8pt}{\makebox(0,0){$\cdot$}}}
\put(811,406){\raisebox{-.8pt}{\makebox(0,0){$\cdot$}}}
\put(814,395){\raisebox{-.8pt}{\makebox(0,0){$\cdot$}}}
\put(817,392){\raisebox{-.8pt}{\makebox(0,0){$\cdot$}}}
\put(820,398){\raisebox{-.8pt}{\makebox(0,0){$\cdot$}}}
\put(823,406){\raisebox{-.8pt}{\makebox(0,0){$\cdot$}}}
\put(827,408){\raisebox{-.8pt}{\makebox(0,0){$\cdot$}}}
\put(830,394){\raisebox{-.8pt}{\makebox(0,0){$\cdot$}}}
\put(833,388){\raisebox{-.8pt}{\makebox(0,0){$\cdot$}}}
\put(836,393){\raisebox{-.8pt}{\makebox(0,0){$\cdot$}}}
\put(839,395){\raisebox{-.8pt}{\makebox(0,0){$\cdot$}}}
\put(842,391){\raisebox{-.8pt}{\makebox(0,0){$\cdot$}}}
\put(845,397){\raisebox{-.8pt}{\makebox(0,0){$\cdot$}}}
\put(848,399){\raisebox{-.8pt}{\makebox(0,0){$\cdot$}}}
\put(851,404){\raisebox{-.8pt}{\makebox(0,0){$\cdot$}}}
\put(854,401){\raisebox{-.8pt}{\makebox(0,0){$\cdot$}}}
\put(857,409){\raisebox{-.8pt}{\makebox(0,0){$\cdot$}}}
\put(860,400){\raisebox{-.8pt}{\makebox(0,0){$\cdot$}}}
\put(863,420){\raisebox{-.8pt}{\makebox(0,0){$\cdot$}}}
\put(866,423){\raisebox{-.8pt}{\makebox(0,0){$\cdot$}}}
\put(869,417){\raisebox{-.8pt}{\makebox(0,0){$\cdot$}}}
\put(872,418){\raisebox{-.8pt}{\makebox(0,0){$\cdot$}}}
\put(875,411){\raisebox{-.8pt}{\makebox(0,0){$\cdot$}}}
\put(878,412){\raisebox{-.8pt}{\makebox(0,0){$\cdot$}}}
\put(881,424){\raisebox{-.8pt}{\makebox(0,0){$\cdot$}}}
\put(884,418){\raisebox{-.8pt}{\makebox(0,0){$\cdot$}}}
\put(887,421){\raisebox{-.8pt}{\makebox(0,0){$\cdot$}}}
\put(891,432){\raisebox{-.8pt}{\makebox(0,0){$\cdot$}}}
\put(894,424){\raisebox{-.8pt}{\makebox(0,0){$\cdot$}}}
\put(897,432){\raisebox{-.8pt}{\makebox(0,0){$\cdot$}}}
\put(900,435){\raisebox{-.8pt}{\makebox(0,0){$\cdot$}}}
\put(903,440){\raisebox{-.8pt}{\makebox(0,0){$\cdot$}}}
\put(906,441){\raisebox{-.8pt}{\makebox(0,0){$\cdot$}}}
\put(909,436){\raisebox{-.8pt}{\makebox(0,0){$\cdot$}}}
\put(912,450){\raisebox{-.8pt}{\makebox(0,0){$\cdot$}}}
\put(915,447){\raisebox{-.8pt}{\makebox(0,0){$\cdot$}}}
\put(918,458){\raisebox{-.8pt}{\makebox(0,0){$\cdot$}}}
\put(921,465){\raisebox{-.8pt}{\makebox(0,0){$\cdot$}}}
\put(924,456){\raisebox{-.8pt}{\makebox(0,0){$\cdot$}}}
\put(927,461){\raisebox{-.8pt}{\makebox(0,0){$\cdot$}}}
\put(930,477){\raisebox{-.8pt}{\makebox(0,0){$\cdot$}}}
\put(933,476){\raisebox{-.8pt}{\makebox(0,0){$\cdot$}}}
\put(936,480){\raisebox{-.8pt}{\makebox(0,0){$\cdot$}}}
\put(939,497){\raisebox{-.8pt}{\makebox(0,0){$\cdot$}}}
\put(942,519){\raisebox{-.8pt}{\makebox(0,0){$\cdot$}}}
\put(945,536){\raisebox{-.8pt}{\makebox(0,0){$\cdot$}}}
\put(948,535){\raisebox{-.8pt}{\makebox(0,0){$\cdot$}}}
\put(951,517){\raisebox{-.8pt}{\makebox(0,0){$\cdot$}}}
\put(955,492){\raisebox{-.8pt}{\makebox(0,0){$\cdot$}}}
\put(958,505){\raisebox{-.8pt}{\makebox(0,0){$\cdot$}}}
\put(961,509){\raisebox{-.8pt}{\makebox(0,0){$\cdot$}}}
\put(964,499){\raisebox{-.8pt}{\makebox(0,0){$\cdot$}}}
\put(967,492){\raisebox{-.8pt}{\makebox(0,0){$\cdot$}}}
\put(970,508){\raisebox{-.8pt}{\makebox(0,0){$\cdot$}}}
\put(973,527){\raisebox{-.8pt}{\makebox(0,0){$\cdot$}}}
\put(976,530){\raisebox{-.8pt}{\makebox(0,0){$\cdot$}}}
\put(979,522){\raisebox{-.8pt}{\makebox(0,0){$\cdot$}}}
\put(982,530){\raisebox{-.8pt}{\makebox(0,0){$\cdot$}}}
\put(985,536){\raisebox{-.8pt}{\makebox(0,0){$\cdot$}}}
\put(988,534){\raisebox{-.8pt}{\makebox(0,0){$\cdot$}}}
\put(991,547){\raisebox{-.8pt}{\makebox(0,0){$\cdot$}}}
\put(994,551){\raisebox{-.8pt}{\makebox(0,0){$\cdot$}}}
\put(997,548){\raisebox{-.8pt}{\makebox(0,0){$\cdot$}}}
\put(1000,549){\raisebox{-.8pt}{\makebox(0,0){$\cdot$}}}
\put(1003,557){\raisebox{-.8pt}{\makebox(0,0){$\cdot$}}}
\put(1006,556){\raisebox{-.8pt}{\makebox(0,0){$\cdot$}}}
\put(1009,561){\raisebox{-.8pt}{\makebox(0,0){$\cdot$}}}
\put(1012,573){\raisebox{-.8pt}{\makebox(0,0){$\cdot$}}}
\put(1015,574){\raisebox{-.8pt}{\makebox(0,0){$\cdot$}}}
\put(1019,581){\raisebox{-.8pt}{\makebox(0,0){$\cdot$}}}
\put(1022,592){\raisebox{-.8pt}{\makebox(0,0){$\cdot$}}}
\put(1025,594){\raisebox{-.8pt}{\makebox(0,0){$\cdot$}}}
\put(1028,587){\raisebox{-.8pt}{\makebox(0,0){$\cdot$}}}
\put(1031,582){\raisebox{-.8pt}{\makebox(0,0){$\cdot$}}}
\put(1034,578){\raisebox{-.8pt}{\makebox(0,0){$\cdot$}}}
\put(1037,575){\raisebox{-.8pt}{\makebox(0,0){$\cdot$}}}
\put(1040,577){\raisebox{-.8pt}{\makebox(0,0){$\cdot$}}}
\put(1043,570){\raisebox{-.8pt}{\makebox(0,0){$\cdot$}}}
\put(1046,579){\raisebox{-.8pt}{\makebox(0,0){$\cdot$}}}
\put(1049,567){\raisebox{-.8pt}{\makebox(0,0){$\cdot$}}}
\put(1052,562){\raisebox{-.8pt}{\makebox(0,0){$\cdot$}}}
\put(1055,558){\raisebox{-.8pt}{\makebox(0,0){$\cdot$}}}
\put(1058,570){\raisebox{-.8pt}{\makebox(0,0){$\cdot$}}}
\put(1061,571){\raisebox{-.8pt}{\makebox(0,0){$\cdot$}}}
\put(1064,581){\raisebox{-.8pt}{\makebox(0,0){$\cdot$}}}
\put(1067,584){\raisebox{-.8pt}{\makebox(0,0){$\cdot$}}}
\put(1070,591){\raisebox{-.8pt}{\makebox(0,0){$\cdot$}}}
\put(1073,595){\raisebox{-.8pt}{\makebox(0,0){$\cdot$}}}
\put(1076,596){\raisebox{-.8pt}{\makebox(0,0){$\cdot$}}}
\put(1079,598){\raisebox{-.8pt}{\makebox(0,0){$\cdot$}}}
\put(1083,605){\raisebox{-.8pt}{\makebox(0,0){$\cdot$}}}
\put(1086,604){\raisebox{-.8pt}{\makebox(0,0){$\cdot$}}}
\put(1089,605){\raisebox{-.8pt}{\makebox(0,0){$\cdot$}}}
\put(1092,612){\raisebox{-.8pt}{\makebox(0,0){$\cdot$}}}
\put(1095,605){\raisebox{-.8pt}{\makebox(0,0){$\cdot$}}}
\put(1098,610){\raisebox{-.8pt}{\makebox(0,0){$\cdot$}}}
\put(1101,617){\raisebox{-.8pt}{\makebox(0,0){$\cdot$}}}
\put(1104,619){\raisebox{-.8pt}{\makebox(0,0){$\cdot$}}}
\put(1107,628){\raisebox{-.8pt}{\makebox(0,0){$\cdot$}}}
\put(1110,628){\raisebox{-.8pt}{\makebox(0,0){$\cdot$}}}
\put(1113,624){\raisebox{-.8pt}{\makebox(0,0){$\cdot$}}}
\put(1116,629){\raisebox{-.8pt}{\makebox(0,0){$\cdot$}}}
\put(1119,628){\raisebox{-.8pt}{\makebox(0,0){$\cdot$}}}
\put(1122,619){\raisebox{-.8pt}{\makebox(0,0){$\cdot$}}}
\put(1125,606){\raisebox{-.8pt}{\makebox(0,0){$\cdot$}}}
\put(1128,619){\raisebox{-.8pt}{\makebox(0,0){$\cdot$}}}
\put(1131,621){\raisebox{-.8pt}{\makebox(0,0){$\cdot$}}}
\put(1134,642){\raisebox{-.8pt}{\makebox(0,0){$\cdot$}}}
\put(1137,641){\raisebox{-.8pt}{\makebox(0,0){$\cdot$}}}
\put(1140,628){\raisebox{-.8pt}{\makebox(0,0){$\cdot$}}}
\put(1143,623){\raisebox{-.8pt}{\makebox(0,0){$\cdot$}}}
\put(1147,617){\raisebox{-.8pt}{\makebox(0,0){$\cdot$}}}
\put(1150,622){\raisebox{-.8pt}{\makebox(0,0){$\cdot$}}}
\put(1153,616){\raisebox{-.8pt}{\makebox(0,0){$\cdot$}}}
\put(1156,619){\raisebox{-.8pt}{\makebox(0,0){$\cdot$}}}
\put(1159,618){\raisebox{-.8pt}{\makebox(0,0){$\cdot$}}}
\put(1162,608){\raisebox{-.8pt}{\makebox(0,0){$\cdot$}}}
\put(1165,603){\raisebox{-.8pt}{\makebox(0,0){$\cdot$}}}
\put(1168,599){\raisebox{-.8pt}{\makebox(0,0){$\cdot$}}}
\put(1171,598){\raisebox{-.8pt}{\makebox(0,0){$\cdot$}}}
\put(1174,592){\raisebox{-.8pt}{\makebox(0,0){$\cdot$}}}
\put(1177,573){\raisebox{-.8pt}{\makebox(0,0){$\cdot$}}}
\put(1180,585){\raisebox{-.8pt}{\makebox(0,0){$\cdot$}}}
\put(1183,582){\raisebox{-.8pt}{\makebox(0,0){$\cdot$}}}
\put(1186,584){\raisebox{-.8pt}{\makebox(0,0){$\cdot$}}}
\put(1189,594){\raisebox{-.8pt}{\makebox(0,0){$\cdot$}}}
\put(1192,598){\raisebox{-.8pt}{\makebox(0,0){$\cdot$}}}
\put(1195,604){\raisebox{-.8pt}{\makebox(0,0){$\cdot$}}}
\put(1198,607){\raisebox{-.8pt}{\makebox(0,0){$\cdot$}}}
\put(1201,618){\raisebox{-.8pt}{\makebox(0,0){$\cdot$}}}
\put(1204,611){\raisebox{-.8pt}{\makebox(0,0){$\cdot$}}}
\put(1207,608){\raisebox{-.8pt}{\makebox(0,0){$\cdot$}}}
\put(1211,614){\raisebox{-.8pt}{\makebox(0,0){$\cdot$}}}
\put(1214,612){\raisebox{-.8pt}{\makebox(0,0){$\cdot$}}}
\put(1217,628){\raisebox{-.8pt}{\makebox(0,0){$\cdot$}}}
\put(1220,627){\raisebox{-.8pt}{\makebox(0,0){$\cdot$}}}
\put(1223,630){\raisebox{-.8pt}{\makebox(0,0){$\cdot$}}}
\put(1226,631){\raisebox{-.8pt}{\makebox(0,0){$\cdot$}}}
\put(1229,626){\raisebox{-.8pt}{\makebox(0,0){$\cdot$}}}
\put(1232,628){\raisebox{-.8pt}{\makebox(0,0){$\cdot$}}}
\put(1235,643){\raisebox{-.8pt}{\makebox(0,0){$\cdot$}}}
\put(1238,640){\raisebox{-.8pt}{\makebox(0,0){$\cdot$}}}
\put(1241,634){\raisebox{-.8pt}{\makebox(0,0){$\cdot$}}}
\put(1244,634){\raisebox{-.8pt}{\makebox(0,0){$\cdot$}}}
\put(1247,623){\raisebox{-.8pt}{\makebox(0,0){$\cdot$}}}
\put(1250,626){\raisebox{-.8pt}{\makebox(0,0){$\cdot$}}}
\put(1253,636){\raisebox{-.8pt}{\makebox(0,0){$\cdot$}}}
\put(1256,648){\raisebox{-.8pt}{\makebox(0,0){$\cdot$}}}
\put(1259,644){\raisebox{-.8pt}{\makebox(0,0){$\cdot$}}}
\put(1262,650){\raisebox{-.8pt}{\makebox(0,0){$\cdot$}}}
\put(1265,672){\raisebox{-.8pt}{\makebox(0,0){$\cdot$}}}
\put(1268,677){\raisebox{-.8pt}{\makebox(0,0){$\cdot$}}}
\put(1271,687){\raisebox{-.8pt}{\makebox(0,0){$\cdot$}}}
\put(1275,692){\raisebox{-.8pt}{\makebox(0,0){$\cdot$}}}
\put(1278,698){\raisebox{-.8pt}{\makebox(0,0){$\cdot$}}}
\put(1281,699){\raisebox{-.8pt}{\makebox(0,0){$\cdot$}}}
\put(1284,701){\raisebox{-.8pt}{\makebox(0,0){$\cdot$}}}
\put(1287,706){\raisebox{-.8pt}{\makebox(0,0){$\cdot$}}}
\put(1290,700){\raisebox{-.8pt}{\makebox(0,0){$\cdot$}}}
\put(1293,706){\raisebox{-.8pt}{\makebox(0,0){$\cdot$}}}
\put(1296,710){\raisebox{-.8pt}{\makebox(0,0){$\cdot$}}}
\put(1299,711){\raisebox{-.8pt}{\makebox(0,0){$\cdot$}}}
\put(1302,715){\raisebox{-.8pt}{\makebox(0,0){$\cdot$}}}
\put(1305,736){\raisebox{-.8pt}{\makebox(0,0){$\cdot$}}}
\put(1308,743){\raisebox{-.8pt}{\makebox(0,0){$\cdot$}}}
\put(1311,746){\raisebox{-.8pt}{\makebox(0,0){$\cdot$}}}
\put(1314,732){\raisebox{-.8pt}{\makebox(0,0){$\cdot$}}}
\put(1317,742){\raisebox{-.8pt}{\makebox(0,0){$\cdot$}}}
\put(1320,741){\raisebox{-.8pt}{\makebox(0,0){$\cdot$}}}
\put(1323,743){\raisebox{-.8pt}{\makebox(0,0){$\cdot$}}}
\put(1326,741){\raisebox{-.8pt}{\makebox(0,0){$\cdot$}}}
\put(1329,737){\raisebox{-.8pt}{\makebox(0,0){$\cdot$}}}
\put(1332,729){\raisebox{-.8pt}{\makebox(0,0){$\cdot$}}}
\put(1335,737){\raisebox{-.8pt}{\makebox(0,0){$\cdot$}}}
\put(1339,738){\raisebox{-.8pt}{\makebox(0,0){$\cdot$}}}
\put(1342,736){\raisebox{-.8pt}{\makebox(0,0){$\cdot$}}}
\put(1345,740){\raisebox{-.8pt}{\makebox(0,0){$\cdot$}}}
\put(1348,740){\raisebox{-.8pt}{\makebox(0,0){$\cdot$}}}
\put(1351,734){\raisebox{-.8pt}{\makebox(0,0){$\cdot$}}}
\put(1354,730){\raisebox{-.8pt}{\makebox(0,0){$\cdot$}}}
\put(1357,735){\raisebox{-.8pt}{\makebox(0,0){$\cdot$}}}
\put(1360,738){\raisebox{-.8pt}{\makebox(0,0){$\cdot$}}}
\put(1363,747){\raisebox{-.8pt}{\makebox(0,0){$\cdot$}}}
\put(1366,743){\raisebox{-.8pt}{\makebox(0,0){$\cdot$}}}
\put(1369,754){\raisebox{-.8pt}{\makebox(0,0){$\cdot$}}}
\put(1372,759){\raisebox{-.8pt}{\makebox(0,0){$\cdot$}}}
\put(1375,771){\raisebox{-.8pt}{\makebox(0,0){$\cdot$}}}
\put(1378,792){\raisebox{-.8pt}{\makebox(0,0){$\cdot$}}}
\put(1381,793){\raisebox{-.8pt}{\makebox(0,0){$\cdot$}}}
\put(1384,801){\raisebox{-.8pt}{\makebox(0,0){$\cdot$}}}
\put(1387,798){\raisebox{-.8pt}{\makebox(0,0){$\cdot$}}}
\put(1390,803){\raisebox{-.8pt}{\makebox(0,0){$\cdot$}}}
\put(1393,822){\raisebox{-.8pt}{\makebox(0,0){$\cdot$}}}
\put(1396,813){\raisebox{-.8pt}{\makebox(0,0){$\cdot$}}}
\put(1399,796){\raisebox{-.8pt}{\makebox(0,0){$\cdot$}}}
\put(1403,800){\raisebox{-.8pt}{\makebox(0,0){$\cdot$}}}
\put(1406,804){\raisebox{-.8pt}{\makebox(0,0){$\cdot$}}}
\put(1409,797){\raisebox{-.8pt}{\makebox(0,0){$\cdot$}}}
\put(1412,788){\raisebox{-.8pt}{\makebox(0,0){$\cdot$}}}
\put(1415,784){\raisebox{-.8pt}{\makebox(0,0){$\cdot$}}}
\put(1418,774){\raisebox{-.8pt}{\makebox(0,0){$\cdot$}}}
\put(1421,770){\raisebox{-.8pt}{\makebox(0,0){$\cdot$}}}
\put(1424,774){\raisebox{-.8pt}{\makebox(0,0){$\cdot$}}}
\put(1427,769){\raisebox{-.8pt}{\makebox(0,0){$\cdot$}}}
\put(1430,775){\raisebox{-.8pt}{\makebox(0,0){$\cdot$}}}
\put(1433,769){\raisebox{-.8pt}{\makebox(0,0){$\cdot$}}}
\put(220,237){\usebox{\plotpoint}}
\multiput(220.00,235.94)(1.651,-0.468){5}{\rule{1.300pt}{0.113pt}}
\multiput(220.00,236.17)(9.302,-4.000){2}{\rule{0.650pt}{0.400pt}}
\multiput(232.00,231.95)(2.695,-0.447){3}{\rule{1.833pt}{0.108pt}}
\multiput(232.00,232.17)(9.195,-3.000){2}{\rule{0.917pt}{0.400pt}}
\put(245,228.17){\rule{2.500pt}{0.400pt}}
\multiput(245.00,229.17)(6.811,-2.000){2}{\rule{1.250pt}{0.400pt}}
\put(257,226.67){\rule{2.891pt}{0.400pt}}
\multiput(257.00,227.17)(6.000,-1.000){2}{\rule{1.445pt}{0.400pt}}
\put(281,226.67){\rule{3.132pt}{0.400pt}}
\multiput(281.00,226.17)(6.500,1.000){2}{\rule{1.566pt}{0.400pt}}
\put(294,228.17){\rule{2.500pt}{0.400pt}}
\multiput(294.00,227.17)(6.811,2.000){2}{\rule{1.250pt}{0.400pt}}
\multiput(306.00,230.61)(2.472,0.447){3}{\rule{1.700pt}{0.108pt}}
\multiput(306.00,229.17)(8.472,3.000){2}{\rule{0.850pt}{0.400pt}}
\multiput(318.00,233.59)(1.378,0.477){7}{\rule{1.140pt}{0.115pt}}
\multiput(318.00,232.17)(10.634,5.000){2}{\rule{0.570pt}{0.400pt}}
\multiput(331.00,238.59)(1.267,0.477){7}{\rule{1.060pt}{0.115pt}}
\multiput(331.00,237.17)(9.800,5.000){2}{\rule{0.530pt}{0.400pt}}
\multiput(343.00,243.59)(1.033,0.482){9}{\rule{0.900pt}{0.116pt}}
\multiput(343.00,242.17)(10.132,6.000){2}{\rule{0.450pt}{0.400pt}}
\multiput(355.00,249.59)(0.758,0.488){13}{\rule{0.700pt}{0.117pt}}
\multiput(355.00,248.17)(10.547,8.000){2}{\rule{0.350pt}{0.400pt}}
\multiput(367.00,257.59)(0.824,0.488){13}{\rule{0.750pt}{0.117pt}}
\multiput(367.00,256.17)(11.443,8.000){2}{\rule{0.375pt}{0.400pt}}
\multiput(380.00,265.59)(0.669,0.489){15}{\rule{0.633pt}{0.118pt}}
\multiput(380.00,264.17)(10.685,9.000){2}{\rule{0.317pt}{0.400pt}}
\multiput(392.00,274.58)(0.600,0.491){17}{\rule{0.580pt}{0.118pt}}
\multiput(392.00,273.17)(10.796,10.000){2}{\rule{0.290pt}{0.400pt}}
\multiput(404.00,284.58)(0.652,0.491){17}{\rule{0.620pt}{0.118pt}}
\multiput(404.00,283.17)(11.713,10.000){2}{\rule{0.310pt}{0.400pt}}
\multiput(417.00,294.58)(0.496,0.492){21}{\rule{0.500pt}{0.119pt}}
\multiput(417.00,293.17)(10.962,12.000){2}{\rule{0.250pt}{0.400pt}}
\multiput(429.00,306.58)(0.543,0.492){19}{\rule{0.536pt}{0.118pt}}
\multiput(429.00,305.17)(10.887,11.000){2}{\rule{0.268pt}{0.400pt}}
\multiput(441.00,317.58)(0.496,0.492){21}{\rule{0.500pt}{0.119pt}}
\multiput(441.00,316.17)(10.962,12.000){2}{\rule{0.250pt}{0.400pt}}
\multiput(453.00,329.58)(0.539,0.492){21}{\rule{0.533pt}{0.119pt}}
\multiput(453.00,328.17)(11.893,12.000){2}{\rule{0.267pt}{0.400pt}}
\multiput(466.00,341.58)(0.496,0.492){21}{\rule{0.500pt}{0.119pt}}
\multiput(466.00,340.17)(10.962,12.000){2}{\rule{0.250pt}{0.400pt}}
\multiput(478.00,353.58)(0.496,0.492){21}{\rule{0.500pt}{0.119pt}}
\multiput(478.00,352.17)(10.962,12.000){2}{\rule{0.250pt}{0.400pt}}
\multiput(490.00,365.58)(0.539,0.492){21}{\rule{0.533pt}{0.119pt}}
\multiput(490.00,364.17)(11.893,12.000){2}{\rule{0.267pt}{0.400pt}}
\multiput(503.00,377.58)(0.496,0.492){21}{\rule{0.500pt}{0.119pt}}
\multiput(503.00,376.17)(10.962,12.000){2}{\rule{0.250pt}{0.400pt}}
\multiput(515.00,389.58)(0.600,0.491){17}{\rule{0.580pt}{0.118pt}}
\multiput(515.00,388.17)(10.796,10.000){2}{\rule{0.290pt}{0.400pt}}
\multiput(527.00,399.58)(0.543,0.492){19}{\rule{0.536pt}{0.118pt}}
\multiput(527.00,398.17)(10.887,11.000){2}{\rule{0.268pt}{0.400pt}}
\multiput(539.00,410.59)(0.728,0.489){15}{\rule{0.678pt}{0.118pt}}
\multiput(539.00,409.17)(11.593,9.000){2}{\rule{0.339pt}{0.400pt}}
\multiput(552.00,419.59)(0.758,0.488){13}{\rule{0.700pt}{0.117pt}}
\multiput(552.00,418.17)(10.547,8.000){2}{\rule{0.350pt}{0.400pt}}
\multiput(564.00,427.59)(0.758,0.488){13}{\rule{0.700pt}{0.117pt}}
\multiput(564.00,426.17)(10.547,8.000){2}{\rule{0.350pt}{0.400pt}}
\multiput(576.00,435.59)(1.033,0.482){9}{\rule{0.900pt}{0.116pt}}
\multiput(576.00,434.17)(10.132,6.000){2}{\rule{0.450pt}{0.400pt}}
\multiput(588.00,441.59)(1.123,0.482){9}{\rule{0.967pt}{0.116pt}}
\multiput(588.00,440.17)(10.994,6.000){2}{\rule{0.483pt}{0.400pt}}
\multiput(601.00,447.60)(1.651,0.468){5}{\rule{1.300pt}{0.113pt}}
\multiput(601.00,446.17)(9.302,4.000){2}{\rule{0.650pt}{0.400pt}}
\put(613,451.17){\rule{2.500pt}{0.400pt}}
\multiput(613.00,450.17)(6.811,2.000){2}{\rule{1.250pt}{0.400pt}}
\put(625,453.17){\rule{2.700pt}{0.400pt}}
\multiput(625.00,452.17)(7.396,2.000){2}{\rule{1.350pt}{0.400pt}}
\put(638,454.67){\rule{2.891pt}{0.400pt}}
\multiput(638.00,454.17)(6.000,1.000){2}{\rule{1.445pt}{0.400pt}}
\put(650,454.67){\rule{2.891pt}{0.400pt}}
\multiput(650.00,455.17)(6.000,-1.000){2}{\rule{1.445pt}{0.400pt}}
\put(662,453.17){\rule{2.500pt}{0.400pt}}
\multiput(662.00,454.17)(6.811,-2.000){2}{\rule{1.250pt}{0.400pt}}
\put(674,451.17){\rule{2.700pt}{0.400pt}}
\multiput(674.00,452.17)(7.396,-2.000){2}{\rule{1.350pt}{0.400pt}}
\multiput(687.00,449.94)(1.651,-0.468){5}{\rule{1.300pt}{0.113pt}}
\multiput(687.00,450.17)(9.302,-4.000){2}{\rule{0.650pt}{0.400pt}}
\multiput(699.00,445.94)(1.651,-0.468){5}{\rule{1.300pt}{0.113pt}}
\multiput(699.00,446.17)(9.302,-4.000){2}{\rule{0.650pt}{0.400pt}}
\multiput(711.00,441.94)(1.797,-0.468){5}{\rule{1.400pt}{0.113pt}}
\multiput(711.00,442.17)(10.094,-4.000){2}{\rule{0.700pt}{0.400pt}}
\multiput(724.00,437.93)(1.267,-0.477){7}{\rule{1.060pt}{0.115pt}}
\multiput(724.00,438.17)(9.800,-5.000){2}{\rule{0.530pt}{0.400pt}}
\multiput(736.00,432.93)(1.267,-0.477){7}{\rule{1.060pt}{0.115pt}}
\multiput(736.00,433.17)(9.800,-5.000){2}{\rule{0.530pt}{0.400pt}}
\multiput(748.00,427.94)(1.651,-0.468){5}{\rule{1.300pt}{0.113pt}}
\multiput(748.00,428.17)(9.302,-4.000){2}{\rule{0.650pt}{0.400pt}}
\multiput(760.00,423.93)(1.378,-0.477){7}{\rule{1.140pt}{0.115pt}}
\multiput(760.00,424.17)(10.634,-5.000){2}{\rule{0.570pt}{0.400pt}}
\multiput(773.00,418.95)(2.472,-0.447){3}{\rule{1.700pt}{0.108pt}}
\multiput(773.00,419.17)(8.472,-3.000){2}{\rule{0.850pt}{0.400pt}}
\multiput(785.00,415.95)(2.472,-0.447){3}{\rule{1.700pt}{0.108pt}}
\multiput(785.00,416.17)(8.472,-3.000){2}{\rule{0.850pt}{0.400pt}}
\put(797,412.17){\rule{2.700pt}{0.400pt}}
\multiput(797.00,413.17)(7.396,-2.000){2}{\rule{1.350pt}{0.400pt}}
\put(810,410.67){\rule{2.891pt}{0.400pt}}
\multiput(810.00,411.17)(6.000,-1.000){2}{\rule{1.445pt}{0.400pt}}
\put(269.0,227.0){\rule[-0.200pt]{2.891pt}{0.400pt}}
\put(834,411.17){\rule{2.500pt}{0.400pt}}
\multiput(834.00,410.17)(6.811,2.000){2}{\rule{1.250pt}{0.400pt}}
\multiput(846.00,413.61)(2.695,0.447){3}{\rule{1.833pt}{0.108pt}}
\multiput(846.00,412.17)(9.195,3.000){2}{\rule{0.917pt}{0.400pt}}
\multiput(859.00,416.59)(1.267,0.477){7}{\rule{1.060pt}{0.115pt}}
\multiput(859.00,415.17)(9.800,5.000){2}{\rule{0.530pt}{0.400pt}}
\multiput(871.00,421.59)(0.874,0.485){11}{\rule{0.786pt}{0.117pt}}
\multiput(871.00,420.17)(10.369,7.000){2}{\rule{0.393pt}{0.400pt}}
\multiput(883.00,428.59)(0.824,0.488){13}{\rule{0.750pt}{0.117pt}}
\multiput(883.00,427.17)(11.443,8.000){2}{\rule{0.375pt}{0.400pt}}
\multiput(896.00,436.59)(0.669,0.489){15}{\rule{0.633pt}{0.118pt}}
\multiput(896.00,435.17)(10.685,9.000){2}{\rule{0.317pt}{0.400pt}}
\multiput(908.00,445.58)(0.543,0.492){19}{\rule{0.536pt}{0.118pt}}
\multiput(908.00,444.17)(10.887,11.000){2}{\rule{0.268pt}{0.400pt}}
\multiput(920.00,456.58)(0.496,0.492){21}{\rule{0.500pt}{0.119pt}}
\multiput(920.00,455.17)(10.962,12.000){2}{\rule{0.250pt}{0.400pt}}
\multiput(932.58,468.00)(0.493,0.536){23}{\rule{0.119pt}{0.531pt}}
\multiput(931.17,468.00)(13.000,12.898){2}{\rule{0.400pt}{0.265pt}}
\multiput(945.58,482.00)(0.492,0.539){21}{\rule{0.119pt}{0.533pt}}
\multiput(944.17,482.00)(12.000,11.893){2}{\rule{0.400pt}{0.267pt}}
\multiput(957.58,495.00)(0.492,0.625){21}{\rule{0.119pt}{0.600pt}}
\multiput(956.17,495.00)(12.000,13.755){2}{\rule{0.400pt}{0.300pt}}
\multiput(969.58,510.00)(0.493,0.536){23}{\rule{0.119pt}{0.531pt}}
\multiput(968.17,510.00)(13.000,12.898){2}{\rule{0.400pt}{0.265pt}}
\multiput(982.58,524.00)(0.492,0.625){21}{\rule{0.119pt}{0.600pt}}
\multiput(981.17,524.00)(12.000,13.755){2}{\rule{0.400pt}{0.300pt}}
\multiput(994.58,539.00)(0.492,0.539){21}{\rule{0.119pt}{0.533pt}}
\multiput(993.17,539.00)(12.000,11.893){2}{\rule{0.400pt}{0.267pt}}
\multiput(1006.58,552.00)(0.492,0.539){21}{\rule{0.119pt}{0.533pt}}
\multiput(1005.17,552.00)(12.000,11.893){2}{\rule{0.400pt}{0.267pt}}
\multiput(1018.00,565.58)(0.539,0.492){21}{\rule{0.533pt}{0.119pt}}
\multiput(1018.00,564.17)(11.893,12.000){2}{\rule{0.267pt}{0.400pt}}
\multiput(1031.00,577.58)(0.543,0.492){19}{\rule{0.536pt}{0.118pt}}
\multiput(1031.00,576.17)(10.887,11.000){2}{\rule{0.268pt}{0.400pt}}
\multiput(1043.00,588.59)(0.758,0.488){13}{\rule{0.700pt}{0.117pt}}
\multiput(1043.00,587.17)(10.547,8.000){2}{\rule{0.350pt}{0.400pt}}
\multiput(1055.00,596.59)(0.950,0.485){11}{\rule{0.843pt}{0.117pt}}
\multiput(1055.00,595.17)(11.251,7.000){2}{\rule{0.421pt}{0.400pt}}
\multiput(1068.00,603.59)(1.267,0.477){7}{\rule{1.060pt}{0.115pt}}
\multiput(1068.00,602.17)(9.800,5.000){2}{\rule{0.530pt}{0.400pt}}
\multiput(1080.00,608.60)(1.651,0.468){5}{\rule{1.300pt}{0.113pt}}
\multiput(1080.00,607.17)(9.302,4.000){2}{\rule{0.650pt}{0.400pt}}
\put(1092,611.67){\rule{2.891pt}{0.400pt}}
\multiput(1092.00,611.17)(6.000,1.000){2}{\rule{1.445pt}{0.400pt}}
\put(822.0,411.0){\rule[-0.200pt]{2.891pt}{0.400pt}}
\put(1117,611.67){\rule{2.891pt}{0.400pt}}
\multiput(1117.00,612.17)(6.000,-1.000){2}{\rule{1.445pt}{0.400pt}}
\put(1129,610.67){\rule{2.891pt}{0.400pt}}
\multiput(1129.00,611.17)(6.000,-1.000){2}{\rule{1.445pt}{0.400pt}}
\put(1141,609.17){\rule{2.500pt}{0.400pt}}
\multiput(1141.00,610.17)(6.811,-2.000){2}{\rule{1.250pt}{0.400pt}}
\put(1153,607.67){\rule{3.132pt}{0.400pt}}
\multiput(1153.00,608.17)(6.500,-1.000){2}{\rule{1.566pt}{0.400pt}}
\put(1104.0,613.0){\rule[-0.200pt]{3.132pt}{0.400pt}}
\put(1178,608.17){\rule{2.500pt}{0.400pt}}
\multiput(1178.00,607.17)(6.811,2.000){2}{\rule{1.250pt}{0.400pt}}
\multiput(1190.00,610.60)(1.797,0.468){5}{\rule{1.400pt}{0.113pt}}
\multiput(1190.00,609.17)(10.094,4.000){2}{\rule{0.700pt}{0.400pt}}
\multiput(1203.00,614.59)(0.874,0.485){11}{\rule{0.786pt}{0.117pt}}
\multiput(1203.00,613.17)(10.369,7.000){2}{\rule{0.393pt}{0.400pt}}
\multiput(1215.00,621.58)(0.600,0.491){17}{\rule{0.580pt}{0.118pt}}
\multiput(1215.00,620.17)(10.796,10.000){2}{\rule{0.290pt}{0.400pt}}
\multiput(1227.58,631.00)(0.492,0.539){21}{\rule{0.119pt}{0.533pt}}
\multiput(1226.17,631.00)(12.000,11.893){2}{\rule{0.400pt}{0.267pt}}
\multiput(1239.58,644.00)(0.493,0.576){23}{\rule{0.119pt}{0.562pt}}
\multiput(1238.17,644.00)(13.000,13.834){2}{\rule{0.400pt}{0.281pt}}
\multiput(1252.58,659.00)(0.492,0.669){21}{\rule{0.119pt}{0.633pt}}
\multiput(1251.17,659.00)(12.000,14.685){2}{\rule{0.400pt}{0.317pt}}
\multiput(1264.58,675.00)(0.492,0.712){21}{\rule{0.119pt}{0.667pt}}
\multiput(1263.17,675.00)(12.000,15.616){2}{\rule{0.400pt}{0.333pt}}
\multiput(1276.58,692.00)(0.493,0.616){23}{\rule{0.119pt}{0.592pt}}
\multiput(1275.17,692.00)(13.000,14.771){2}{\rule{0.400pt}{0.296pt}}
\multiput(1289.00,708.58)(0.496,0.492){21}{\rule{0.500pt}{0.119pt}}
\multiput(1289.00,707.17)(10.962,12.000){2}{\rule{0.250pt}{0.400pt}}
\multiput(1301.00,720.59)(0.669,0.489){15}{\rule{0.633pt}{0.118pt}}
\multiput(1301.00,719.17)(10.685,9.000){2}{\rule{0.317pt}{0.400pt}}
\multiput(1313.00,729.60)(1.651,0.468){5}{\rule{1.300pt}{0.113pt}}
\multiput(1313.00,728.17)(9.302,4.000){2}{\rule{0.650pt}{0.400pt}}
\put(1325,732.67){\rule{3.132pt}{0.400pt}}
\multiput(1325.00,732.17)(6.500,1.000){2}{\rule{1.566pt}{0.400pt}}
\put(1166.0,608.0){\rule[-0.200pt]{2.891pt}{0.400pt}}
\multiput(1350.00,734.60)(1.651,0.468){5}{\rule{1.300pt}{0.113pt}}
\multiput(1350.00,733.17)(9.302,4.000){2}{\rule{0.650pt}{0.400pt}}
\multiput(1362.00,738.58)(0.539,0.492){21}{\rule{0.533pt}{0.119pt}}
\multiput(1362.00,737.17)(11.893,12.000){2}{\rule{0.267pt}{0.400pt}}
\multiput(1375.58,750.00)(0.492,0.798){21}{\rule{0.119pt}{0.733pt}}
\multiput(1374.17,750.00)(12.000,17.478){2}{\rule{0.400pt}{0.367pt}}
\multiput(1387.58,769.00)(0.492,0.669){21}{\rule{0.119pt}{0.633pt}}
\multiput(1386.17,769.00)(12.000,14.685){2}{\rule{0.400pt}{0.317pt}}
\multiput(1399.00,785.60)(1.651,0.468){5}{\rule{1.300pt}{0.113pt}}
\multiput(1399.00,784.17)(9.302,4.000){2}{\rule{0.650pt}{0.400pt}}
\multiput(1411.00,789.58)(0.652,0.491){17}{\rule{0.620pt}{0.118pt}}
\multiput(1411.00,788.17)(11.713,10.000){2}{\rule{0.310pt}{0.400pt}}
\multiput(1424.58,799.00)(0.492,0.539){21}{\rule{0.119pt}{0.533pt}}
\multiput(1423.17,799.00)(12.000,11.893){2}{\rule{0.400pt}{0.267pt}}
\put(1338.0,734.0){\rule[-0.200pt]{2.891pt}{0.400pt}}
\end{picture}